\documentclass[graybox, envcountchap]{svmult}

\usepackage{mathptmx}        
\usepackage{amsmath}
\usepackage{amssymb}
\usepackage{color}
\usepackage{helvet}          
\usepackage{courier}         
\usepackage{dirtree}

\usepackage{makeidx}        
\usepackage{graphicx}        
\usepackage{subfig}

\usepackage{multicol}        
\usepackage[bottom]{footmisc}

\usepackage{hyperref}        
\hypersetup{colorlinks=true,urlcolor=blue}

\usepackage[misc]{ifsym}

\makeindex             

\usepackage[normalem]{ulem}

\newcommand{\be}{\begin{equation}}
\newcommand{\ee}{\end{equation}}
\newcommand{\bea}{\begin{eqnarray}}
\newcommand{\eea}{\end{eqnarray}}

\begin{document}


\title{Black holes in asymptotically safe gravity and beyond}
\author{Astrid Eichhorn and Aaron Held}
\institute{Astrid Eichhorn (\Letter) \at CP3-Origins, University of Southern Denmark, Campusvej 55, 5230 Odense M, Denmark, \email{eichhorn@cp3.sdu.dk}
\and Aaron Held \at Theoretisch-Physikalisches Institut, Friedrich-Schiller-Universit\"at Jena,
Max-Wien-Platz 1, 07743 Jena, Germany, 
The Princeton Gravity Initiative, Jadwin Hall, Princeton University,
Princeton, New Jersey 08544, U.S.,
\email{aaron.held@uni-jena.de}}
%
%
\maketitle

\vspace*{-6em}

\abstract{
Asymptotically safe quantum gravity is an approach to quantum gravity that achieves formulates a standard quantum field theory for the metric. Therefore, even the deep quantum gravity regime, that is expected to determine the true structure of the core of black holes, is described by a spacetime metric.
\\
The essence of asymptotic safety lies in a new symmetry of the theory -- quantum scale symmetry -- which characterizes the short-distance regime of quantum gravity. It implies the absence of physical scales. Therefore, the Newton coupling, which corresponds to a scale, namely the Planck length, must vanish asymptotically in the short-distance regime. This implies a weakening of the gravitational interaction, from which a resolution of classical spacetime singularities can be expected.
\\
In practise, properties of black holes in asymptotically safe quantum gravity cannot yet be derived from first principles, but are constructed using a heuristic procedure known as Renormalization Group improvement. The resulting asymptotic-safety inspired black holes have been constructed both for vanishing and for nonvanishing spin parameter. They are characterized by (i) the absence of curvature singularities, (ii) a more compact event horizon and photon sphere, (iii) a second (inner) horizon even at vanishing spin and (iv) a cold remnant as a possible final product of the Hawking evaporation.
\\
Observations can start to constrain the quantum-gravity scale that can be treated as a free parameter in asymptotic-safety inspired black holes. 
For slowly-spinning black holes, constraints from the Event Horizon Telescope and X-ray observations can only constrain quantum-gravity scales far above the Planck length. In the limit of near-critical spin, asymptotic-safety inspired black holes may ``light up" in a way the next-generation Event Horizon Telescope may be sensitive to, even for a quantum-gravity scale equalling the Planck length. Finally, a connection to gravitational-wave observations of the ringdown phase can currently only be established under very strong theoretical assumptions, due to a lack of a dynamical equation to which asymptotic-safety inspired black holes constitute a solution.
}


\tableofcontents


%
%
\newpage

\section{Invitation: Black holes, quantum scale symmetry and probes of quantum gravity}
In General Relativity (GR), black holes harbor several unphysical properties: first, they contain a curvature singularity. The singularity renders black-hole spacetimes geodesically incomplete. Second, they contain a Cauchy horizon, unless their spin is set exactly to zero, which is not very likely to be astrophysically relevant. The Cauchy horizon leads to a breakdown of predictive evolution.
\\
Therefore, it is not a question whether General Relativity breaks down for black holes. It is only a question, where and how it breaks down and what it is substituted by.

In this chapter, we explore the possibility that an asymptotically safe quantum theory of gravity provides a fundamental description of gravity and we focus on potential consequences for black holes.  Asymptotically safe quantum gravity is an attractive contender for a quantum theory of gravity, because it stays as close as possible to GR in the sense that (i) the gravitational degrees of freedom are carried by the metric \footnote{Attempts at formulating asymptotically safe gravity in terms of, e.g., the vielbein \cite{Harst:2012ni}, the vielbein and the connection \cite{Daum:2010qt}, a unimodular metric \cite{Eichhorn:2013xr} or a generalized connection \cite{Gies:2022ikv} also exist; however, metric gravity is by far the most explored of these options, see \cite{Percacci:2017fkn,Reuter:2019byg} for textbooks and \cite{Eichhorn:2017egq,Pereira:2019dbn,Bonanno:2020bil,Reichert:2020mja,Pawlowski:2020qer,Eichhorn:2022jqj} for reviews. The other options all start from theories which are classically equivalent to GR.}, (ii) the rules of standard (quantum) field theories still apply and (iii) as a consequence of (i) and (ii), a continuum spacetime picture continues to hold, where spacetime is a differentiable, Lorentzian manifold with a metric. Therefore, it is expected that black holes exist in asymptotically safe gravity, instead of being substituted by black-hole mimickers without event horizon. This immediately gives rise to the following questions: (i) What is the fate of the classical curvature singularity? (ii) What is the fate of the Cauchy horizon? (iii) Are all effects of quantum gravity hidden behind the event horizon and therefore in principle undetectable? In short, as we will review in this chapter, the answers are: (i) the singularity may be resolved completely or weakened and geodesic completeness may be achieved; (ii) the Cauchy horizon's fate is unknown, but current models contain a Cauchy horizon, raising questions of instability; (iii) in asymptotically safe quantum gravity, singularity resolution goes hand in hand with effects at all curvature scales, with their size determined by an  undetermined parameter of the theory.  If this parameter, and thus the onset of observable deviations, are \emph{not} tied to the Planck scale, current astrophysical observations can start to place constraints on black-hole spacetimes motivated by asymptotically safe quantum gravity.

The main idea that this chapter relies on, is that quantum scale symmetry, which is realized in asymptotically safe gravity, ``turns off" the gravitational interaction at high curvature scales, resolving spacetime singularities.
In fact, asymptotic safety is a theory that fits well into a highly successful paradigm in fundamental physics, namely the use of symmetries. In asymptotic safety, quantum scale symmetry determines the high-curvature regime. Scale symmetry implies that there cannot be any distinct physical scales. Quantum scale symmetry is a bit more subtle, because it need not be realized at all scales: at low energy (or curvature) scales, quantum scale symmetry is absent, and distinct physical scales exist (e.g., masses of various elementary particles). Conversely, beyond a critical high energy scale, quantum scale symmetry is present, and no distinct physical scales can exist in that regime.\\
Classically, gravity comes with a scale, namely the Planck mass. In asymptotically safe gravity, the Planck mass is present at low curvature scales (where quantum scale symmetry is absent), but is absent at high curvature scales (where quantum scale symmetry is present). This translates into a dynamical weakening of gravity at high curvature scales, suggesting a resolution of spacetime singularities.
 \\

In Sec.~\ref{sec:weakening}, we open our discussion with a heuristic argument why quantum scale symmetry should lead to resolution or weakening of singularities. In Sec.~\ref{sec:construction} we then discuss, how asymptotic-safety inspired spacetimes are constructed through the method of Renormalization Group improvement. We also elaborate on ambiguities and potential pitfalls of the method. In Sec.~\ref{sec:spherical-BHs} we discuss the resulting spherically symmetric spacetimes and in Sec.~\ref{sec:spinning-BHs} we focus on axisymmetric, asymptotic-safety inspired spacetimes, highlighting universal results, such as singularity resolution, which are independent of some of the ambiguities of the Renormalization Group improvement. In Sec.~\ref{sec:formation}, we go beyond stationary settings and discuss gravitational collapse, as well as the formation of black holes in high-energy scattering processes. Finally, in Sec.~\ref{sec:towards-observation} we connect to observations, opening with a discussion on the scale of quantum gravity, on which detectability of quantum-gravity imprints depends. In Sec.~\ref{sec:principledparameterized}, we broaden our viewpoint beyond asymptotic safety and discuss, how Renormalization Group improved black holes fit into the principled-parameterized approach to black holes beyond GR. In Sec.~\ref{sec:summary} we summarize the current state of the art and point out the open challenges and future perspectives of the topic.

We aim to be introductory and pedagogical, making this review suitable for non-experts. Thus, all sections also contain ``further reading" subsections, where we briefly point out literature that goes beyond our discussion.

\section{A heuristic argument for singularity resolution from quantum scale symmetry}\label{sec:weakening}

\emph{...where we discuss how quantum scale symmetry, because it implies the absence of physical scales, forces the Newton coupling to vanish or the Planck mass to diverge at asymptotically small distances. The resulting weakening of gravity is expected to resolve (or at least weaken) classical curvature singularities. This can be viewed as a consequence of the dynamical decoupling of transplanckian degrees of freedom, which is expected to occur if quantum scale symmetry is realized in gravity.
}
\\

 \begin{figure}[!t]
\begin{center}
\includegraphics[width=\linewidth]{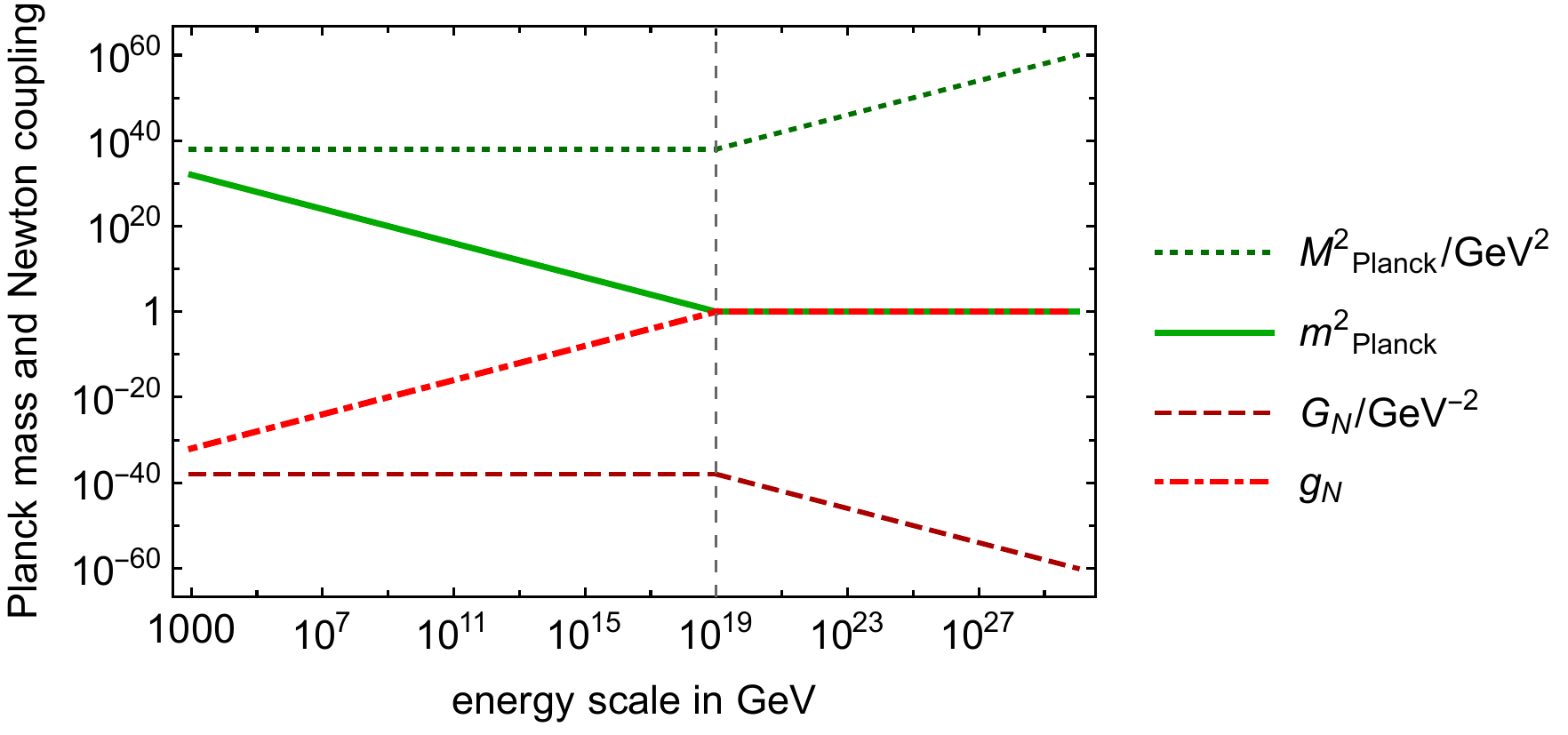}
\end{center}
\caption{\label{fig:scaling} We show the dimensionfull Planck mass $M_{\rm Planck}$, which is constant in the classical gravity regime below $10^{19}\, \rm GeV$, and scales quadratically above. The dimensionless Planck mass $m_{\rm Planck}$ becomes constant there, as does the dimensionless Newton coupling $g_N$, whereas the dimensionfull Newton coupling $G_N$ decreases quadratically. Thus, quantum scale symmetry translates into a dynamical decoupling of gravity, because the associated mass scale diverges and the corresponding interaction vanishes. As a consequence, gravitational modes beyond the classical Planck scale are dynamically removed from the theory. We argue that therefore curvature invariants are limited in asymptotically safe gravity and the limiting value is determined by the classical Planck scale, which is the transition scale to the quantum scale invariant regime.}
\end{figure}

\subsubsection*{Scale symmetry and the absence of scales}
Quantum scale symmetry is a form of scale symmetry. Scale symmetry says that no distinct physical scales can exist, i.e., the theory must be invariant under scale transformations, which is a form of self-similarity. Starting off with a theory that contains distinct physical mass scales, i.e., which is not scale-symmetric, one can arrive at scale symmetry in two distinct ways: one can either set all masses to zero or to infinity.
The difference between classical and quantum scale symmetry is that in a theory with classical symmetry, all mass scales are set to zero. In contrast, in a theory with quantum scale symmetry, which does not feature classical scale symmetry, the dimensionless counterparts of mass scales are finite.\footnote{In technical language, we state here that a theory with quantum scale symmetry but without classical scale symmetry has at least one non-vanishing canonically relevant or irrelevant coupling and not just canonically marginal ones. Due to its canonical dimension, the coupling implicitly defines a mass scale.} Therefore, the dimensionfull masses diverge in the small-distance regime of a quantum scale symmetric theory, cf.~Fig.~\ref{fig:scaling}. Here, we already assume that the theory is not quantum scale symmetric at all scales, but that quantum scale symmetry only determines the short-distance behavior. In contrast, above a critical distance scale (or below a critical mass scale), the theory leaves the quantum scale symmetric regime and distinct physical scales (e.g., masses for elementary particles) emerge.\\
Quantum scale symmetry has consequences for the degrees of freedom of the theory: simply put, because mass scales diverge, one may expect degrees of freedom to decouple. If a theory exhibits quantum scale symmetry only above a critical mass scale, then only those degrees of freedom at higher energies decouple.\footnote{There are exceptions to this argument, namely when a theory has vanishing mass parameters and the diverging mass scale is related to a relevant interaction. In that case quantum scale symmetry corresponds to a strongly-coupled regime.}
 
\subsubsection*{Interlude: Terminology}
On a technical level, asymptotic safety is a statement about the scale-dependence of couplings in the action of a quantum field theory (QFT). Formally, this scale dependence is described by the flow with Renormalization Group (RG) scale $k$ and affects not just the Newton coupling $G_N$, but all possible couplings $C_i$ allowed by the fundamental symmetries of the action.
The scale-dependence of their dimensionless counterparts $c_i = C_i\,k^{-d_i}$ is governed by the $\beta$-functions of the QFT, i.e., by
\begin{align}
	\label{eq:betas}
	k\partial_k\,c_i(k)=\beta_{c_i}(\lbrace u_j\rbrace)\;.
\end{align}
Asymptotic safety occurs at fixed points of the RG flow corresponds at which
\begin{align}
	\beta_{c_i}(\lbrace u_{j\ast}\rbrace) = 0\;,
\end{align} 
hence the flow vanishes and quantum scale symmetry is achieved. If the respective QFT exhibits such an asymptotically safe scale-invariant regime at high energies, the limit $k\rightarrow\infty$ can be taken without divergences and the respective QFT is fundamental.\\
Of the couplings in the theory, only those that are RG relevant correspond to free parameters. Those are the couplings for which quantum fluctuations favor a departure from scale symmetry. If an asymptotically safe theory has only a finite number of such relevant parameters (for which there is a general argument), then the theory is UV finite with just a finite number of free parameters that determines the low-energy physics. This generalizes the perturbative notion of renormalizability (which is not available for Einstein gravity) to a nonperturbative form of renormalizability.

\subsubsection*{Quantum scale symmetry and weakening of gravity}
In gravity, this is relevant for the Newton coupling. Building a classically scale symmetric theory of the gravitational field, one would put the Planck mass to zero. In a quantum scale symmetric theory, the dimensionless Newton coupling $g_N$ becomes constant, as does the dimensionless Planck mass. The dimensionfull Newton coupling $G_N$ thus scales quadratically with the inverse distance, and the dimensionfull Planck mass scales quadratically with the distance, see Fig.~\ref{fig:scaling}. The transition scale to the scale-symmetric regime is expected to be the classical Planck mass, see~Sec.~\ref{sec:scale-of-qG} for potential caveats to this expectation.

A diverging Planck mass means that the gravitational interaction decouples. The same can be inferred from the Newton coupling going to zero: the gravitational interaction becomes weaker at lower distances, such that the small-distance limit of quantum gravity is one where all degrees of freedom decouple. This ensures that divergences in the theory, be they in scattering cross-sections or in spacetime curvature invariants, are softened or even removed completely. For scattering cross-sections, the idea was originally put forward by Weinberg, who suggested that asymptotic safety ensures finiteness of cross-sections.

In turn, quantum scale symmetry, or the decoupling of the high-energy degrees of freedom, has consequences for the form of the solutions: because high-energy modes are absent from the theory, there is a limit to the maximum value of curvature one can reach: if we decompose the metric into eigenmodes of the Laplacian (with respect to some background metric), then all those eigenmodes with eigenfrequencies above a critical value decouple dynamically, and only the low-frequency modes are still present. Accordingly, if we consider a spacetime, which we write in terms of the basis of eigenmodes of the Laplacian, quantum scale symmetry implies that the high-frequency modes must be removed. Therefore, the curvature cannot exceed a limiting value, which is set by the transition scale to the scale-symmetric regime.

\subsubsection*{Singularity resolution from destructive interference in the path integral}
There is a second, independent argument to support the expectation that asymptotically safe quantum gravity lifts curvature singularities. The argument was first made in a cosmological context \cite{Lehners:2019ibe} and then adapted to the black-hole context \cite{Borissova:2020knn}. It is based on the Lorentzian path integral
\be
Z = \int \mathcal{D}g_{\mu\nu}\, e^{i S_{\rm AS}[g_{\mu\nu}]},
\ee
in which spacetime configurations interfere destructively with ``nearby" configurations, if their action diverges. It is expected that a Riemann-squared (or, equivalently, Weyl-squared) term (as well as higher-order terms in the Riemann tensor) are present in $S_{\rm AS}$, see, e.g., \cite{Benedetti:2009rx,Gies:2016con,Knorr:2021slg,Falls:2020qhj}. These terms diverge when evaluated on classical, singular black-hole spacetimes. In turn, this results in destructive interference  of such singular spacetimes with ``neighboring" configurations.
Therefore, one may expect that singular black-hole geometries are not part of the asymptotically safe path integral.\footnote{Incidentally, this also implies that a ``folk theorem" about the violation of global symmetries from quantum gravity may not be applicable to asymptotically safe quantum gravity.} As a result, the spacetime structure of black holes in asymptotically safe quantum gravity may be expected to be regular.\\
In this argument, scale symmetry does not explicitly appear and one may thus wonder whether the two arguments we have given for singularity resolution are related. They are related, because the presence of higher-order curvature terms appear to be required by scale symmetry; a theory with just the Einstein-Hilbert action is not asymptotically safe, according to the state-of-the-art in the literature.

\subsection*{Further reading}

\begin{itemize} 
\item {\bf State of the art in asymptotically safe quantum gravity:}
\\
Recent reviews on asymptotically safe quantum gravity are \cite{Pereira:2019dbn,Eichhorn:2020mte,Reichert:2020mja,Pawlowski:2020qer,Bonanno:2020bil}.  
\cite{Eichhorn:2020mte,Reichert:2020mja} are lecture notes that serve as an introduction to the topic. The books \cite{Percacci:2017fkn, Reuter:2019byg} provide introductions to the topic and in-depth expositions of recent results.
\cite{Pereira:2019dbn} reviews asymptotically safe quantum gravity and explains the tentative connection to tensor models, which, if they have a universal continuum limit, could be in the same universality class, i.e., have a continuum limit that yields the physics of asymptotically safe gravity.
\cite{Pawlowski:2020qer} focuses on the distinction between the background metric and the fluctuation field, which is introduced by the gauge fixing as well as the infrared regulator term. 
\cite{Bonanno:2020bil} addresses open challenges of asymptotically safe gravity, also responding to the questions raised in \cite{Donoghue:2019clr}.
Asymptotically safe gravity-matter systems are reviewed in \cite{Eichhorn:2018yfc} and an update is provided in \cite{Eichhorn:2022jqj}.
\end{itemize}

\section{Constructing asymptotic-safety inspired spacetimes}
\label{sec:construction}
\emph{...where we explain how a classical theory or its solutions can be upgraded to asymptotic-safety inspired solutions by accounting for the scale-dependence of the theories' couplings. In this RG-improvement procedure, the key physics input is the identification of the RG scale -- which determines up to which length scale quantum effects are present -- with a scale of the system, e.g., a curvature scale.\\
Physically, RG improvement is based on a decoupling of modes in the path integral, that occurs, when physical scales act as a low-energy cutoff in the path integral.}
\\

Ideally, to derive the structure of black holes in asymptotic safety, the following program should be implemented: first, the asymptotically safe fixed point must be determined with sufficiently high precision and in a Lorentzian regime. Both of these points are open challenges, although progress towards precision is being made, see, e.g., \cite{Falls:2018ylp,Falls:2020qhj}, as well as progress towards Lorentzian signature, see, e.g., \cite{Fehre:2021eob}.
\\
Second, the effective action of the theory must be calculated, such that all quantum fluctuations are integrated out. The resulting effective action will contain various higher-order curvature operators, the coefficients of which are functions of the free parameters (the relevant couplings) of the theory.
\\ 
Third, the effective equations of motion must be derived by applying the variational principle to the effective action. Finally, black-hole solutions to these effective equations of motion must be found.

This program is currently not (yet) feasible.
Instead, a simpler program, based on the method of RG improvement, is implemented, which does not strictly derive the black-hole solutions in asymptotic safety. Instead, it gives rise to asymptotic-safety inspired black holes.
Generally, it is expected that these asymptotic-safety inspired black holes may capture some of the salient features of black holes in full asymptotic safety.

\subsection{Renormalization-Group improvement and the decoupling mechanism}
\label{sec:flat-space-RG-improvement}

\emph{...where we review two classic flat-space examples of RG improvement.}
\\

RG improvement is a method in which a classical solution of the theory is taken; the coupling constants in that solution are replaced by their running counterparts and the Renormalization Group scale is replaced by a physical scale of the solution that is studied.

The expectation that RG improvement captures the effects of a scale-invariant regime in quantum gravity is based on the success of RG improvement in flat-space quantum field theories, which we briefly review here.

The method of RG improvement has been applied to many field theories. The case of massless quantum electrodynamics is particularly instructive and has, therefore, been used in~\cite{Bonanno:1998ye} as an analogy to motivate the first application of RG-improvement in the context of black-hole spacetimes. In this case, the electromagnetic coupling $e$ is the analogue of the Newton coupling $G_N$. The Coulomb potential 
\be
V_\text{cl} = e^2/(4\pi r)
\ee
serves as a simple flat-space analog of the gravitational potential.
First,
one replaces the electromagnetic coupling $e$ by its RG-scale dependent counterpart (evaluated at one loop in perturbation theory)
\be
e^2(k) =e^2(k_0) \frac{1}{1-\frac{e^2(k_0)}{6\pi^2} \ln\left(k/k_0 \right)},
\ee
where $e^2(k_0)$ is the low-energy value and $k_0$ an infrared scale, corresponding to an inverse distance $k_0 = r_0^{-1}$.
Next, the RG scale is identified with the only physical scale in the classical potential, which is the distance $r$ between the two charged particles, i.e., $k\sim 1/r$. As a result, one recovers the well-known Uehling correction to the Coulomb potential, as obtained by a calculation of the one-loop effective potential~\cite{Uehling:1935},
\be
V_{\rm RG-improved}(r) = \frac{e^2(r_0^{-1})}{4\pi r} \left(1+ \frac{e^2(r_0^{-1})}{6\pi^2} \ln\left(\frac{r_0}{r} \right) + \mathcal{O}(e^4) \right),
\ee
where a series expansion in small $e^2(r_0^{-1})$ is done in the last step. 

Beyond this simple example, an extensively studied case is the RG improvement of the scalar potential \cite{Coleman:1973jx}, e.g., for the Higgs scalar in the Standard Model, but also beyond. It is an instructive case, because the RG improvement is done not in terms of some external, fixed scale, as for the electrodynamic potential. Instead, the field itself is used as a scale. This is much closer in spirit to what we will do in gravity, where the physical scale necessarily depends on the field, i.e., the metric.
Here, the classical scalar potential for the massless scalar $\phi$
\be
V_{\rm scalar\, cl}(\phi) =\frac{ \lambda} {4}\phi^4
\ee
is RG-improved in the same two-step procedure. First, the classical coupling constant $\lambda$ is replaced by its RG-scale dependent counterpart $\lambda(k)$, which at one loop reads
 \be
 \lambda(k) = \frac{\lambda(k_0)}{1-\frac{9}{8\pi^2}\ln\left(\frac{k}{k_0} \right)}.
 \ee
 Second, the RG-scale $k$ is identified with the field value itself, i.e., $k\sim\phi$. The result captures the leading-order logarithmic quantum corrections to the classical potential. The scale identification $k\sim\phi$ is unambiguous in the massless case, because the field $\phi$ is the only dimensionful quantity in the classical potential.
\\

In the above, we have used $k$ for the RG scale. This notation is geared towards the functional Renormalization Group, where the scale $k$ appears as an infrared cutoff in the path integral.\footnote{In the above, one-loop examples, the scale-dependence of couplings on $k$ agrees with that on $\mu$, the RG scale that appears in perturbative renormalization.} Field configurations with momenta below $k$ are thereby suppressed. Thus, $k$ initially serves as a ``book-keeping" scale that sorts the field configurations in the path integral and enables a ``step-wise" calculation of the path integral.
As exposed in detail in \cite{Borissova:2022mgd}, $k$ acquires a physical meaning in settings with physical IR scales. For instance, in the above example of the Coulomb potential, field configurations with momenta lower than the inverse distance between the charges are not among the virtual quanta that are being exchanged. Thus, $r^{-1} = k$ acts as an IR cutoff.
Similarly, in the case of the RG-improved Higgs potential, the propagator around any nonzero field value $\phi$ acquires an effective mass term $\sim \lambda \phi^2$, which suppresses quantum fluctuations with momenta lower than this mass.

More generally, the success of RG improvement is based on the decoupling of low-energy modes: if a physical scale $s$ is present that acts as an IR cutoff, then one does not need to evaluate the path integral to $k<s$, but can stop at $k \approx s$. The effect of quantum fluctuations is then encoded in the scale-dependence of the couplings on $k$, which is equated to $s$. Thereby, if $s$ is high, only UV modes contribute, whereas, if $s$ is low, most modes in the path integral contribute.
\\

To apply the method of RG improvement to gravitational solutions -- black holes in particular -- the starting point is no longer a scalar potential but a classical spacetime which is a solution to GR. This brings several added challenges, which are related to the fact that these solutions are typically expressed in a covariant form (i.e., as a spacetime metric) and not in an invariant form; and that the scales of the system usually involve the field itself, i.e., are constructed from the metric.
Before turning to specific applications, we review (i) the scale dependence and (ii) the choice of scale identification.
Finally, we also discuss pitfalls and, in particular, the altered role of coordinate-dependence.

\subsection{Scale dependence of gravitational couplings}

\emph{...where we review the scale dependence of the gravitational coupling.}
\\

In analogy to the scale dependence of the electromagnetic coupling $e(k)$ in the previous section, RG improvement of gravitational systems starts by the replacement of classical gravitational couplings with their scale-dependent counterparts.
In explicit calculations, we focus on the Newton coupling $G_N$ and on the simplest possible interpolation between the quantum scale-invariant regime in the UV and the ``classical'' regime in the IR\footnote{We put ``classical" in quotation marks, because it is not the 
$\hslash \rightarrow 0$ 
regime -- in nature, 
$\hslash \rightarrow 0$ 
is at best an approximately observable limit -- instead, the IR is the setting in which all quantum fluctuations in the path integral are present, and we are simply probing the effective action of the theory in its low-curvature regime.}, cf.~Sec.~\ref{sec:weakening} and Fig.~\ref{fig:scaling}, i.e.,
\begin{align}
	G_N(k) = \frac{G_0}{1+ (G_0/g_{N,\ast})\,k^2}.
	\label{eq:runningNewton}
\end{align}
Indeed, this scale dependence can formally be derived as an approximate solution to the Einstein-Hilbert truncation of the functional RG flow between asymptotic safety and a classical low-curvature regime~\cite{Bonanno:2000ep}.
Here, $G_0$ denotes the measured low-curvature value of the dimensionful Newton coupling. This value is approached for $k\rightarrow0$. In the limit $k\rightarrow\infty$, the interpolation scales like $G_N(k)\sim g_{N,\, \ast}/k^2$. The coefficient $g_{N,\ast}$ is the fixed-point value of the dimensionless Newton coupling  $g_N = G_N\, k^2$ in the asymptotically safe regime. It also determines the transition scale between classical scaling $G_N = \rm const$ and asymptotically safe scaling $G_N \sim k^{-2}$. This transition scale is typically expected to coincide with the Planck scale. However, because asymptotic safety has further relevant parameters, the onset of quantum gravity effects may well be shifted away from the Planck scale, towards lower energies. We will discuss this possibility more extensively in Sec.~\ref{sec:towards-observation}, where it motivates a comparison of RG-improved black holes with observations, while treating $g_{N,\, \ast}$ as a free parameter, to be constrained by observations.\footnote{Note that such constraints on $g_{N,\, \ast}$ should not be misinterpreted as actual constraints on the fixed-point value. The latter is a non-universal quantity that is not directly accessible to measurements. It makes its way into observable quantities in the RG improvement procedure, because the procedure is an approximate one. Observations constrain a physical scale, namely the scale at which fixed-point scaling sets in. This scale depends on various couplings of the theory, but in our simple approximation it is only set by the fixed-point value of the Newton coupling.}

One typically includes only the scale-dependence of the Newton coupling.\footnote{More complete studies with higher-order gravitational couplings would have to start from black-hole solutions in corresponding higher-order theories, which are typically not known at all or only for small or vanishing spin.} 
One may therefore expect that the RG improvement captures leading-order quantum effects in the semi-classical regime, but might no longer be fully adequate in the deep quantum regime. In other words, such asymptotic-safety inspired black-hole spacetimes may work well for comparisons with observations, which access horizon-scale physics. In contrast, the deep interior of such asymptotic-safety inspired black holes may still feature pathologies (Cauchy horizons, unresolved singularities, wormholes) which may only be resolved only once further couplings are included. 

\subsection{Scale identification for gravitational solutions}
\emph{...where we review the different choices for the scale identification in the RG improvement of gravitational solutions. }
\\

The RG scale dependence of the Newton coupling is a robust result of asymptotic safety. The other physical input for the RG-improvement -- namely the scale identification -- is less robust, because one may a priori identify the RG scale with quite distinct notions of scale.
In the context of quantum-improved black-hole spacetimes, several different scale identifications have been put forward in the literature. In order to produce meaningful results, it is key that $k^2$ is identified with an invariant quantity (and not, e.g., a coordinate distance).
We will discuss some possible choices below and compare them throughout the following applications.
\begin{itemize}
	\item
	The arguably most obvious scale identification of the RG scale $k$ is with a curvature scale. Such a scale identification follows the expectation that quantum-corrections to black holes are larger, the larger the local curvature scale. One thus identifies $k^2$ with the local value of a curvature scalar $K_i$ such that
	\begin{align}
	\label{eq:scale-id_curvature}
	k^2 
	=\widetilde{\xi}
	K_i^{n_i}\;.
	\end{align}
	Herein, $\widetilde{\xi}$ determines a dimensionless number of order 1.
	The index $i$ indicates that there can be multiple inequivalent curvature scalars. The respective exponent $n_i$ is uniquely fixed by dimensional analysis, because no other scale should enter the identification Eq.~\eqref{eq:scale-id_curvature}.
	In highly symmetric situations, such as for Schwarzschild spacetime, all choices of curvature scalars $K_i\equiv K\;\forall\,i$ are equivalent,  see also Sec.~\ref{sec:spherical-BHs} below.

	For cases with multiple distinct curvature scalars, one may identify the RG scale with the root mean square of all distinct curvature scalars, i.e.,
	\begin{align}
	\label{eq:scale-id_curvature-average}
	k^2 
	= \widetilde{\xi}
	\left(\sum_i (K_i^2)^{n_i}\right)^{1/2}\;.
	\end{align}	
	This also solves another potential issue of Eq.~\eqref{eq:scale-id_curvature} which arises if curvature scalars change sign. Such a sign-change happens, for example, in the near-horizon region of the Kerr spacetime, cf.~Sec.~\ref{sec:spinning-BHs}. A scale identification like Eq.~\eqref{eq:scale-id_curvature} can thus result in complex-valued $k^2$, which is not possible for the RG scale. Taking an absolute value in Eq.~\eqref{eq:scale-id_curvature} avoids this, but introduces non-smooth behavior in $k$ and thus eventually in the RG-improved solution. 
	The scale identification in Eq.~\eqref{eq:scale-id_curvature-average} avoids both issues since the sum of all curvature scalars only vanishes in the approach to an asymptotically flat region of the spacetime.\\
	By plugging in the explicit expression for $\left(\sum_i (K_i^2)^{n_i}\right)^{1/2}$ in terms of the coordinates $x^{\mu}$, one obtains an expression $k^2 = k^2(x^{\mu})$. For instance, for Schwarzschild spacetime, one obtains $k^2 \sim r^{-3}$. Because the spacetime may not be geodesically complete for $r\geq 0$, one has to consider negative $r$ as well. The correct identification (which ensures that $k^2 \geq 0$ always holds), is then in terms of $|r|$, not $r$.
	\item
	Another scale identification is with the (radial) geodesic distance $d$ from the center of the black hole to an asymptotic observer, i.e., 
	\begin{align}
	\label{eq:scale-id_geodesic-distance}
	k^2 
	= \widetilde{\xi}
	\frac{1}{d^2}\;.
	\end{align}
For Schwarzschild spacetime, this assignment is unique,
but for less symmetric settings such as Kerr spacetime it depends on the geodesic path connecting the center of the spacetime to an asymptotic observer. Another key difference to the scale identification with local curvature is that the geodesic distance $d$ is a nonlocal quantity of the classical spacetime. This identification is closest in spirit to the RG-improvement of the Coulomb potential discussed in Sec.~\ref{sec:flat-space-RG-improvement}. However, it is less obvious that the geodesic distance of a (timelike) observer sets an IR cutoff in the path integral.\\
For this scale identification, similar considerations regarding $r<0$ apply as for the first scale identification.
	
	\item
	Finally, in the presence of matter fields (e.g., to describe gravitational collapse), one can identify the RG scale with the trace of the stress-energy tensor~\cite{Bonanno:2016dyv, Bonanno:2017zen}, i.e.,
	\begin{align}
		k^4 
		=\widetilde{\xi}
		g^{\mu\nu}T_{\mu\nu}\;.
	\end{align}
	For matter models with vanishing pressure, this scale identification reduces to $k^4\sim\rho$, with $\rho = T^{00}$ the energy density. Similarly, in settings with a traceless energy-momentum tensor, one may choose $T^{00}$ for the identification.
	In contrast to the trace $g^{\mu\nu}T_{\mu\nu}$, however, this is not a coordinate independent scale identification because $T^{00}$ does not transform like a scalar.
\end{itemize}
 
Which scale identification is most suited may ultimately also depend on the questions one is interested in and which physical assumptions one wants to implement. 
For instance, a scale identification with temperature (as in \cite{Borissova:2022jqj}) appears most ``natural'' when one is interested in thermal properties of a black hole.
In contrast, a scale identification with local curvature scalars matches the effective-field-theory assumption that all quantum-gravity modifications are tied to local curvature scales.
\\
All scale identifications follow from dimensional analysis under the assumption that no second scale is relevant, i.e., one identifies the physical scale with an appropriate power of $k$, with only $\xi$ allowed as the constant of proportionality.
These dimensionless factors are either treated as free parameters, or determined by additional considerations: For instance, \cite{Bonanno:2000ep} argues to fix these dimensionless factors by matching to perturbative calculations in the context of effective field theory~\cite{Donoghue:1993eb,Donoghue:1994dn}.
Generically one may expect these dimensionless factors to be of order unity.

In spite of these differing choices, there are some results about RG improved black holes that are universal, whereas others depend on the scale identification. We will highlight both below.

\subsection*{Further reading}

\begin{itemize}
\item{\bf RG improvement based on Hawking temperature:}
\\
In the spirit of black-hole thermodynamics, it has been advocated~\cite{Borissova:2022jqj} to use the temperature $T$ of a black hole, i.e., to identify
	\begin{align}
		k 
		= \widetilde{\xi}\,k_B\,T\;.
		\label{eq:scale-id_temperature}
	\end{align}	
	with $k_B$ the Boltzmann constant.
Such a scale identification follows the intuition from thermal field theory where the temperature can act akin to an infrared cutoff.
\item
{\bf RG improvement of equations of motion or action:}
\\
Departing from the more standard method of RG improvement at the level of 
the classical solution (i.e., the potential in electrodynamics and the black-hole geometry in the gravitational case), one may also RG improve at the level of the equations of motion~\cite{Bonanno:2001hi,Kofinas:2015sna} or even the action~\cite{Reuter:2003ca,Reuter:2004nx,Borissova:2022jqj}. There is an expectation that, if results from all three methods agree, the result is robust, see, e.g., \cite{Borissova:2022mgd}. However, we caution that there is no a priori reason to expect that RG improving the action gives a result that is closer to the full result than if one RG improves the solution.
\item {\bf Coordinate dependence of the RG improvement procedure:}
\\
One potential pitfall in the RG-improvement of gravitational systems is coordinate dependence. This added difficulty can be understood by comparing the RG-improvement of the Coulomb potential in electromagnetism, cf.~Sec.~\ref{sec:flat-space-RG-improvement} and the RG-improvement of gravitational spacetimes: For the flat-space RG-improvement, the Coulomb potential as a starting point is itself a coordinate-invariant (scalar) quantity. In contrast, for the gravitational RG-improvement, the metric as a starting point does not transform as a scalar under coordinate transformations. This leads to a dependence of the RG improvement on the choice of coordinates in the classical metric~\cite{Held:2021vwd}.
For instance, using coordinates in which a spinning black-hole metric has coordinate singularities, RG improvement may lead to curvature singularities which are not present in other horizon-penetrating coordinate choices, cf.~Sec.~\ref{sec:spinning-BHs}.
One potential remedy is to perform the RG-improvement at the level of quantities which transform as scalars under coordinate transformations, see~\cite{Held:2021vwd}.
\end{itemize}

\section{Spherically symmetric black holes}
\label{sec:spherical-BHs}
\emph{...where we analyze the properties of asymptotic-safety inspired spherically symmetric spacetimes, focusing on the fate of the classical singularity and of the inner and outer event horizon. We also review results on black-hole thermodynamics.}

\subsection{Construction of asymptotic-safety inspired black holes}

\emph{...where we construct the RG improved spherically symmetric line element and discuss the universality of different  scale identifications.}
\\

To obtain an asymptotic-safety inspired, spherically symmetric black hole, we start with the Schwarzschild metric in Schwarzschild coordinates $t, r, \theta, \phi$,
\be
ds^2 = - f(r) dt^2 +f(r)^{-1}dr^2 +r^2 d\Omega^2,\label{eq:dsSchw}
\ee
with
\be
f(r) = 1- \frac{2G_N\, M}{r}.
\ee
The high degree of symmetry ensures that the choice of Schwarzschild coordinates is equivalent to the choice of horizon-penetrating Eddington-Finkelstein coordinates, cf.~\cite{Held:2021vwd} for an explicit discussion.

We follow the two-step procedure of RG-improvement as outlined in Sec.~\ref{sec:construction}: First, we upgrade the classical Newton coupling $G_N$ to its RG-scale dependent counterpart $G_N(k)$, as given in Eq.~\eqref{eq:runningNewton}. Second, we fix the scale identification. As we will see, the local scale identification with curvature (cf.~Eq.~\ref{eq:scale-id_curvature-average} and~\cite{Bonanno:1998ye}), and the non-local scale identification with the radial geodesic distance (cf.~Eq.~\eqref{eq:scale-id_geodesic-distance} and~\cite{Bonanno:2000ep}) are equivalent. This equivalence is, once again, due to the high degree of symmetry of the Schwarzschild geometry. Therefore, there are universal results for RG improved spherically symmetric results.

To make the scale identification with local curvature explicit, we determine the curvature invariants. When evaluated in Schwarzschild spacetime, all curvature invariants built from the Ricci tensor vanish and all others are equivalent to the Kretschmann scalar
\begin{align}
\label{eq:Kretschmann-Schw}
K = R_{\mu\nu\kappa \lambda}R^{\mu\nu\kappa \lambda} = \frac{48 G_N^2\, M^2}{r^6}\;.
\end{align}
The Kretschmann scalar is manifestly positive.
Hence, no average or absolute value needs to be taken and we find the scale identification
\begin{align}
	\label{eq:scale-id_Schw}
	k^2 = \widetilde{\xi}\,\sqrt{K} = \widetilde{\xi}\, \frac{G_0\, M}{\sqrt{r^6}}\;,
\end{align}
where we have included an a priori unknown constant of proportionality $\widetilde{\xi}$.

Accordingly, the RG improved line-element is given by Eq.~\eqref{eq:dsSchw} with
\be
f(r) = 1- \frac{2G_N(k)\, M}{r} = 1- \frac{2 G_0\, M}{r \left(1 + \xi \frac{G_0^2\, M}{\sqrt{r^6}} \right)}\;,
\label{eq:frRGimp}
\ee
where we have absorbed the fixed-point value for the dimensionless Newton coupling $g_{N,\ast}$, cf.~Eq.~\eqref{eq:runningNewton} and the constant of proportionality $\widetilde{\xi}$ arising in the scale identification, cf.~Eq.~\eqref{eq:scale-id_Schw}, into a single dimensionless constant 
\be
\xi = \sqrt{48}\widetilde{\xi}/g_{N,\ast}.
\ee
Incidentally, this agrees with the mass function of a Hayward black hole \cite{Hayward:2005gi} for $r>0$. 

The free parameter $\xi$ determines the scale at which quantum-gravity effects become sizable: the larger $\xi$, the lower the curvature scale at which quantum-gravity effects are sizeable. For $\xi=1$, quantum-gravity effects effectively set in at the Planck scale, because $f(r) \approx 1- 2 G_0 M/r$ for any $r$ greater than the Planck length. For astrophysical black holes, the curvature radius at the horizon is far above the Planck length. Thus, $\xi \approx 10^{95}$ would be required to achieve $\mathcal{O}(1)$ modifications at horizon scales~\cite{Held:2019xde}. (This estimate assumes a supermassive black holes with mass $M\sim 10^9 M_\odot$.) 
As we discuss in Sec.~\ref{sec:scale-of-qG}, such large choices of $\xi$ may be relevant for black holes.

In the following, we analyze the above RG-improved Schwarzschild spacetime from the inside out: we start at the core and investigate the fate of the curvature singularity and of geodesic completeness; then we move on to the inner and outer horizon and the photon sphere.

\subsection{Singularity-resolution}
\emph{...where we show that the asymptotic-safety inspired spherically symmetric black hole is regular with finite curvature invariants and we also discuss the status of geodesic completeness.}
\\

The asymptotic-safety inspired spherically symmetric black hole does not have a curvature singularity.  This can be seen by calculating the curvature invariants\footnote{We use the term `curvature invariants' to refer to Riemann invariants, i.e., scalars built from contractions of any number of Riemann tensors and the metric. This does not include derivative invariants, i.e., those which involve additional covariant derivatives.} of the RG-improved metric.
The RG-improved spacetime is no longer a vacuum solution of GR\footnote{Heuristically, one can think of the non-zero energy-momentum tensor as a contribution on the \emph{left-hand-side} of the generalized Einstein equations, i.e., a contribution from higher-order curvature terms which arise in asymptotically safe gravity.} and thus curvature invariants involving the Ricci tensor can be non-vanishing.
In a general 4-dimensional spacetime, there are up to  14 curvature invariants which can be polynomially independent. To capture all possible (degenerate) cases at once, a complete set of 17 curvature invariants is required~\cite{1991JMP....32.3135C, 1997GReGr..29..539Z, 2002nmgm.meet..831C}.
(At most) four of these are polynomially independent for the RG-improved spacetime at hand~\cite[App.~D]{Held:2021vwd}. These four polynomially independent invariants may be chosen as $R$, $R_{\mu\nu}R^{\mu\nu}$, $C_{\mu\nu\rho\sigma}C^{\mu\nu\rho\sigma}$, and $R_{\mu\nu}R_{\rho\sigma}C^{\mu\nu\rho\sigma}$, where $R$ is the Ricci scalar, $R_{\mu\nu}$ is the Ricci tensor and $C_{\mu\nu\rho\sigma}$ is the Weyl tensor. They take the form
\begin{align}
	R &= \frac{2 M\left(r\,G_N'' + 2\,G_N'\right)}{r^{2}}\;,
	\\
	R_{\mu\nu}R^{\mu\nu} &= \frac{2 M^2\left(r^2 \left(G_N''\right)^2 + 4\left(G_N'\right)^2\right)}{r^{4}}\;,
	\\
	C_{\mu\nu\rho\sigma}C^{\mu\nu\rho\sigma} &= \left(\frac{2M \left(r \left(r G_N''-4 G_N'\right)+6 G_N\right)}{\sqrt{3}r^{3}}\right)^2\;,
	\\
	R_{\mu\nu}R^{\mu\nu}C^{\mu\nu\rho\sigma} &=
	\frac{1}{3}\left(\frac{M \left(r G_N''-2 G_N'\right)}{r^2}\right)^2
	\left(\frac{2M \left(r \left(r G_N''-4 G_N'\right)+6 G_N\right)}{\sqrt{3}r^{3}}\right)\;.
\end{align}
Primes denote radial derivatives, i.e., $G_N'(r) = \partial G_N(r)/\partial r$ and $G_N''(r) = \partial^2 G_N(r)/\partial r^2$. From these expressions, we can draw conclusions about the presence/absence of a curvature singularity, depending on the behavior of the scale-dependent Newton coupling near $r=0$, i.e., depending on the leading exponent $n$ in $G_N(r)\sim r^n + \mathcal{O}(r^{n+1})$:
\begin{itemize}
	\item 
	If $n=3$, the curvature invariants take finite values at $r=0$.
	\item
	If $n>3$, all curvature invariants vanish and the spacetime is flat in the center.
	\item
	If $n<3$, the curvature singularity remains, although it is weakened for $0<n<3$ compared to the Schwarzschild case.
\end{itemize}

\begin{figure}[!t]
\begin{center}
\includegraphics[width=0.7\linewidth]{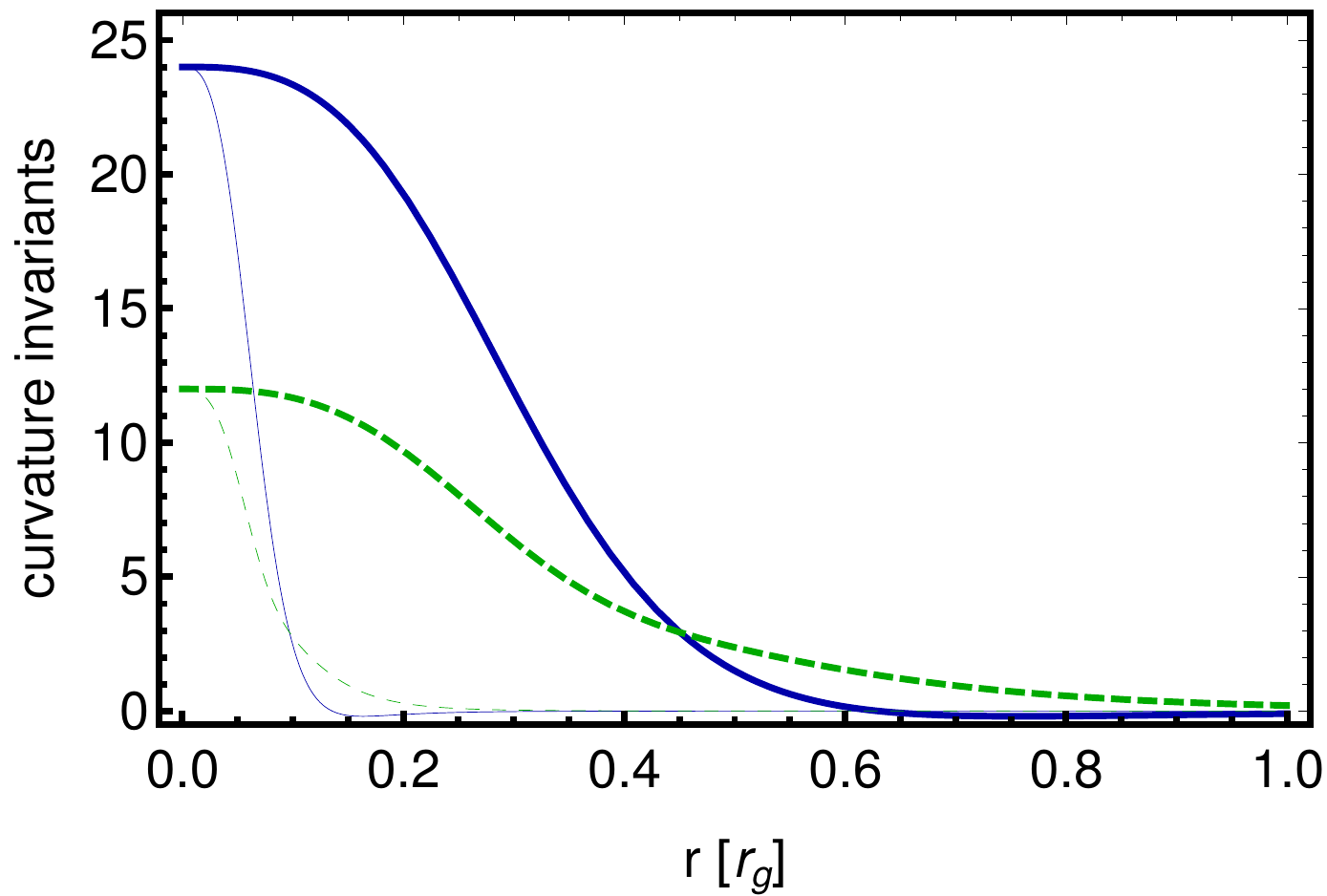}
\end{center}
\caption{\label{fig:spherical}
We show the behavior of the curvature scalars $R$ (blue continuous) and $\sqrt{R_{\mu\nu}R^{\mu\nu}}$ (green dashed) on the RG-improved Schwarzschild geomerty, cf.~Eq.~\eqref{eq:scale-id_Schw}.
The thick lines are for $M=1$, the thin lines for $M=10$; in all cases 
$G_0=1$
and $\xi=1$.
}
\end{figure}

From Eq.~\eqref{eq:scale-id_Schw}, we see that the RG-improved Newton coupling realizes the critical behavior, i.e., scales like $G_N(r)\sim r^3 + \mathcal{O}(r^4)$ for small $r$.
In other words, RG-improvement suggests that asymptotically safe quantum gravity limits the maximal value that each curvature invariant can attain, and therefore these invariants tend to universal, i.e., mass-independent, limits,
\begin{align}
R\overset{r \rightarrow 0}{\rightarrow}\frac{24}{G_0 \xi},
\quad
R_{\mu\nu}R^{\mu\nu}\overset{r \rightarrow 0}{\rightarrow} \frac{144}{G_0^2 \xi^2},
\quad
C_{\mu\nu\rho\sigma}C^{\mu\nu\rho\sigma}\overset{r \rightarrow 0}{\rightarrow} 0,
\quad
R_{\mu\nu}R^{\mu\nu}C^{\mu\nu\rho\sigma}\overset{r \rightarrow 0}{\rightarrow} 0,
\end{align}
see also Fig.~\ref{fig:spherical}. 
This is exactly what one expects from the dynamical decoupling of the high-energy modes of the metric.
It implies that those modes with high eigenvalues of a (suitably chosen) Laplacian, which must be present in order to generate high values of the curvature, are absent. Therefore, the curvature radius is limited, and the limiting value is determined by  $1/\sqrt{G_0\xi}$, i.e., by the quantum-gravity scale.
\\

In contrast, the black-hole mass determines the radius at which effects become sizable: At fixed $\xi$, the departure of $R$ from zero becomes of order one at smaller $r/r_S$, if $M$ is made smaller, with $r_S$ the classical Schwarzschild radius.\\

We conclude that RG-improvement suggests that an asymptotically safe scaling regime resolves the curvature singularity at the center of Schwarzschild spacetime. 

Resolution of curvature singularities does not imply geodesic completeness, because the two are independent requirements. It nevertheless turns out that the RG improved black hole is geodesically complete as well. This can be seen by following \cite{Zhou:2022yio}, where it is pointed out that the Hayward black hole is geodesically incomplete for $r \geq 0$ and therefore needs to be extended to $r<0$. This then also applies to RG-improved spherically symmetric black holes. In the $r<0$ region, geodesics terminate at a curvature singularity for the Hayward black hole. However, in that region, the RG-improved black hole differs from the Hayward black hole: As we pointed out in Sec.~\ref{sec:construction}, a sensible scale identification, which ensures that $k^2>0$, forces us to use $|r|$ instead of $r$. This, as has first been pointed out by \cite{Torres:2017gix}, is sufficient to remedy the curvature singularity at $r<0$ and also avoids a pole in the geodesic equation that was found in \cite{Zhou:2022yio}. Thus, the maximal extension of RG improved spherically symmetric black holes is geodesically complete.\\
Another independent requirement is completeness of the spacetime for accelerating observers, which, to the best of our knowledge, has not yet been checked for these black holes.

\subsection{Spacetime structure}
\label{sec:sph-sym-spacetime-structure}
\emph{...where we explain how the regularization by quantum gravity affects the spacetime at all scales, such that the black hole is more compact than its classical counterpart and also has a more compact photon sphere, i.e., casts a smaller shadow.}
\\

The spacetime is not only modified in the center of the black hole, but everywhere, although the size of the modifications decreases with the (geodesic) distance from the center. Whether or not modifications lead to detectable effects depends on the value of $\xi$, which determines at what scale quantum-gravity effects are sizable. For now we discuss the effects at a qualitative level, and comment on detectability later on.
\\
Because the Newton coupling is smaller than its classical counterpart, gravity is weaker in the quantum regime. Thus, the horizon, the photon sphere (and any special surface in the spacetime) are more compact. At a first glance, this may appear counterintuitive, because the converse is true for compact objects (e.g., stars): they are \emph{less} compact, if gravity is weaker. However, there is an important difference between compact objects without horizon and black holes: in a horizonless compact object, such as a neutron star, some form of pressure balances the gravitational force and this balance determines the location of the surface. In a black hole, the horizon is not a material surface and its location is determined by gravity alone: the expansions of in- and outgoing null geodesic congruences are negative (i.e., the lightcone tilts ``inwards"), once gravity is locally strong enough. Thus, a weakening of gravity results in a more compact horizon -- in the limit of vanishing gravitational force, the size of the horizon would collapse to zero.

\begin{figure}[!t]
\begin{center}
\includegraphics[width=0.7\linewidth]{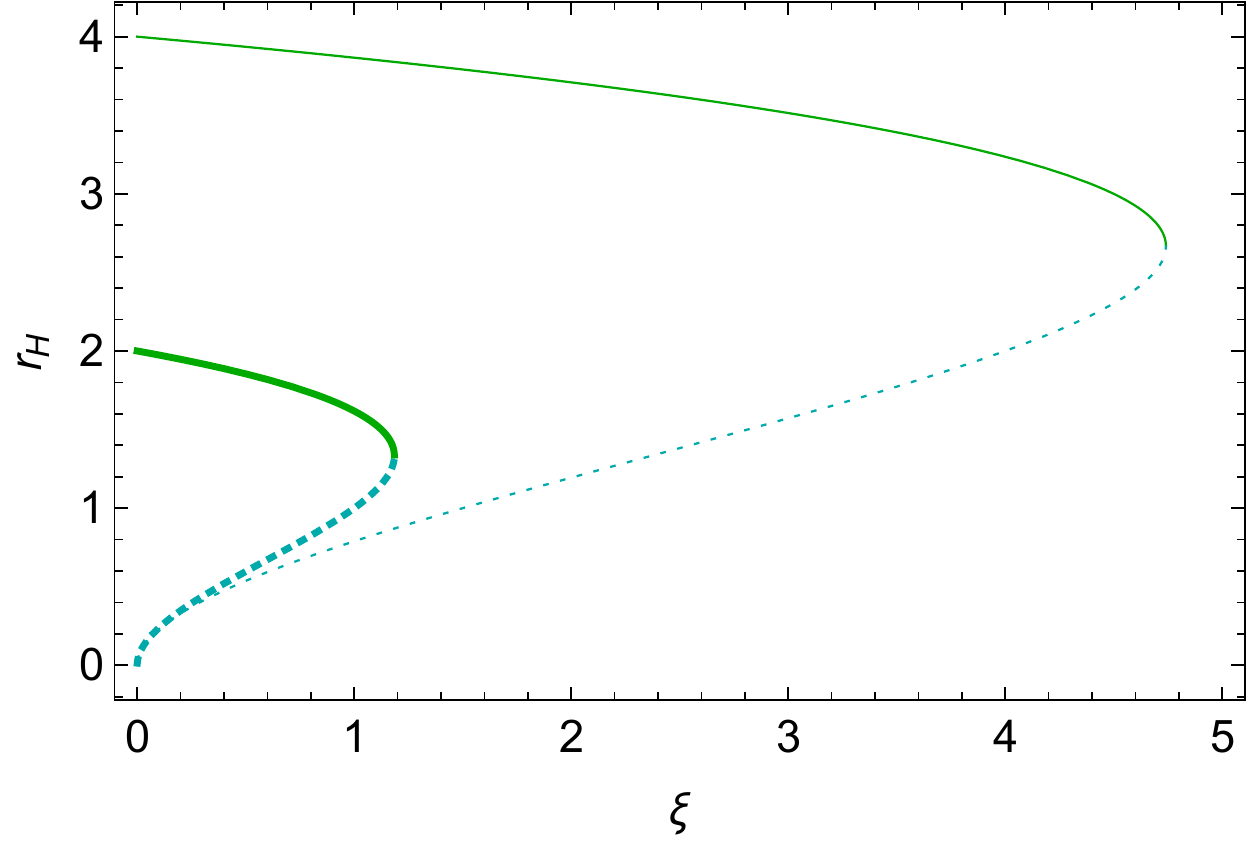}
\end{center}
\caption{\label{fig:rH} We show the location of the outer (green continuous) and inner (cyan dotted) horizon as a function of $\xi$ for $M=1$ (thick lines) and $M=2$ (thin lines) and $G_0=1$.}
\end{figure}

To determine the location of the horizon, we proceed as in the Schwarzschild case:
The horizon respects the Killing symmetries of the spacetime, i.e., it is a function of $r$ and $\theta$ only and can thus be described by the condition $h(r, \theta)=0$. Because the horizon is a null surface, its normal is a null vector, i.e., $n^{\mu} = \partial^{\mu}h(r, \theta)$ is null:
\be
0 = g^{\mu\nu}n_{\mu}n_{\nu}= g^{\mu\nu} \left(\partial_{\mu}h(r, \theta) \right) \left(\partial_{\nu}h(r, \theta) \right).\label{eq:horizon}
\ee
For the Schwarzschild spacetime, $g^{r\theta}=0$, which continues to hold at finite $\xi$. Additionally, the horizon may not violate spherical symmetry, and thus $\partial_{\theta}h(r, \theta)=0$. This leaves us with
\be
g^{rr} = 0. \label{eq:horcondsph}
\ee
For the RG improved spacetime, the condition reads
\be
0=1- \frac{2 G_0\, M}{r \left(1+\xi \frac{G_0^2\, M}{r^3} \right)}.
\ee
This condition has two or no real and positive solutions. The black hole therefore has an outer and inner horizon. At large values of the parameter $\xi$, the two horizons can merge and leave behind a horizonless object, cf.~Fig.~\ref{fig:rH}. The larger the mass of the black hole, the larger the value of $\xi_{\rm crit}$ for which the horizons merge,
\be
\xi_{\rm crit}=\frac{32}{27}G_0\, M^2.
\ee

The inner horizon is a Cauchy horizon. This is not only problematic when setting up initial-value-problems in the spacetime, it even suggests the presence of an instability. This instability has been discussed extensively in the literature~\cite{Carballo-Rubio:2018pmi,Carballo-Rubio:2019nel,Bonanno:2020fgp,Carballo-Rubio:2021bpr,DiFilippo:2022qkl,Bonanno:2022jjp,Carballo-Rubio:2022kad},
and is the subject of another chapter in this book, to which we refer the interested reader~\cite{Bonanno:2022rvo}.
\\

To bridge the gap between theoretical studies and observations, we investigate the photon sphere, which determines the size of the black-hole shadow. We explicitly prove that the photon sphere is more compact for $\xi >0$ than it is for $\xi=0$. This serves as one example of a general result, namely that all special hypersurfaces in a black-hole spacetime (event horizon, photon sphere, (ergosphere for finite spin)) are more compact~\cite{Eichhorn:2022oma}.
 
The photon sphere is the border of the spacetime region within which infalling photons inevitably fall into the event horizon, i.e., it corresponds to the innermost circular photon orbit. The radius of this orbit can be determined by using four constants of motion, namely the energy $E$ and all three components of the angular momentum $L$.
We thereby obtain three equations for generic null geodesics parameterized by the affine parameter $\lambda$:
\bea
\left(\frac{dr}{d\lambda}\right)^2 = - V_r(r),\quad
\frac{d\phi}{d\lambda}= \frac{L}{r^2},\quad
\frac{dt}{d\lambda}= \frac{E}{f(r)},
\eea
where the effective radial potential is
\be
V_r(r) = - \left(E^2 - f(r)\frac{L^2}{r^2}\right).
\ee
For circular geodesics, $dr/d\lambda=0$ for all $\lambda$, which entails that
\bea
V_r(r)=0,\quad\quad
\frac{d}{dr}V_r(r)=0.\label{eq:effpotcond2}
\eea
The final equation from which the photon sphere $r_{\gamma}$ is determined is the explicit form of Eq.~\eqref{eq:effpotcond2}:
\bea
0&=& f'(r=r_{\gamma}) -  \frac{2 f(r= r_{\gamma})}{r_{\gamma}}.\label{eq:photspherecond}
\eea
From here, one can show that the photon sphere is more compact for $\xi>0$ than in the classical Schwarzschild case. To that end, we write $f(r)$ in terms of its classical part and the quantum correction,
\be
f(r) = 1- \frac{2G_0\, M}{r} f_{\rm qm} (r),
\ee
where
\be
f_{\rm qm}(r) = \frac{1}{1+ \xi \frac{G_0^2\, M}{r^3}}.
\ee
For our derivation, the critical properties of $f_{\rm qm}(r)$ are i) that it is everywhere smaller than one, because it weakens gravity everywhere and ii) that its derivative is positive, because the quantum effects fall off with the radial distance. Then, Eq.~\eqref{eq:photspherecond} becomes
\bea
0 = - \frac{G_0\, M}{r_{\gamma}}f_{\rm qm}'(r= r_{\gamma}) - \frac{1}{r_{\gamma}} + 3 \frac{G_0\, M}{r_{\gamma}^2}f_{\rm qm}(r= r_{\gamma}),\label{eq:photsphercond2}
\eea
from where $r_{\gamma} = 3 G_0\, M$ follows for $\xi=0$. For $\xi>0$, one can rewrite Eq.~\eqref{eq:photsphercond2} by using the two conditions on $f_{\rm qm}(r)$:
\be
r_{\gamma} = \frac{3 G_0\, M\, f_{\rm qm}(r=r_{\gamma})}{1+G_0\, M\, f'_{\rm qm}(r=r_{\gamma})} < \frac{3G_0\, M}{1+ G_0\, M\, f_{\rm qm}'(r=r_{\gamma})}< 3 G_0\, M.
\ee
Therefore, the photon sphere of an asymptotic-safety inspired, spherically symmetric black hole is smaller than for its Schwarzschild counterpart with the same mass.

In addition, Eq.~\eqref{eq:photsphercond2} has more than one solution: in concert with an inner horizon, an inner circular orbit appears. In contrast to the outer one, it has different stability properties and is therefore not relevant for images of black holes. The two photon spheres also annihilate at a critical value of $\xi_{\rm crit,\, 2}$, which, however, is larger than the critical value for horizon-annihilation, $\xi_{\rm crit}$. Therefore, within the interval $\xi \in [1.185,1.808] G_0 M^2$ there is a horizonless object that generates photon rings in its image \cite{Eichhorn:2022oma}.

\subsection{Thermodynamics}
\emph{...where we review calculations of black-hole entropy and black-hole evaporation in asymptotic-safety inspired black holes. We highlight that the evaporation process may end in a remnant.}
\\

We have seen that regularity implies a second inner horizon and thus a critical mass $M_\text{crit}$ (for a fixed quantum-gravity scale $\xi$) at which the inner and outer horizon coincide. As we will see below, this modifies the Bekenstein-Hawking temperature of the black-hole and thus the evaporation process.
While the original discussion was given in the context of RG-improved black holes~\cite{Bonanno:2000ep}, the following applies to spherically-symmetric regular black holes more generally.

The Bekenstein-Hawking temperature arises generically from quantum field theory on a curved background spacetime which contains a (black-hole) horizon. It thus applies to any asymptotically flat, spherically symmetric line element
\begin{align}
	ds = - f(r) dt^2 + \frac{1}{f(r)}dr^2 + r^2d\Omega\;,
\end{align}
for which the largest root of $f(r=r_h)=0$ implies a horizon at $r=r_h$.
In a setting which does not account for the backreaction of the quantum fields, and is thus independent of the gravitational dynamics,
the presence of an event horizon implies Hawking radiation with an associated temperature
\begin{align}
	\label{eq:BH-temperature}
	T_\text{BH} = \frac{f'(r_h)}{4\pi}\;.
\end{align}
For very large asymptotic-safety inspired black holes with 
\be
M\gg M_\text{crit}= \frac{3^{\frac{3}{2}}
}{\sqrt{32} G_0}\sqrt{\xi}
\ee
the exterior spacetime approximates the classical Schwarzschild spacetime very well and thus recovers the classical temperature $T_\text{BH} = 1/(8\pi G M)$. As the regular black hole evaporates, its temperature first grows, as in the case without RG-improvement. However, as the evaporation continues, the black hole approaches the critical mass $M_\text{crit}$ at which the inner and outer horizon merge, cf.~Fig.~\ref{fig:rH}. At this point, $f'(r_h)|_{M_\text{crit}}=0$, and thus the temperature of the RG-improved black hole must vanish. Inbetween the temperature thus peaks and then decreases to zero.
\\
This has a profound implication, namely, that a vanishing temperature at finite mass $M_\text{crit}$ suggests a stable end state of Hawking evaporation of an asymptotic-safety inspired black hole~\cite{Bonanno:2000ep}. An application of the Boltzmann law suggests that it takes infinite proper time for the asymptotic-safety inspired black hole to cool down to its finite remnant state. It has not been investigated, whether these remnants pose problems, as is generically expected for remnants \cite{Chen:2014jwq}, such as the overproduction problem. It has to be stressed, though, that the pair-production of black hole remnants (usually argued to lead to instabilities), depends on the dynamics of the theory, which is not yet fully understood in the case of asymptotically safe gravity.
As in the classical-spacetime case, one can also assign an entropy to evaporating black holes \cite{Bonanno:2000ep,Falls:2010he,Falls:2012nd,Borissova:2022mgd}. 

Both the classical and the RG-improved scenario for black-hole evaporation make use of a calculation for Hawking radiation on a fixed background spacetime and thus explicitly neglect backreaction of the radiation onto the black-hole spacetime. In the classical case, where the late-time temperature diverges, this approximation must necessarily break down when the black hole evaporates to sufficiently sub-Planckian mass. In the RG-improved case, the issue of backreaction could be less severe, because the temperature remains bounded throughout the entire evaporation process. However, the asymptotically safe dynamics for quantum fields on curved backgrounds contain non-standard non-minimal interactions (schematically of the form ``kinetic term $\times$ curvature"), cf.~\cite{Eichhorn:2022jqj,Eichhorn:2022gku} for recent reviews. Their effect on the Hawking evaporation system has not been taken into account yet.

\subsection*{Further reading} 
\begin{itemize}
\item {\bf Schwarzschild (Anti-) deSitter black holes:}
\begin{enumerate}
\item[(i)] In \cite{Koch:2013owa}, the authors perform an RG improvement of Schwarzschild-de Sitter black holes, where they also include the running of the cosmological constant. Using the same scale-identification as in the Schwarzschild case then results in a singular spacetime, because the cosmological constant scales with $k^2$. Given that this system features several distinct scales, such a scale identification may no longer be appropriate. Whether a different scale identification should be performed, or whether the RG improvement procedure is too simple to properly deal with the situation with several distinct scales is an open question.
\item[(ii)] In \cite{Adeifeoba:2018ydh}, the authors include the scaling of the cosmological constant away from the fixed-point regime, obtaining a non-singular black-hole spacetime when conditions on the scaling exponents hold.
\item[(iii)] In \cite{Torres:2017ygl} the author instead considers unimodular gravity, where the cosmological constant arises as a constant of integration at the level of the equations of motion and which may also be asymptotically safe \cite{Eichhorn:2013xr,Eichhorn:2015bna}. Therefore, it is not part of the RG-improvement, but remains classical. Accordingly, the resulting black holes are regular.
\end{enumerate}
\item {\bf Leading-order quantum-gravity correction:}
\\
In \cite{Bonanno:2000ep}, the authors use a result for $G_N(k)$ which includes additional scale-dependence around the transition scale and thereby reproduces the leading-order quantum-gravity correction at large $r$.
\item {\bf Iterated RG improvement:}
\\
In \cite{Platania:2019kyx}, the author iterates the RG improvement, starting with Eq.~\eqref{eq:frRGimp} as the first step of an iteration procedure. The final result of this procedure is the Dymnikova line element \cite{Dymnikova:1992ux}. The same iteration procedure was used to describe the formation and evaporation of black holes in \cite{Borissova:2022mgd}.
\item {\bf Evaporation of a Schwarzschild black hole:}
\\
The evaporation of a Schwarzschild black hole due to Hawking radiation has been effectively described by a Vaidya metric in which the mass function decreases with time~\cite{Hiscock:1980ze}. In~\cite{Bonanno:2006eu}, the authors RG-improve said Vaidya metric as a model to incorporate the quantum fluctuations in the late-time evolution of Hawking evaporation. In agreement with the discussion above, they find that the final state of the evaporation process is a Planck-sized cold remnant.
\item {\bf Black holes beyond four dimensional spacetime:}
\\
Going beyond the four-dimensional spacetime setting, black solutions have been RG-improved in \cite{Litim:2013gga}. The authors consider Myers-Perry-black holes, which are higher-dimensional rotating solutions of higher-dimensional GR. Their study finds that the ring singularity is softened enough to achieve geodesic completeness. They also find a minimum black-hole mass, related to the weakening of gravity through asymptotic safety: below a critical mass, gravity is no longer strong enough to form an event horizon.
\end{itemize}

\section{Spinning asymptotic-safety inspired black holes}
\label{sec:spinning-BHs}
\emph{...where we consider spinning asymptotic-safety inspired black holes and explore the fate of the classical curvature singularity, the fate of the Cauchy horizon, the event horizon, the ergosphere and the photon spheres.}
\\

Spinning black holes have a much more intricate structure than their spherically symmetric counterparts. This 
leads to
several ambiguities in the RG improvement procedure, which are mostly irrelevant in the spherically symmetric case and that we discuss below. 

Despite its complexity, the Kerr spacetime can be uniquely characterized by a single -- remarkably simple -- complex invariant
\begin{align}
	\label{eq:complex-inv-Kerr}
	\mathcal{C} = \frac{G_0\,M}{(r - i\,a\,\cos(\theta))^3}\;,
\end{align}
where $r$ is the radial coordinate and $\theta$ the polar angle with respect to the black hole's spin axis.\footnote{The definition of the coordinates $(r,\,\theta)$ is equivalent in many of the standard coordinate systems for Kerr spacetime and, in particular, agrees with the $(r,\,\theta)$ as defined in ingoing Kerr and Boyer-Lindquist coordinates.}
The more standard Weyl invariants follow directly from $\mathcal{C}$ as
\begin{align}
	C_{\mu\nu\rho\sigma}C^{\mu\nu\rho\sigma} &= 
	\text{Re}\left[48\,\mathcal{C}^2\right]\;,
	\\
	C_{\mu\nu\rho\sigma}\overline{C}^{\mu\nu\rho\sigma}&= 
	\text{Im}\left[48\,\mathcal{C}^2\right]\;,
\end{align}
where $C_{\mu\nu\rho\sigma}$ and $\overline{C}_{\mu\nu\rho\sigma} = 1/2\,\epsilon_{\mu\nu \kappa\lambda }C^{\kappa\lambda }_{\phantom{\kappa\lambda }\rho\sigma}$ denote the Weyl tensor and the dual Weyl tensor.
All other Riemann invariants are either polynomially dependent or vanish.

\begin{figure}[!t]
	\begin{center}
		\includegraphics[width=0.48\linewidth]{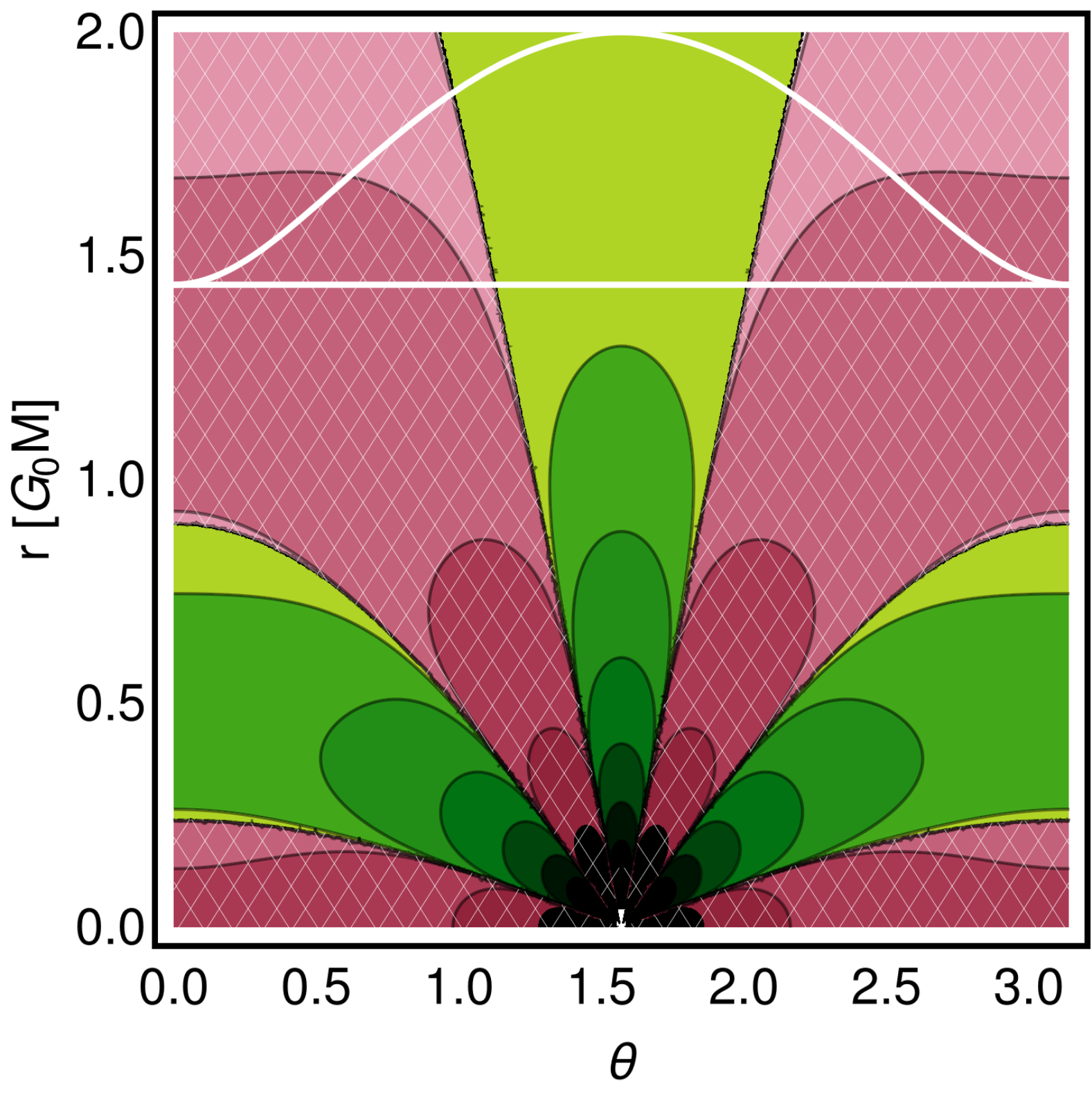}
		\hfill
		\includegraphics[width=0.48\linewidth]{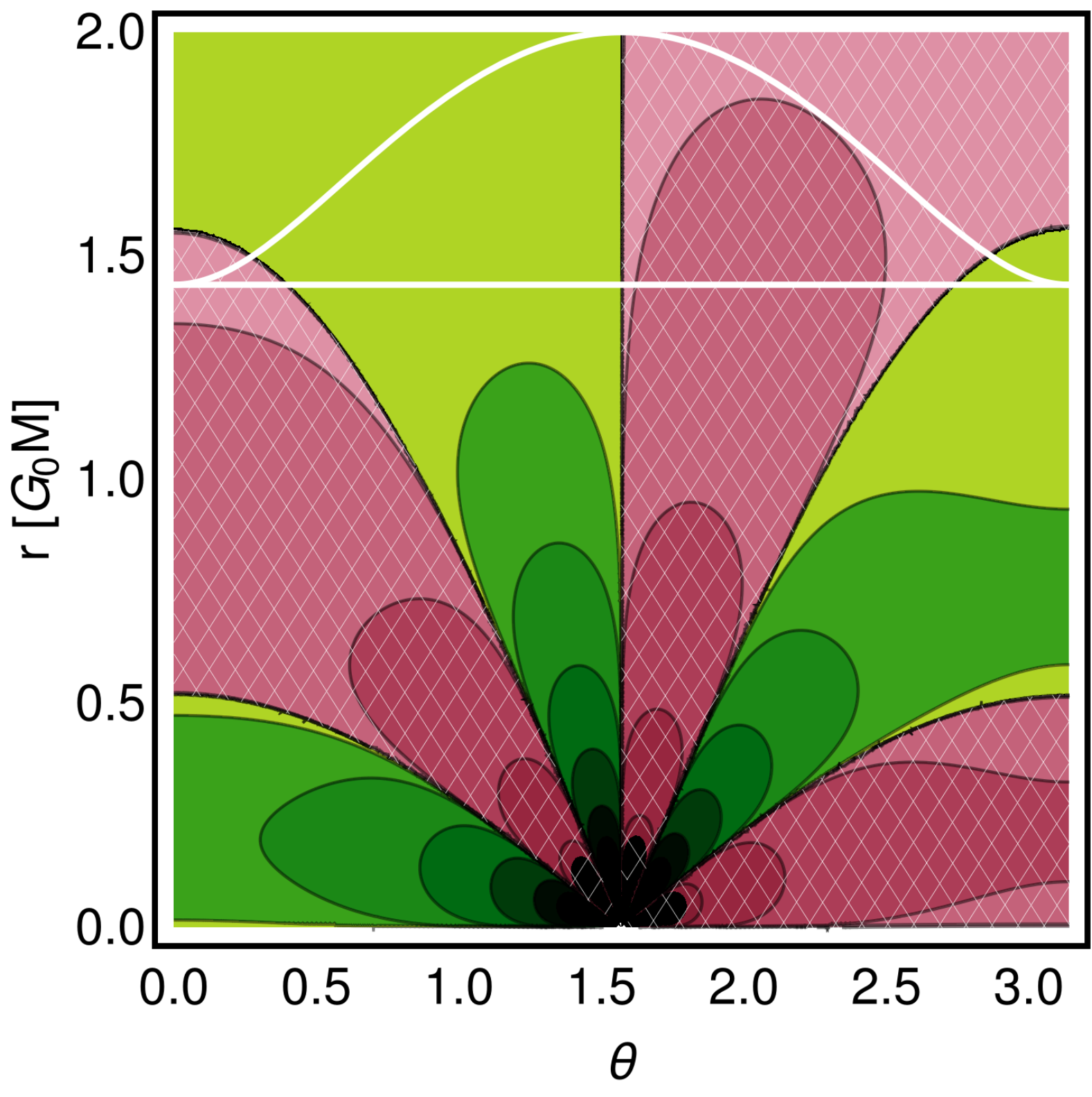}
	\end{center}
	\caption{\label{fig:spinning-Kerr-invariants} 
	We show contour plots of the two polynomially independent real invariants $C_{\mu\nu\rho\sigma}C^{\mu\nu\rho\sigma}$ (left-hand panel) and $C_{\mu\nu\rho\sigma}\overline{C}^{\mu\nu\rho\sigma}$ (right-hand panel) of Kerr spacetime. Increasingly positive values are shaded in increasingly dark shades of green (non-hatched). Increasingly negative values are shaded in increasingly dark shades of red (hatched). The two thick white lines indicate the location of the event horizon (lower line) and the ergosphere (upper line).}
\end{figure}

The key distinction to Schwarzschild spacetime is the angular dependence on $\theta$. This angular dependence has consequences for the construction of spinning RG improved black holes. 
\\
First, in contrast to Schwarzschild spacetime, Kerr spacetime is characterized by two, not one, real invariants.
\\
Second, because of their angular dependence, these two real invariants can change sign -- even in spacetime regions external to the event horizon, cf.~Fig.~\ref{fig:spinning-Kerr-invariants}.
\\
Third, this angular dependence is also mirrored in many of the key physical aspects that distinguish Kerr spacetime from Schwarzschild spacetime: for instance, the ergosphere and thus frame dragging is most prominent close to the equatorial plane (cf.~outer white line in Fig.~\ref{fig:spinning-Kerr-invariants}), and the singularity is a ring singularity, confined to $\theta=\pi/2$ (cf.~dark-shaded region in Fig.~\ref{fig:spinning-Kerr-invariants}).

All of these effects vanish in the continuous non-spinning limit for which Kerr spacetime reduces to Schwarzschild spacetime. This is made explicit by noting that the complex invariant $\mathcal{C}$ turns real and its square converges to the Kretschmann scalar, i.e., $\mathcal{C}^2\stackrel{a\rightarrow0}{\longrightarrow}R_{\mu\nu\rho\sigma}R^{\mu\nu\rho\sigma}$, in the Schwarzschild limit, cf.~Eqs.~\eqref{eq:complex-inv-Kerr} and \eqref{eq:Kretschmann-Schw}.
\\

The above understanding of Kerr spacetime indicates that any RG-improvement that captures the key difference between spinning and non-spinning black-hole spacetimes needs to incorporate angular dependence. In the following, we will first discuss the key difficulties that come with this added complexity: In Sec.~\ref{sec:spinning_dep-on-scale-id}, we show that the choice of scale identification now impacts physical conclusions. In particular, we summarize which conclusions appear robust and which other ones do not. We also argue why we expect a scale identification with local curvature to give the most physical results.\\
We also point out that horizon-penetrating coordinates are crucial to avoid naked singularities which arise from coordinate singularities through the RG improvement.
We will then specify to an RG-improvement based on local curvature invariants and horizon-penetrating coordinates and examine the resulting physical aspects of the resulting RG-improved Kerr spacetime.

\subsection{Universal and non-universal aspects of RG improvement}
\label{sec:spinning_dep-on-scale-id}
\emph{...where we demonstrate that the RG improvement of axisymmetric spacetimes is less universal than in spherically symmetric spacetimes. Starting from Kerr spacetime in horizon-penetrating coordinates, we discuss which conclusions about the RG-improved spacetime depend on the choice of scale identification and, in contrast, which conclusions remain universal.}
\\

We start out from Kerr spacetime in horizon-penetrating (ingoing Kerr) coordinates $(u, r, \chi=\cos(\theta), \phi)$:
\bea
\label{eq:regular-spinning-BH-metric-ingoing-Kerr}
ds^2 &=&-\frac{r^2-2G_N\,M r +a^2 \chi^2}{r^2+a^2 \chi^2}du^2 +2\,du\, dr - 4\frac{G_N\, M a r}{r^2+a^2\chi^2}\left(1-\chi^2 \right) du\, d\phi
\nonumber\\
&{}&+ \frac{1-\chi^2}{r^2+a^2\chi^2}\left(\left(a^2+r^2\right)^2 - a^2\left(r^2-2G_N\,Mr+a^2 \right)\cdot \left(1-\chi^2 \right) \right)d\phi^2
\nonumber\\
&{}&- 2a\left(1-\chi^2 \right)dr\, d\phi + \frac{r^2+a^2\chi^2}{1-\chi^2}d\chi^2\;.
\eea
For the classical case, i.e., for constant Newton coupling $G_N\equiv G_0$, this metric is equivalent to the above representation of Kerr spacetime in terms of curvature invariants, cf.~\eqref{eq:complex-inv-Kerr}. At the level of the metric, RG-improvement is implemented in the same way as for the non-spinning case, i.e., by the two-step replacement (cf.~Sec.~\ref{sec:construction})
\begin{align}
	G_N 
	\stackrel{\text{scale dependence}}{\longrightarrow}
	G_N(k)
	\stackrel{\text{scale identification}}{\longrightarrow}
	G_N(k(r,\,\chi))\;.
\end{align}
For now, we need not specify to any particular choice of scale dependence and scale identification, because one can determine all of the curvature invariants of the RG-improved spacetime for general $G_N(r,\,\chi)$ ~\cite{Held:2021vwd}. Under such an RG-improvement, the complex invariant of Kerr spacetime reads
\begin{align}
	\label{eq:complex-inv-RG-improved-Kerr}
	\mathcal{C} = 
	\frac{
		(r-i a \chi ) \left[
			r (r-i a \chi )MG_N'' 
			-2(2 r+i a \chi )MG_N'
		\right]
		+6(r+i a \chi )MG_N
	}{
		6\,(r-i a \chi )^{3} (r+i a \chi )
	}\;.
\end{align}
Here and in the following, primes are used to denote derivatives with respect to the radial coordinate.
We remind the reader that $C_{\mu\nu\rho\sigma}C^{\mu\nu\rho\sigma} = \text{Re}\left[48\,\mathcal{C}^2\right]$ and $C_{\mu\nu\rho\sigma}\overline{C}^{\mu\nu\rho\sigma}= \text{Im}\left[48\,\mathcal{C}^2\right]$.
Since the RG-improved spinning spacetime is no longer Ricci flat, it is characterized by a second complex invariant
\begin{align}
	\label{eq:complex-inv-2-RG-improved-Kerr}
	\mathcal{C}_2 = 
	\mathcal{C}\times\left[\frac{
		M\left(
			r\left(r^2+a^2 \chi ^2\right)G_N''
			-2(r^2-a^2 \chi^2 )G_N'
		\right)
	}{
		\left(r^2+a^2 \chi ^2\right)^2
	}\right]^2\;,
\end{align}
which in turn is related to the mixed curvature invariants 
\bea
R^{\mu\nu}R^{\rho\sigma}C_{\mu\nu\rho\sigma} &=& \text{Re}\left[4\,\mathcal{C}_2\right],\\
R^{\mu\nu}R^{\rho\sigma}\overline{C}_{\mu\nu\rho\sigma} &=& \text{Im}\left[4\,\mathcal{C}_2\right]. 
\eea
In addition, there are two polynomially independent\footnote{It has not been proven that the displayed invariants are polynomially independent, hence there could, in principle, be further polynomial relations among them.
} Ricci invariants, i.e.,
\begin{align}
	g^{\mu\nu}R_{\mu\nu} &= \frac{
		2 \left(
				r \left(r^2+a^2 \chi ^2\right) MG_N''
				+2 a^2 \chi ^2MG_N'
			+\left(2 r^2 MG_N'\right)
		\right)
	}{
		\left(r^2+a^2 \chi^2\right)^{2}
	}\;,
	\\
	R_{\mu\nu}R^{\mu\nu} &= \frac{
		2 \left(
			\left(
				r \left(r^2+a^2 \chi ^2\right) MG_N''
				+2 a^2 \chi ^2MG_N'
			\right)^2
			+\left(2 r^2 MG_N'\right)^2
		\right)
	}{
		\left(r^2+a^2 \chi^2\right)^{4}
	}\;.
\end{align}
All other curvature invariants of the general RG-improved spacetime can be expressed as polynomials of the above invariants~\cite[App.~D]{Held:2021vwd}. It is quite remarkable that no curvature invariants depend on angular $\chi$-derivatives of the Newton coupling $G_N(r,\chi)$.
\\

With this complete polynomially independent set of curvature invariants at hand, we are in a position to derive generic conditions on $G_N(r,\chi)$ which are required to remove the curvature singularity of Kerr spacetime.
We observe that the only potential curvature singularity occurs when $\chi=0$ and $r=0$.
Just like for the spherically-symmetric case in Sec.~\ref{sec:spherical-BHs}, the removal of this singularity depends on the leading exponent $n$ in the expansion of $G_N(r,\chi)$ close to $\chi=0$ and $r=0$, i.e., $G_N(r,\chi)\sim r^n + \mathcal{O}(r^{n+1}\chi)$. Once again, we find that:
\begin{itemize}
	\item 
	If $n=3$, the curvature invariants are finite at $r=0$. (There is a subtlety about uniqueness of the limit under exchange of $r\rightarrow0$ and $\chi\rightarrow0$ which we will discuss below.)
	\item
	If $n>3$, all curvature invariants vanish and the spacetime is locally flat in the center.
	\item
	If $n<3$, the curvature singularity remains, although it is weakened for $0<n<3$.
\end{itemize}

We can compare this to the outcome of the different scale identification procedures introduced in Sec.~\ref{sec:construction}. In contrast to the spherically-symmetric case, we will see that the choice of scale identification makes a difference when applied to the case of Kerr spacetime.

We start with the scale-identification with radial geodesic distance which is closest in spirit to the flat-spacetime RG-improvement of the Coulomb potential, cf.~Eq.~\eqref{eq:scale-id_geodesic-distance}. 

Ref.~\cite{Reuter:2010xb} uses a radial path in the spacetime, 
\be
k= k(r, \theta) = \left(\int_0^r d\bar{r}\, \sqrt{\Big|\frac{\bar{r}^2 + a^2 \cos^2 \theta}{\bar{r}^2 + a^2 - 2m \bar{r}} \Big|} \right).\label{eq:scaleidReuterTuiran}
\ee
The integral can be performed, e.g., in the equatorial plane, $\theta =\pi/2$, which is in fact the choice the authors work with. It is thereby assumed that quantum-gravity effects have the same size at all values of $\theta$ for a given $r$. The resulting scaling of the Newton coupling is
\be
G_N(r) \sim \frac{G_0}{4 
\xi} r^4 +\mathcal{O}(r^5)\;.
\ee
It is non-local in a sense that it uses a quantity defined at $\theta=\pi/2$ also away from $\theta=\pi/2$.

A similar non-local approximation has been made for the scale-identification with curvature invariants.
Ref.~\cite{Pawlowski:2018swz} uses the Kretschmann scalar in the equatorial plane, i.e., 
\be
k=\xi \left(K(r, \theta=\pi/2)) \right)^{\frac{1}{4}}= \left(\frac{48 G_0^2\, M^2}{r^6}\right)^{\frac{1}{4}}.\label{eq:scaleidStock}
\ee
Also here the assumption is that quantum-gravity effects have the same size at all angles. In fact, one can see that the scale identification effectively neglects all effects due to the breaking of spherical symmetry, i.e., the Kretschmann scalar in the equatorial plane is proportional to the one in a non-spinning Schwarzschild spacetime. The resulting scaling of the Newton coupling is
\be
G_N(r) \sim \frac{1}{4 \sqrt{3}\xi\, M} r^3 + \mathcal{O}(r^4).\label{eq:scalingStock}
\ee
The scale-identification with the curvature scales of an equivalent Schwarzschild black hole thus results in the critical scaling that is necessary to remove the ring singularity of Kerr spacetime.

In contrast to the above, local physics in the Kerr spacetime depends on the angle with respect to the spin axis, cf.~discussion around Fig.~\ref{fig:spinning-Kerr-invariants}. Thus, any scale identification that means to capture this local physics will necessarily need to take into account $\theta$-dependence. 

This was first discussed in~\cite{Held:2019xde}, where the $\theta$-dependence of the Kretschmann scalar was (partially) taken into account. 
The reasoning
starts from the observation that the singularity in Kerr spacetime is a ring singularity, such that curvature invariants only diverge in the equatorial plane. Accordingly, curvature invariants such as the Kretschmann scalar show an angular dependence, cf.~Eq.~\eqref{eq:complex-inv-Kerr}.
Accordingly, the curvature is larger, i.e., closer to the Planck scale, in the equatorial plane than at other values of $\theta$. One may thus argue that quantum-gravity effects are overestimated by the choices in \cite{Reuter:2010xb,Pawlowski:2018swz}. Instead, the angular dependence of Eq.~\eqref{eq:complex-inv-Kerr} should be taken into account.
The angular dependence of curvature invariants is more comprehensively accounted for in \cite{Eichhorn:2021iwq}, see also Sec.~\ref{sec:spinning-line-element} below. This results in the same scaling as Eq.~\eqref{eq:scalingStock} in the equatorial plane.

In contrast to all choices above, at larger distances, relevant, e.g., for black-hole thermodynamics, \cite{Litim:2013gga} does not use any information on the geometry at high curvature, neither local nor non-local, but identifies $k \sim r$ based on dimensional grounds, where $r$ is the radial coordinate in Boyer-Lindquist coordinates. The resulting scaling is
\be
G_N(r) \sim \frac{G_0}{\xi}r^2 + \mathcal{O}(r^4).
\ee
As one might expect from such a large-distance approximation, this scaling is insufficient to remove the curvature singularity.

Irrespective of the choice of scale identification, or at least among the choices discussed above, we find the following universal features.
\begin{itemize}
\item 
A resolution of the curvature singularity, cf.~Tab.~\ref{tab:fateofsingularity}.
\item a more compact event horizon than in the classical case.
\item the possibility of a horizonless spacetime, obtained through a merging of outer and inner horizon, when quantum effects are made large.
\end{itemize}

\begin{table}[!t]
\begin{tabular}{c|c|c|c|c|c|}
$\rm{Weyl}^2$ at $r,\chi \rightarrow 0$&$\rm{Weyl}^2$ at $\chi,r \rightarrow 0$  &Ricci at $r, \chi \rightarrow 0$ & Ricci at $\chi,r \rightarrow 0$ & scale-identification& Ref.\\ \hline\hline
0& 0& 0& 0 & Eq.~\eqref{eq:scaleidReuterTuiran} & \cite{Reuter:2010xb}\\ \hline
0 & 0 &  $2 \sqrt{3}/(
\, \xi)$ &0& $k \sim \sqrt{K(\theta=\pi/2)}$ & \cite{Pawlowski:2018swz}\\ \hline
0 & $-1/(
\xi)^2$ & $2 \sqrt{3}/(
\xi)$ & 0 & Eq.~\eqref{eq:scaleidspinning} & \cite{Eichhorn:2021iwq}\\ \hline
\end{tabular}
\caption{\label{tab:fateofsingularity} We list the behavior of two curvature invariants for the various scale identifications. The singularity is resolved if a diffeomorphism invariant quantity is used in the scale identification, such as a radial path or a curvature invariant.
}
\end{table}

In the following, we focus on the local scale identification. This brings us close in spirit to effective field theory, where quantum effects lead to higher-order curvature terms. These become important at a given spacetime point, when the local curvature exceeds a critical value. In the same vein, we expect higher-order curvature terms in the effective action for asymptotic safety to become important, when the local curvature exceeds a critical value. To mimick this, we choose a local curvature scale for the scale identification in Sec.~\ref{sec:spinning-line-element} below.

To conclude this section, we point out that  Boyer-Lindquist type coordinates with coordinate singularities can be used in the non-local scale identifications, where $G_N = G_N(r)$. In contrast, the local scale identification, in which $G_N= G_N(r, \theta
)
$ is not compatible with Boyer-Lindquist type coordinates. The reason is that the coordinate singularity of the Kerr spacetime in Boyer-Lindquist coordinates can be upgraded to a curvature singularity in the presence of $G_N= G_N(r, \theta
)
$. In \cite{Held:2021vwd}, it is discussed more generally, that RG improvement can result in such pathologies, if coordinates are not chosen with due care.

\subsection{The line element for the locally Renormalization-Group improved Kerr spacetime}
\label{sec:spinning-line-element}

\emph{...where we discuss the  scale identification with local curvature in detail and derive the resulting axisymmetric spacetime.}
\\

We will work in horizon-penetrating coordinates, to avoid 
turning coordinate singularities into unphysical curvature singularities~\cite{Held:2021vwd}.
Our starting point is the Kerr metric in ingoing Kerr coordinates, cf.~Eq.~\eqref{eq:regular-spinning-BH-metric-ingoing-Kerr}.

To capture the angular dependence of the spinning black-hole spacetime, we perform a scale identification based on the local curvature invariants of the Kerr spacetime. The Kerr spacetime is fully characterized by the single complex invariant $\mathcal{C}$, cf.~Eq.~\eqref{eq:complex-inv-Kerr}. Thus, we identify
\begin{align}
\label{eq:scaleidspinning}
	k^2 = \sqrt{48}\,\widetilde{\xi}\,|\mathcal{C}|
	\equiv
	\widetilde{\xi}\,\left(
		\left(C_{\mu\nu\rho\sigma}C^{\mu\nu\rho\sigma}\right)^2
		+ \left(C_{\mu\nu\rho\sigma}\overline{C}^{\mu\nu\rho\sigma}\right)^2
	\right)^{1/4}\;.
\end{align}

This implements the more general scale identification with the root-mean-square of all polynomially independent curvature invariants discussed in Sec.~\ref{sec:construction}. Moreover, it smoothly connects to the spherically symmetric case in the limit of $a\rightarrow 0$, cf.~Sec.~\ref{sec:spherical-BHs}.

With the approximate scale-dependence as given in Eq.~\eqref{eq:runningNewton}, we find the RG-improved Newton coupling
\begin{align}
	\label{eq:GN-local-spinning}
	G_N(r,\theta) = 
	\frac{G_0}{
		1
		+ \frac{G_0^2 M
		\,\xi}{
			\left(r^2 + a^2\,\cos(\theta)^2\right)^{3/2}
		}
	}
	\;,
\end{align}
where we have used the re-definition $\xi = \sqrt{48}\widetilde{\xi}/g_{N,\ast}$, as in the spherically symmetric case.
This functional dependence characterizes the respective RG-improved Kerr black hole which can be represented either by using Eq.~\eqref{eq:GN-local-spinning} by the metric (Eq.~\eqref{eq:regular-spinning-BH-metric-ingoing-Kerr}) or equivalently by the curvature invariants discussed in Sec.~\ref{sec:construction}.

\subsection{Spacetime structure and symmetries}
\emph{...where we discuss that the event horizon of a spinning, asymptotic-safety inspired black hole is more compact than its classical counterpart. Further, we highlight that for such a black hole, the Killing and the event horizon do not agree for the local scale-identification. Moreover, its symmetry properties differ from Kerr: circularity, which is an additional isometry of Kerr spacetime, is not realized for this black hole.}

Working our way from asymptotic infinity to the regular core\footnote{Curvature invariants are finite, but not single-valued at $r \rightarrow 0, \chi \rightarrow 0$.
For $r>0$, the spacetime is not geodesically complete and needs an extension to $r<0$. There, the dependence on $r^2$ (instead of $r$) in Eq.~\eqref{eq:GN-local-spinning} ensures the absence of curvature singularities, see \cite{Torres:2017gix}.} of the asymptotic-safety inspired black hole, we encounter the photon shell\footnote{In the spherically-symmetric case, closed photon orbits can only arise at a 2-dimensional surface at fixed radius, cf.~Sec.~\ref{sec:sph-sym-spacetime-structure} -- hence the name photon sphere. For stationary axisymmetric spacetimes and, in particular, for Kerr spacetime~\cite{1973blho.conf..215B, 2003GReGr..35.1909T}, several classes of closed photon orbits can occur which cover an extended 3-dimensional region -- hence, the name photon shell.}, the ergosphere, and then the event horizon. While all these surfaces/regions persist, they are more compact than for the respective Kerr black hole. The reason is the same as in the spherically symmetric case: the quantum effects weaken gravity and therefore a black hole of a larger mass would be needed to exert the same effects on timelike and null geodesics; first, introducing a rotation for timelike geodesics at the ergosphere, then, preventing stable photon orbits at the photon sphere and finally, redirecting outgoing null geodesics towards smaller radii. Thus, if the asymptotic mass $M$ is held fixed, these surfaces are all located at smaller radii than in the classical case.

The persistence of  the photon shell and the ergosphere is relevant for phenomenology: Supermassive black holes may launch jets through the Blandford-Znajek process, which requires the ergosphere to be present. The photon shell comprises the region in spacetime in which unstable closed photon orbits are possible and hence results in photon rings in black-hole images, which may be detectable with very-large-baseline interferometry \cite{Broderick:2022tfu}.
As we discuss below, the geodesic motion in the RG-improved black-hole spacetime is no longer separable. Thus, we cannot present analytical results for closed photon orbits in the photon shell. A similar proof of increased compactness goes through in the equatorial plane as it does for the spherically symmetric case, cf.~Sec.~\ref{sec:sph-sym-spacetime-structure}; at arbitrary $\theta$, increased compactness can be shown analytically. The increase in compactness is not uniform; but, due to frame-dragging, larger on the prograde side; and also larger in the equatorial plane than away from it.
In Sec.~\ref{sec:towards-observation}, we briefly discuss the qualitative effects resulting from numerical studies.
\\

\begin{figure}[!t]
\centering
\includegraphics[width=0.7\linewidth]{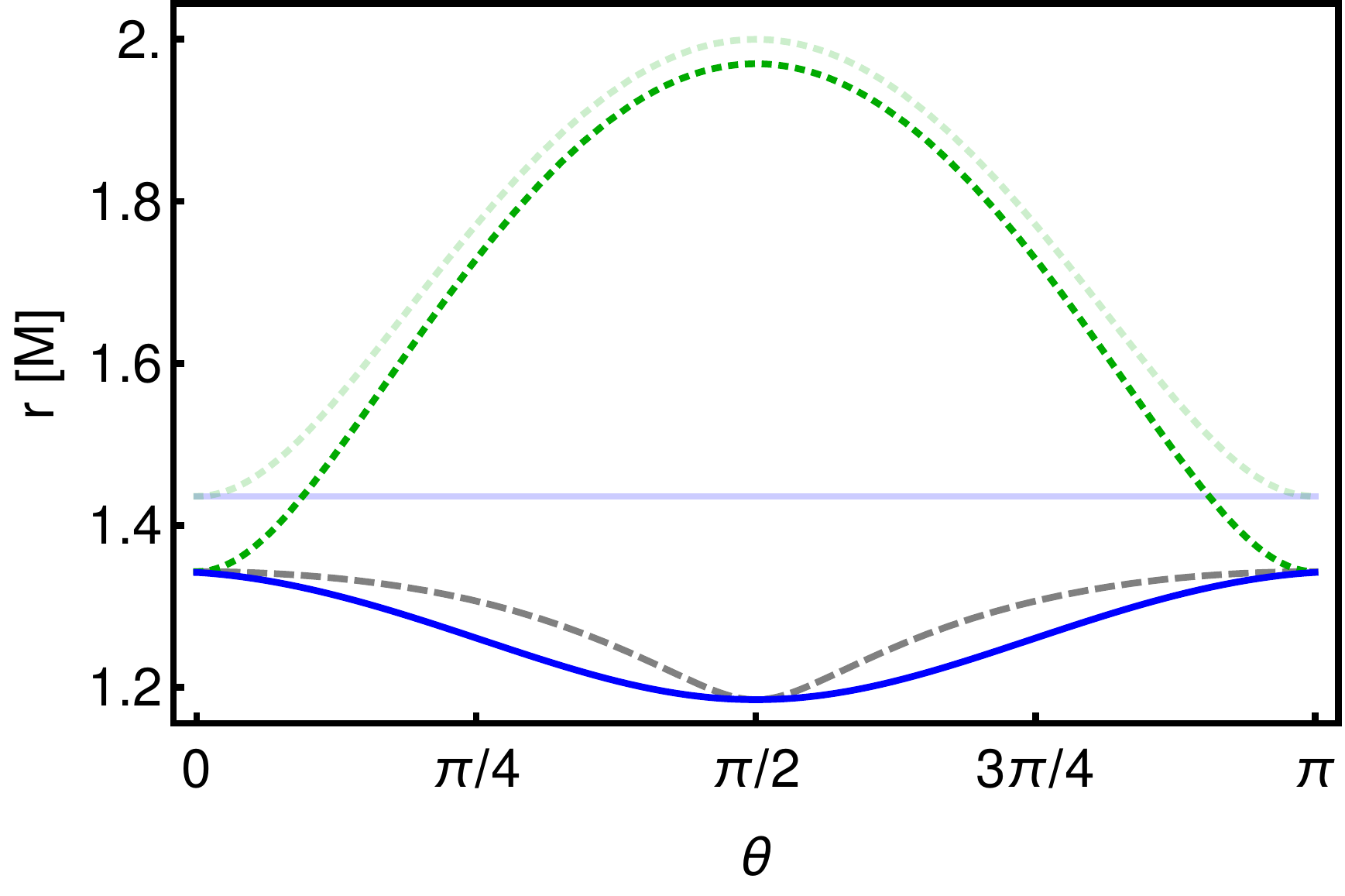}
\caption{
\label{fig:spinningsurfaces}
We show the ergoregion (green dotted), the Killing horizon (gray dashed), and the event horizon (blue continuous) of the RG-improved black hole, cf.~Sec.~\ref{sec:spinning-line-element} for $a=0.9\,M$ and near-critical deviation parameter $\xi=0.13$. In lighter shading and for reference, we also show the Killing/event horizon (blue continuous) and the ergosphere (dotted green) of the corresponding Kerr black hole.}
\end{figure}

The overall increase in radial compactness depends on $\chi=\cos(\theta)$. This is because the quantum-gravity effects, i.e., the weakening of gravity, is always strongest in the equatorial plane. Therefore all surfaces are most compact at $\chi=0$, cf.~Fig.~\ref{fig:spinningsurfaces}. For instance, the horizon exhibits a ``dent", i.e,. the deviation between the classical and the quantum horizon is largest at $\chi=0$, cf.~Fig.~\ref{fig:spinningsurfaces}.

There is an additional effect that relates to the event horizon: in the Kerr spacetime, the event horizon is also a Killing horizon, so that a constant surface gravity can be defined. This is no longer the case for the asymptotic-safety inspired black hole, see \cite{Eichhorn:2021iwq}.\footnote{Consequences for black hole thermodynamics have not been explored yet.} The event horizon, described by Eq.~\eqref{eq:horizon}, which results in $g^{rr}=0$ for the Kerr spacetime, features an angular dependence, necessitating a numerical solution of the horizon equation. Specifically, the horizon equation for the spinning asymptotic-safety inspired black hole is
\be
g^{rr}(r= H(\theta),\theta)+ g^{\theta\theta} (r= H(\theta),\theta) \left(\frac{dH}{d\theta} \right)^2=0.
\ee
The Killing horizon in turn is described by the condition
\be
g_{u\phi}^2-g_{uu}g_{\phi\phi}=0.
\ee
One can confirm that these are the same condition in the Kerr spacetime, but not in the RG improved spacetime based on the local improvement procedure.

The inner horizon which is a Cauchy horizon persists; thus, the RG improvement can only remove some of the pathologies of the Kerr spacetime. Because the RG-improvement only takes the quantum-gravity effect on the most relevant curvature coupling, i.e., the Newton coupling, into account, one may expect that a more complete treatment is more accurate at large curvatures; thus, the line element may be a fair description of a black hole in asymptotically safe gravity at large enough $r$, but miss effects which may, e.g., resolve the Cauchy horizon (and the $r<0$ region).
\\

The RG improvement also modifies the spacetime symmetries. The two Killing symmetries, stationarity and axisymmetry are left intact. The only way to break them would be to build an artificial dependence on time and/or azimuthal angle into the RG improvement.\\
However, Kerr spacetime has a less obvious symmetry, which is called circularity. 
Circularity is an isometry of Kerr spacetime which is easiest to see in Boyer-Lindquist coordinates $(t, r, \theta, \phi)$: under a simultaneous mapping of $t \rightarrow -t$ and $\phi \rightarrow -\phi$, the spacetime is invariant. This is no longer the case for the RG improved black hole (unless one works with a non-local scale identification as in Eq.~\eqref{eq:scaleidStock}).\footnote{Because Killing symmetries can be made manifest in the spacetime metric, RG improvement can always be made to respect Killing symmetries. However, any symmetry that is not a Killing symmetry, but instead only expressible as a condition on the Riemann tensor, need not be respected by the RG improvement.} 
The more general way of testing circularity is by testing the following conditions on the Ricci tensor, see \cite{Papapetrou:1966zz}:
\bea
\xi_1^{[\mu} \xi_2^{\nu}\nabla^{\kappa}\xi_1^{\lambda]} &=& 0 \mbox{ at at least one point,}\notag\\
\xi_2^{[\mu} \xi_1^{\nu}\nabla^{\kappa}\xi_2^{\lambda]} &=& 0 \mbox{ at at least one point,}\notag\\
\xi_1^{\mu} R_{\mu}^{\,\, [\nu}\xi_2^{\kappa}\xi_1^{\lambda]} &=& 0 \mbox{ everywhere,}\notag\\
\xi_2^{\mu} R_{\mu}^{\,\, [\nu}\xi_1^{\kappa}\xi_2^{\lambda]} &=& 0 \mbox{ everywhere} \label{eq:circular}.
\eea
Because the Killing vector for axisymmetry vanishes on the axis of rotation, the first two conditions are always satisfied. The latter two conditions are nontrivial and impose a condition on the Ricci tensor. Because spinning black holes are vacuum spacetimes in GR, all four conditions hold there. 
In contrast, one should not expect the Ricci tensor to be special (in the sense of respecting the above symmetry) in a quantum theory any more -- at least not if effective field theory reasoning, which determines the size of deviations from the Kerr spacetime through local curvature invariants, holds.\footnote{See, however, \cite{Xie:2021bur} for a proof that circularity holds for black holes which are perturbatively connected to the Kerr solution.}\\
Finally, there is also no generalized Carter constant, i.e., geodesic motion is not separable, and energy and angular momentum are the only conserved quantities. This is because the spacetime no longer features a Killing tensor, as it does in GR, where this tensor gives rise to the Carter constant.\\
As a consequence, RG improved spinning black holes are not included in the parameterizations of black-hole spacetimes in \cite{Johannsen:2011dh,Johannsen:2013szh}, which assumes a generalized Carter constant; nor in \cite{Konoplya:2016jvv}, which assumes circularity, but requires a more general parameterization \cite{Delaporte:2022acp}.

\subsection{Horizonless spacetimes}
\label{sec:horizonless}
\emph{...where we discuss how increasing $\xi$ leads to a loss of horizon at $\xi_{\rm crit}(a)$. For near-critical spin, $\xi_{\rm crit}(a)$ can be made very small, such that Planck-scale modifications of the black-hole spacetime suffice to dissolve the horizon. }

The inner and outer horizon approach each other, when $\xi$ increases. Intuitively, this follows from the fact that the limit $\xi \rightarrow \infty$ is a gravity-free limit, i.e., the line element becomes that of Minkowski spacetime. The transition between  $\xi=0$ (a Kerr black hole) and  $\xi\rightarrow \infty$ (Minkowski spacetime) must therefore include a mechanism in which the event horizon of the Kerr spacetime is resolved. Mathematically, this can occur when the solutions to the horizon equation become complex, which they can only do as a pair. This requires them to become equal just before they become complex, i.e., the inner and outer horizon annihilate.

The critical value of $\xi$, at which the annihilation occurs depends on the spin parameter $a$. For a classical Kerr black hole, the horizons annihilate at $|a|=M$. Close to $|a| =M$, where the horizons are close to each other (i.e., at values of $r$ which are close), the tiny increase in compactness that is caused by a Planck-scale $\xi$ is enough to trigger an annihilation of the horizons. More specifically, if we set $|a| = M\left(1- \delta a \right)$, we obtain
\be
\delta a= \frac{\xi}{2} \frac{G_0}{M^2} + \mathcal{O}(\xi^2).
\ee
For Planck-scale effects, i.e., $\xi=1$, the black hole therefore has to be very close to the critical spin value. It is an open question whether or not this is achievable in settings beyond GR and specifically asymptotically safe gravity, see \cite{Li:2013sea,Jiang:2020mws,Yang:2022yvq} for general studies for regular black holes.

\subsection*{Further reading}
\begin{itemize}
	\item
	{\bf Impact of the cosmological constant:}
	\\
	Ref.~\cite{Pawlowski:2018swz} also considers Kerr-(A)dS black holes and includes the cosmological constant in the RG improvement. Just like in the spherically symmetric case, the scaling of the cosmological constant $\Lambda \sim k^2$ dominates the large-curvature behavior and reintroduces a curvature singularity. Again, as in the case of spherical symmetry, performing the RG improvement in this way appears to be insufficient to capture the expected quantum effects. This again highlights the limitations of the RG improvement procedure in settings with several physical scales, where one needs to distinguish carefully which scales matter for the quantum effects and which do not.
	\item
	{\bf Change of coordinates after RG improvement:}
	\\
	For the nonlocal scale-identifications, a coordinate transformation from Boyer-Lindquist coordinates to horizon-penetrating coordinates is straightforward and mimicks the classical coordinate transformation, see, e.g., \cite{Reuter:2010xb}. In contrast, the coordinate transformation is more involved for the local scale identification, see \cite{Delaporte:2022acp}, and leads to a line element which has additional non-vanishing components in comparison to the classical case. This is directly linked to the breaking of a spacetime symmetry, circularity, by the quantum-gravity effects, see the discussion above.
\end{itemize}

\section{Formation of asymptotic-safety inspired black holes}\label{sec:formation}
\emph{...where we discuss the formation of asymptotic-safety inspired black holes. First, we analyze
the spacetime structure of RG-improved black holes formed from gravitational collapse. Second, we discuss whether the formation of black holes in high-energy scattering processes is to be expected in asymptotically safe gravity.}

\subsection{Spacetime structure of gravitational collapse}
\emph{...where we review the spacetime structure and the fate of the classical singularity in asymptotic-safety inspired gravitational collapse.}
\\

Up to here, we have looked at black holes as ``eternal'' objects, i.e., as idealized stationary solutions to gravitational theories. As such, they are merely of academic interest, or, at best constitute approximations to time-dependent systems. 
However, black holes gain physical importance because we expect them to form as the generic final state of gravitational collapse -- at least in GR.

In GR, singularity theorems~\cite{Penrose:1964wq} imply that gravitational collapse ends in the formation of geodesic singularities.
At the same time, gravitational collapse generically results in the formation of black-hole horizons, which has led to the weak cosmic censorship conjecture in GR~\cite{Penrose:1969pc}. In colloquial terms, weak cosmic censorship states that, given physical and generic\footnote{
The word `physical' refers to physically realistic matter models~\cite{Wald:1997wa, Penrose:1999vj}, see also the discussion below Eq.~\eqref{eq:mass-function-Vaidya}. 
The word `generic' is important since non-generic initial conditions result in counter-examples to weak cosmic censorship~\cite{Choptuik:1992jv}, even in restricted sectors (spherically symmetric GR with a real massless scalar field) for which the mathematical conjecture has been proven~\cite{Christodoulou:1999a}.
}
initial conditions, the dynamics of GR ensures that all singularities are hidden behind horizons with respect to asymptotic observers, cf.~\cite{Christodoulou:1999} for a mathematical formulation. 

Weak cosmic censorship in GR may suggest a form of ``quantum gravity censorship": if regions of very high/diverging curvature are generically hidden behind horizons, then so are the effects of quantum gravity (at least, where they are large). Models of gravitational collapse which lead to naked singularities within GR are thus of particular interest for quantum gravity, because they may lead to regions of spacetime with large quantum-gravity effects, and not shielded from asymptotic observers.
\\

A simple model for the exterior spacetime of gravitational collapse is given by the ingoing Vaidya metric~\cite{Vaidya:1951zza, Vaidya:1951zz, Vaidya:1966zza}
\begin{align}
	ds^2 = -f(r,v)\,dv^2 + 2\,dv\,dr + r^2\,d\Omega^2
	\quad\text{with}\quad
	f(r,v) = 1-\frac{2\,m(v)}{r}\;.
\end{align}
For $m(v)=M$, the Vaidya spacetime reduces to Schwarzschild spacetime in ingoing Eddington-Finkelstein coordinates. For general $m(v)$, a calculation of the Einstein tensor confirms that this metric is a solution to the Einstein field equations with stress-energy tensor
\begin{align}
	T_{\mu\nu} = \frac{\dot{m}(v)}{4\pi\,r^2} l_\mu\,l_\nu\;,
\end{align}
where $l_\mu=-\partial_\mu\,v$ is tangent to ingoing null geodesics. This stress-energy tensor describes so-called null dust, i.e., a pressure-less fluid with energy density $\rho = \dot{m}/(4\pi\,r^2)$ and ingoing fluid 4-velocity $l_\mu$. For null dust which satisfies the weak (or null) energy condition, the energy density $\rho\geq 0$ must not be negative, and we see that the mass function $m(v)$ must not decrease.

The Vaidya spacetime -- as well as other null-dust collapse models -- admit initial conditions which form spacelike singularities that always remain hidden behind horizons. However, they also permit initial conditions which form naked singularities, visible to (at least some) asymptotic observers, cf.~\cite{Joshi:2011rlc} for review.
For instance, the simple choice of a linearly increasing mass function
\begin{align}
\label{eq:mass-function-Vaidya}
	m(v) = \begin{cases}
		0 & v < 0 \\
		\lambda\,v & 0 \leqslant v \leqslant \overline{v} \\
		\overline{m} & v>\overline{v}
	\end{cases}
\end{align}
results in a naked curvature singularity if $\lambda\geqslant 1/(16\,G_0)$ and has been one of the earliest models discussed in the context of cosmic censorship in GR. At this point, it is important that pressureless fluids may not be considered as describing physically relevant initial data since they can form density singularities even in flat spacetime, cf.~\cite{Wald:1997wa,Penrose:1999vj} 
for a more complete discussion.
\\

Irrespective of whether or not the above null-dust collapse-models are considered as physically relevant counterexamples to the cosmic-censorship conjecture, they may be used as toy models for potential classical cosmic-censorship violations.
Thus, we
ask whether quantum effects can remove the naked singularity. We highlight, that (i) if, in GR, physically relevant cosmic censorship violations exist, and (ii) if the naked singularity is lifted by quantum gravity effects, then the result is a quantum-gravity region not shielded from asymptotic observers by a horizon. Depending on the properties of such a region, its effects may be detectable, e.g., in electromagnetic radiation emitted from or traversing the region.\footnote{Note that there are spacetimes which satisfy weak cosmic censorship in GR, but their RG-improved counterparts do not satisfy ``quantum gravity censorship": For instance, a Kerr-black hole, with sub-, but near-critical spin is an example, since its RG-improved counterpart can be horizonless, cf.~\cite{Eichhorn:2022bbn} and Sec.~\ref{sec:horizonless}.}\\

Spherically-symmetric RG-improved gravitational collapse of null-dust models has been investigated in~\cite{Casadio:2010fw, Torres:2014gta, Torres:2014pea, Torres:2015aga, Bonanno:2016dyv, Bonanno:2017zen}.
The specific example of the Vaidya spacetime has been RG-improved in~\cite{Bonanno:2016dyv, Bonanno:2017zen}. In this work, the running Newton coupling has been approximated by Eq.~\eqref{eq:runningNewton}. Moreover, the energy density $\rho$ of the null dust in the Vaidya metric was used to set the RG scale $k$: On dimensional grounds, this implies
\begin{align}
	k^4 \sim \rho
\end{align}
for the scale identification. 
This results in a radially and advanced-time dependent cutoff,
\be
k^2=k(r,v) = \widetilde{\xi}\left(\frac{\dot{m}(v)}{4 \pi r^2} \right)^{\frac{1}{4}},
\ee
leading to the RG improved lapse function
\be
f(r,v)=1- \frac{2G_0\, m(v)}{r+\frac{G_0}{\sqrt{4\pi}}\xi^2 \sqrt{\dot{m}(v)}}.
\ee
For these choices, the antiscreening effect of RG improvement is not strong enough to fully remove the curvature singularity. The resulting RG-improved Vaidya metric still contains a naked, although now integrable, curvature singularity~\cite{Bonanno:2016dyv}. The authors also highlight that quantum effects even seem to broaden the range of initial conditions for which the formed singularity is naked.

A different null-dust collapse-model (Lemaitre-Tolman-Bondi~\cite{1933ASSB...53...51L,tolman1934effect,10.1093/mnras/107.5-6.410}) has been RG-improved in~\cite{Torres:2014gta}, while also matching the interior solution to an exterior RG-improved Schwarzschild spacetime. Notably, \cite{Torres:2014gta} uses the scale identification with geodesic distance (originally proposed in~\cite{Bonanno:2000ep}). In that case, the central curvature singularity (naked or hidden) is removed by the RG improvement.

While the two null-dust collapse-models are distinct and an explicit comparative study remains to be performed, we expect that a key difference is the use of distinct scale identifications. It is a universal result that RG-improvement weakens the curvature singularities which arise in simple null-dust models of gravitational collapse, cf.~also~\cite{Casadio:2010fw}. Whether or not the curvature singularities are fully lifted, seems to depend on the scale identification, just as it does for stationary black holes.

\subsubsection*{Further reading}
\begin{itemize}
\item
{\bf RG improvement of gravitational collapse based on decoupling:}
\\
The RG-improvement of spherically-symmetric gravitational collapse has recently been investigated~\cite{Borissova:2022mgd} in the context of the decoupling mechanism~\cite{Reuter:2003ca} and with the iterative RG-improvement developed in~\cite{Platania:2019kyx}, to find a self-consistent RG improved model that describes a black hole from its formation through gravitational collapse to its evaporation through Hawking radiation. The study finds that the classical curvature singularity of a Vaidya-Kuroda-Papapetrou spacetime is weakened, but not fully lifted.
\end{itemize}
\subsection{Formation of black holes in high-energy scattering}
\emph{...where we discuss the expectation that scattering at transplanckian center-of-mass energies must necessarily result in a black hole and point out that it may not be realized in asymptotically safe gravity.}
\\

In GR, confining an energy density or mass to a spacetime region smaller than the associated Schwarzschild radius is expected to lead to black-hole formation. Thorne's hoop conjecture accordingly says that if one can localize particles to within their Schwarzschild radius, a black hole will form. If one uses a particle's de Broglie wavelength as the radius of a region within which the particle is localized, black-hole formation is expected to set in at the Planck scale. This argument, which has also been studied numerically in \cite{East:2012mb}, appears to be at odds with the idea of asymptotic safety: if indeed transplanckian scattering inevitably leads to black holes, then black holes dominate the high-energy spectrum of a theory. In turn, because more massive black holes have larger area, there is a UV-IR-duality in the theory: the deeper one tries to probe in the UV, the further in the IR one ends up. If such a picture is correct, there does not appear to be any room for asymptotic safety at transplanckian scales.

We propose, see also \cite{Basu:2010nf,Bonanno:2020bil} that this apparent problem can be resolved as follows: to decide whether or not a black hole forms, it is not enough to state that the energy is transplanckian, because transplanckian energies are reached both in classical and in quantum processes. That black holes form in transplanckian, classical processes is not in doubt -- after all, astrophysical black holes all have masses corresponding to highly transplanckian energies. Such processes are characterized by an impact parameter much larger than the Planck length. In contrast, we are interested in a regime where energies are transplanckian and impact parameters subplanckian, because this is the quantum regime. However, because it is the quantum regime, it is incorrect to apply GR, and therefore it is a priori unclear whether or not a black hole forms.

The outcome of scattering processes at transplanckian center-of-mass energy and subplanckian impact parameter depends on the dynamics of quantum gravity. To obtain some intuition for what the outcome could be in asymptotically safe quantum gravity, we perform an RG improvement of the hoop conjecture. To that end, we simply take the classical Schwarzschild radius, $R_{S,\, \rm cl} = \sqrt{2 G_{N} M}$ (with $c=1$), and upgrade $G_{N}$ to its scale-dependent counterpart, $G_{N}= G_{N}(k)$. We subsequently identify $k = \xi/b$, where $b$ is the impact parameter and $\xi$ is a number of order one. We thereby obtain an RG-improved $R_S$ that we can compare to its classical counterpart. If $R_S(b) \geq R_{S,\, \rm cl}$, then the classical argument underestimates the impact parameter, at which black holes form. Conversely, if $R_S(b) < R_{S,\, \rm cl}$, then the classical argument overestimates the impact parameter, at which black holes form. Because $G_{N}(k<M_{\rm Planck})= \rm const$, the classical and the quantum estimate agree for superplanckian impact parameter. For subplanckian impact parameter, $G_{N}(k> M_{\rm Planck}) \sim k^{-2} \sim 
{b^2}
$ implies that the RG-improved $R_S$ shrinks linearly with decreasing impact parameter:
\be
R_S(b) = \sqrt{2M}\sqrt{\frac{G_0}{1+ \xi \frac{G_0}{b^2}}
}
\approx 
\sqrt{\frac{2M}{\xi}}b
\ee
In this regime, the critical radius at which black-hole formation occurs is therefore smaller than suggested by the classical estimate. Therefore, the classical hoop conjecture does not generically apply in this simple RG-improved setup. Whether this simple argument captures the relevant gravitational dynamics in asymptotic safety is of course an open question.

Whether or not black holes form is therefore currently an unanswered question. The weakening of gravity associated with asymptotic safety implies that the answer may be negative, because, simply put, the Schwarzschild radius decreases faster (as a function of impact parameter) than the impact parameter itself.

\section{Towards observational constraints}
\label{sec:towards-observation}
\emph{...where we first discuss theoretical expectations on $\xi$ and argue that, irrespective of theoretical considerations, any observational avenue to put constrains on deviations from GR, should be explored. In this spirit, we discuss how different types of observations, both gravitational and electromagnetic, can be used to constrain asymptotic-safety inspired black holes.}
\\

Observational insight into quantum gravity is scarce. Nevertheless, it is crucial that theoretical progress should always be confronted with observation. Therefore, in Sec.~\ref{sec:emobs} and \ref{sec:gravobs}, we discuss how the constructed asymptotic-safety inspired black holes can be confronted with recent electromagnetic and gravitational-wave observations of astrophysical black holes and which part of the parameter space may be accessed by current observations. 

By construction, the previously discussed asymptotic-safety inspired black holes derived from an RG-improvement of the Newton coupling introduce a single new physics scale which is tied to $\xi$. All observational constraints will thus effectively constrain the scale $\xi M_{\rm Planck}$. In Sec.~\ref{sec:scale-of-qG}, we discuss how $\xi$ relates to theoretical expectations about the scale of quantum gravity.

In the context of observational constraints, it is important to keep in mind that beyond GR, black-hole uniqueness theorems need not hold. A simple example is Stelle gravity, in which there is more than one spherically-symmetric black-hole solution \cite{Lu:2015psa,Lu:2015cqa,Pravda:2016fue,Podolsky:2019gro} (although in this case thermodynamics~\cite{Lu:2017kzi} and linear instabilities~\cite{Held:2022abx} suggest that only one solution is stable at a given set of parameters). In particular, beyond GR, if there is no black-hole uniqueness, it may well depend on the formation history, which metric describes a black hole, and supermassive and solar-mass-black holes may be described by different metrics.\footnote{Our simple RG-improvement procedure can of course not account for this possibility, because it starts from a unique starting point, namely a Kerr black hole.} Constrains from different populations of black holes can therefore only be merged under the added assumption of black hole uniqueness.

\subsection{The scale of quantum gravity}
\label{sec:scale-of-qG}
\emph{...where we discuss theoretical expectations on the scale of quantum gravity relevant for black hole physics. We argue that a twofold strategy is called for, in which (i) we remain agnostic with respect to theoretical considerations and explore how far observations can probe and (ii) we remain conservative with respect to theoretical considerations and investigate whether Planck-scale modifications can be enhanced to make quantum-gravity effects significant.}
\\

It is often assumed that quantum-gravity effects are negligible in astrophysical black holes. The argument is based on the low value of the curvature at the horizon, compared to the Planck scale. However, the Planck scale is actually based on a simple dimensional analysis that uses no information whatsoever on the dynamics of quantum gravity. Therefore, it is conceivable that in a given quantum-gravity theory, effects are present at other scales; and, indeed, examples exist in quantum gravity theory.
\\
Within asymptotic safety, the relevant scale is the transition scale, at which the scaling of $G_N$ (and/or further gravitational couplings) changes. In pure-gravity calculations which include only the Newton coupling and cosmological constant, this scale roughly agrees with the Planck scale. This implies $\xi \approx 1$, which in turn means that
modifications on horizon scales are tiny, unless the spin is close to criticality, see \cite{Eichhorn:2022bbn}.
Given this initial expectation that $\xi \approx 1$, we argue that a twofold strategy is called for: First, we explore whether effects tied to $\xi \approx 1$ can be enhanced to impact scales that can be probed. We find that spin can act as a lever arm in this context. Second, we challenge this initial theoretical expectation; on the one hand by reflecting on freedom within the theory and on the other hand by remaining agnostic and exploring which range of $\xi$ can be constrained with current observations.
\\
Within the theory, there may be scope for $\xi>1$.
First, the transition scale to the fixed-point regime is known to depend on the matter content of the theory; and can be significantly lower, if a large number of matter fields is present, see \cite{Dona:2013qba, deAlwis:2019aud}.
\\
Further, asymptotic safety has more than one relevant parameter. Each relevant parameter can be translated into a scale. Specifically, there is evidence for three relevant parameters \cite{Falls:2020qhj}: the Newton coupling (which sets the Planck scale), the cosmological constant (which sets the corresponding low-energy scale) and a superposition of two curvature-squared couplings. The last one can set a scale significantly different from the Planck scale; e.g., in \cite{Gubitosi:2018gsl}, the free parameter is used to obtain Starobinsky inflation, which translates into a scale four orders of magnitude below the Planck mass. Instead of accommodating Starobinsky inflation, that same free parameter can be used to lower the scale at which a departure from Einstein gravity is significant even further.\footnote{An inclusion of such higher-order couplings for black holes would require the construction of spinning black-hole solutions in higher-curvature gravity, followed by an RG improvement. Due to the technical complexity of this task, it has not been attempted.}

There is also an argument based solely on GR, independent of the specifics of a quantum gravity theory, that suggests that quantum-gravity effects are not confined to the core of black holes, and may indeed become important at scales where the local curvature of the Kerr spacetime is still low: this argument considers perturbations on top of the Kerr spacetime, which experience  an infinite  blueshift at the Cauchy horizon and thereby destabilize the Cauchy horizon so that its curvature increases and a singularity forms. This suggests that quantum-gravity effects are important at the Cauchy horizon -- a location where the spacetime curvature of the Kerr spacetime is still relatively low.\footnote{This reasoning does not invalidate effective field theory, because not all terms in the effective action remain low. Because perturbations become blueshifted, the kinetic terms describing perturbations of scalars/fermions/vectors/metric fluctuations become large, signalling the breakdown of effective field theory and the need for a quantum theory of gravity.}

The above reasoning, and in fact, any other determination of the scale at which quantum gravity is significant, is so far purely theoretical. Thus, it is critical to place the determination of this scale on a different footing, i.e., an observational one.
In the last few years, new observational opportunities have become available, including most importantly LIGO and the EHT. This enables us to place observational constraints on $\xi$ without relying purely on theoretical considerations.

\subsection{Constraints from electromagnetic signatures}
\label{sec:emobs}

\emph{... where we explain how data from electromagnetic observations could now and in the future be used to constrain the scale of quantum gravity in asymptotic safety. We also highlight the importance of nonlocal observables such as the presence/absence of an event horizon.}
\\

The effects introduced by the RG improvement can be recast as an effective mass function $M(r,\chi)$, much like for other classes of regular black holes in the literature. Because the angular dependence is subleading, we discuss only the radial dependence here.
A reconstruction of this mass function requires observations that probe the spacetime at several distinct distance scales: for instance, at large distances $r\gg r_g$, where $r_g$ is the gravitational radius, $M$ approaches a constant; at shorter distances, $M$ is always smaller than this asymptotic constant. For two supermassive black holes, M $87^\ast$ and Sag $A^\ast$, measurements at two different distances are available: for each of these supermassive black holes, $M$ can be inferred from stellar orbits at $r \gg r_g$, and $M$ can also be inferred from the diameter of the black-hole shadow imaged by the Event Horizon Telescope. If $\xi$ is very large, then these two measurements are expected to be significantly different from each other. Given that for both supermassive black holes, they agree within the errors of the respective measurements, constraints on $\xi$ can be derived. In practise, these constraints are of the order $\xi \simeq 10^{95}$ for M $87^{\ast}$, \cite{Held:2019xde}, see also \cite{Kumar:2019ohr}.
\\

\begin{figure}[t!]
\centering
\includegraphics[width=0.49\linewidth]{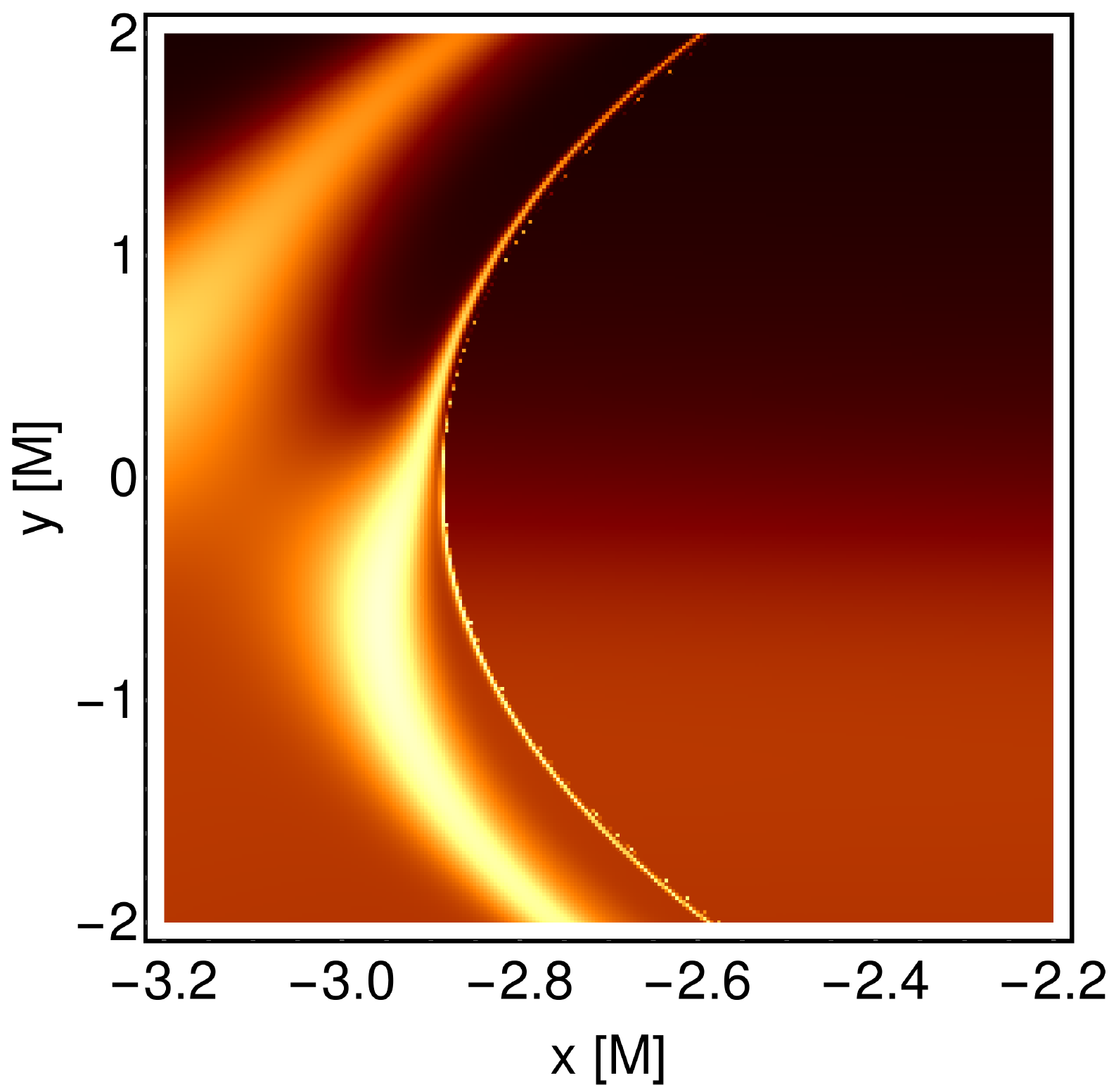}
\includegraphics[width=0.49\linewidth]{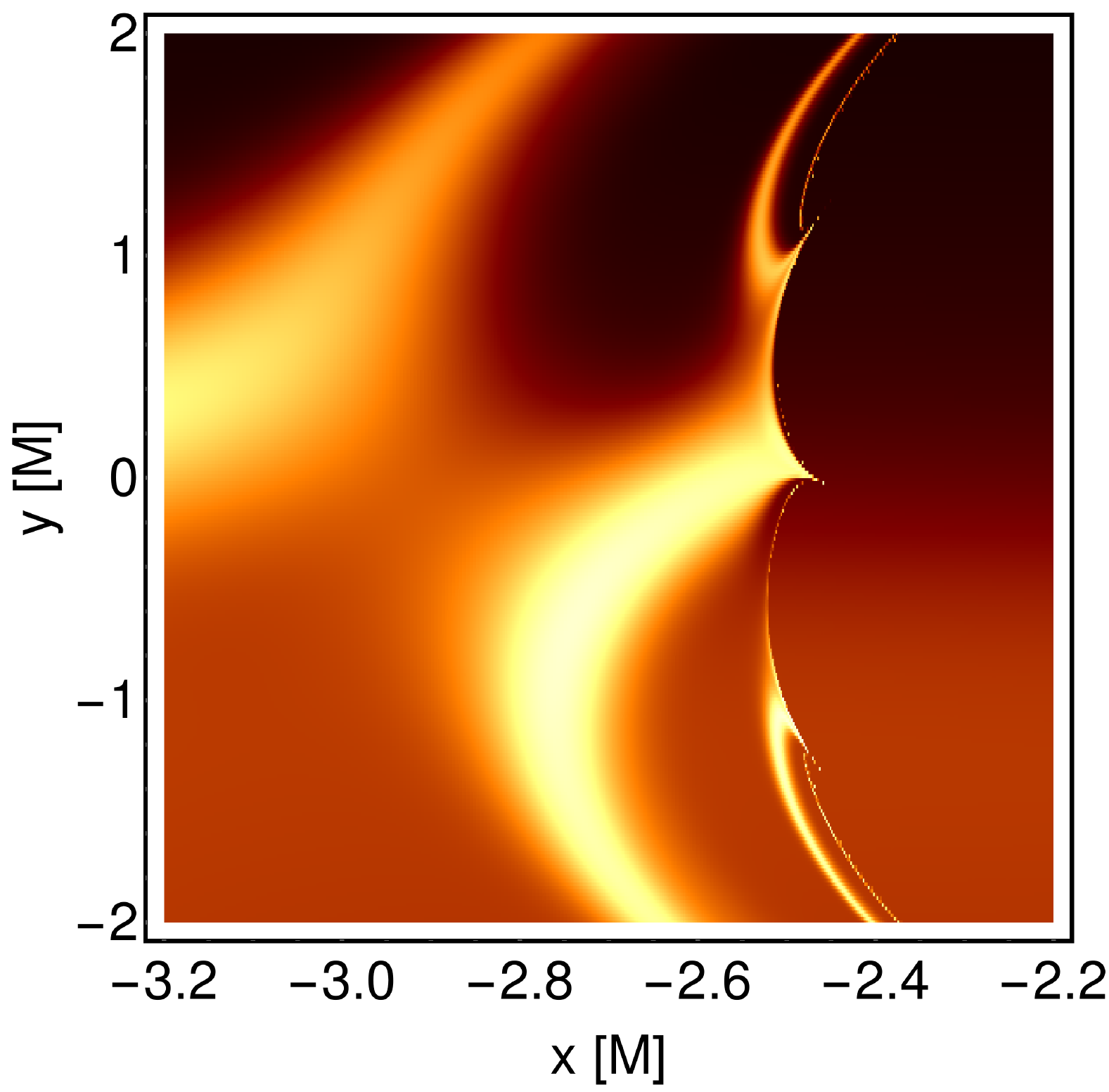}
\caption{\label{fig:image} 
We compare a detailed view of the prograde shadow boundary of a Kerr black hole (left-hand panel) with that of the corresponding RG-improved black hole, cf.~Sec.~\ref{sec:spinning-line-element}. We choose $a=0.9\,M$ and compare the Kerr case ($\xi=0$) with a near-critical deviation parameter $\xi=0.131$. The latter value is chosen such that the differences between both images are maximized. In both cases, the black hole is viewed at near-edge-on inclination $\theta_\text{obs}=9\pi/20$, where $\theta_\text{obs}$ is the angle between the black-hole spin axis and the vector pointing towards the observer, and the emission arises from an a geometrically thin disk model, cf.~\cite[`slow model' in Tab.~1]{Eichhorn:2022bbn}.
}
\end{figure}

Going beyond spherical symmetry, additional potential signatures in the shadow image are expected if $\xi$ is large. This expectation arises because idealized calculations of the shadow boundary show a deviation of its shape from the corresponding shape in the Kerr spacetime, see Fig.~\ref{fig:image}.
This deviation is due to the same physics which cause a ``dent" in the horizon: in the equatorial plane, all special surfaces of the black hole (horizon, ergosphere, photon shell) experience the highest increase in compactness in the equatorial plane. There is therefore a ``dent" in all these surfaces. In turn, a ``dent" in the photon shell is most visible on the prograde image side, where frame dragging causes geodesics to approach the event horizon more closely than on the retrograde side.
In addition, such a ``dent" implies that, if the inclination is not edge on (i.e., orthogonal to the spin axis), the shadow boundary is not reflection symmetric about the horizontal axis through the image.
These features are tied to the angular dependence in $G_N(r,\theta)$ and thus to the local choice of scale identification, cf.~Sec.~\ref{sec:spinning-line-element}, which violates the circularity conditions. They are not present for nonlocal RG-improvements which result in $G_N(r)$.

A final effect relates to the photon rings, i.e., the higher-order lensed images of the accretion disk. These higher-order images approach the shadow boundary, with the more highly lensed images found further inwards. The distance between these photon rings depends on the spacetime and on the astrophysics, i.e., characteristics of the accretion disk. However, for a given accretion disk, the photon rings are more separated, the larger $\xi$ is. This can be inferred, because higher-order rings are generated by null geodesics which experience a larger effective mass. It can be confirmed by explicitly calculating simulated images for a given model of an accretion disk, as, e.g., in \cite{Eichhorn:2021etc}. This effect is present for both the local and the nonlocal RG-improvement since it arises from the overall increase in compactness.
\\

Spinning black holes also give access to much lower values of $\xi$: For the spin parameter close to criticality, $|a| \lesssim M$, the two horizons of the Kerr spacetime are very close to each other. In GR, they disappear at $|a|=M$, upon which a naked singularity is left behind. Thus, no physical process can spin up a black hole in GR to $|a| = M$, unless cosmic censorship in GR is violated.\\
In the RG-improved black hole, the two horizons disappear at $|a| = M \left(1-\delta a \right)$, where 
$\delta a= \frac{\xi}{2} \frac{G_0}{M^2} + \mathcal{O}(\xi^2)$
\cite{Eichhorn:2022bbn}. Upon the disappearance of the horizons, a non-singular spacetime is left behind. Thus, there is no restriction on $|a|$ from a generalized cosmic censorship conjecture.\footnote{Whether or not a regular black hole can be overspun is not settled; see the discussion and references in \cite{Eichhorn:2022bbn}.} The image of such a spacetime contains the ``standard" photon rings which also characterize images of black holes. In addition, it contains a series of inner photon rings, which are constituted by geodesics which are blocked by the horizon in the images of black holes. This additional intensity in the inner region of black hole images may be detectable by the next-generation Event Horizon Telescope, \cite{Eichhorn:2022fcl}.
\\

Additional observational constraints on $\xi$ arise from spectroscopy. Emission from the vicinity of the black hole is redshifted on its way to a far-away observer. The gravitational redshifts depends on the spacetime. Therefore, the gravitational redshift of spectral lines can be used to constrain deviations of the spacetime from the Kerr spacetime. Just like for images of black holes, these accretion-disk spectra depend not just on spacetime-properties but also on astrophysics, i.e., properties of the accretion disk. Therefore, constraints on deviations from GR also depend on assumptions about the accretion disks.
\\
For RG-improved black holes, this has been studied in \cite{Zhang:2018xzj,Zhou:2020eth}. X-ray spectroscopy is possible, e.g., for known X-ray binaries. Because their masses are lower than that of the supermassive black holes observed by the EHT, the resulting constraints on $\xi$ are stronger and \cite{Zhou:2020eth} obtains $\xi \simeq 10^{77}$.
\\

\subsection{Towards constraints from gravitational-wave signatures}
\label{sec:gravobs}

\emph{...where we discuss what it would take to constrain asymptotic-safety inspired black holes from gravitational-wave observations.}
\\

In contrast to the previous section, \ref{sec:emobs}, constraining gravity theories from gravitational waves, e.g., through LIGO observations, or future gravitational-wave interferometers as well as pulsar-timing arrays, requires knowledge of a dynamics, not just of a background spacetime. For asymptotically safe gravity, the full dynamics is currently unknown. Nevertheless, one can make contact with LIGO observations, specifically the ringdown phase of a merger signal, by making a series of assumptions. First, one assumes that quasinormal mode (QNM) frequencies deviate from GR as a function of $\xi$, which controls the scale of quantum gravity. Second, one assumes that QNM frequencies for tensor modes and the eigenfrequencies of scalar (or Dirac or vector) field oscillations have a similar dependence on $\xi$. Third, one assumes that the relevant equation to determine the eigenfrequencies for a scalar field is the Klein-Gordon-equation on the background of an asymptotic-safety inspired black hole. The second and third assumption place strong constraints on the underlying dynamics; e.g., it is assumed that scalar fields do not have sizable nonminimal couplings. One may well question whether these strong assumptions are justified. We leave it as an important future line of research to calculate QNM frequencies from the full dynamics in asymptotically safe gravity in order to check whether the assumptions were justified.

Under these assumptions, the dependence of the QNM spectrum on $\xi$ was studied in \cite{Liu:2012ee,Li:2013kkb,Rincon:2020iwy,Konoplya:2022hll}. The RG improvement of the spacetime results in higher real part of the fundamental mode, but lower imaginary part. The oscillations are therefore less dampened, i.e., the RG improved black hole is a better oscillator than its classical counterpart. The size of the deviations depends on $\xi$ and on the mass, and increases with decreasing mass. For the smallest black holes\footnote{For smaller mass, there is no longer an event horizon.}, for which $M \sim 3.5 M_{\rm Planck}$, the deviation is found to be about 20\%.
In \cite{Konoplya:2022hll}, it was further found that while the fundamental mode only deviates very slightly from its value in Schwarzschild spacetimes for astrophysical black holes, larger deviations may occur for overtones.

Going beyond the study of just scalar-field perturbations, one could take the corresponding equations for tensor perturbations, the Regge-Wheeler-equation and the Zerilli equation, and solve for the quasinormal-mode frequencies on the background spacetime of an asymptotic-safety inspired black hole. However, proceeding in this way relies on an assumption that one already knows to be false, namely that the perturbation equations are the same as in GR. This is known to be incorrect, because higher-order curvature terms are known to exist. In the presence of higher-order curvature terms, the equations of motion may still be recast in the form of the Einstein equations, with an effective energy-momentum tensor that contains the higher-order terms. However, for quasinormal modes, the second variation of the action is required, i.e., perturbations on top of the equations of motion. Thus, it is no longer equivalent to work with the full equations of motion or to work with the Einstein equations plus an effective energy-momentum tensor. The lack of knowledge about the full dynamics is therefore a major obstacle to make contact with gravitational wave observations.

\section{Generalization beyond asymptotic safety: The principled-parameterized approach to black holes}\label{sec:principledparameterized}
\emph{...where we generalize from RG improved black-hole spacetimes to families of black holes which incorporate a set of fundamental principles. They arise within the principled-parameterized approach to black holes which aims to connect fundamental principles to properties of black-hole spacetimes, while also being general enough to cover whole classes of theories beyond GR.}
\\

The method of RG improvement can be generalized, resulting in the principled-parameterized approach to black holes, introduced in \cite{Eichhorn:2021iwq,Eichhorn:2021etc}. In this approach, fundamental principles are built into phenomenological models of black-hole spacetimes. For instance, in the case of asymptotically safe gravity, these fundamental principles are quantum scale symmetry for the couplings and a locality principle in the RG improvement. By generalizing the approach beyond asymptotic safety, we obtain families of black-hole spacetimes, with free parameters or even free functions.
\\

This principled-parameterized approach lies inbetween two other approaches to black-hole spacetimes beyond GR, and unites their strengths.
\\
First, a principled approach starts from a specific theory beyond GR, in which black-hole solutions are computed. Their properties, e.g., with regards to their images, are then calculated and can ultimately be compared with data. As an advantage, this approach connects the fundamental principles of a specific theory directly with properties of the black-hole spacetimes. As a disadvantage, a comprehensive exploration of black-hole spacetimes beyond GR is in practise impossible, given how difficult it is to find rotating solutions to theories beyond GR.
\\
Second, a parameterized approach starts from the most general parameterization of black-hole spacetimes. Their properties can then be computed, in some cases with the general functions that parameterize the spacetime, in others, for specific choices. As an advantage, this approach is comprehensive and can in principle cover any theory beyond GR that has a black-hole spacetime as a solution. As a disadvantage, the connection to fundamental principles is lost.\\

The principled-parameterized approach, proposed in \cite{Eichhorn:2021iwq,Eichhorn:2021etc} and further developed in \cite{Delaporte:2022acp,Eichhorn:2022oma}, starts from fundamental principles which are built into black-hole spacetimes at a heuristic level, such that families of spacetimes, parameterized by free functions and/or free parameters, satisfy these principles. Thereby, a connection between properties of spacetimes and fundamental principles is kept. At the same time, the approach is more comprehensive, in principle covering whole classes of theories at once.\\

As an example, a useful set of principles may be:
\begin{enumerate}
\item[(i)] \emph{Regularity}: Gravity is weakened at high curvatures, such that curvature singularities are resolved.
\item[(ii)] \emph{Locality}: Spacetimes are modified locally, such that the size of deviations from the Kerr spacetime depends on the size of curvature invariants in the Kerr spacetime.
\item[(iii)] \emph{Simplicity}: The modifications of the spacetime accommodate (i) and (ii) in the simplest possible way, such that no more than a single new-physics scale is introduced.
\end{enumerate}
The first principle is expected to cover quantum gravity theories, but may also be demanded of classical modifications of gravity. The second principle arises from an effective-field-theory approach to beyond-GR theories, where modifications to the dynamics are given by higher powers in curvature and therefore the Kerr spacetime is expected to remain an (approximate) solution at sufficiently low curvature scales, but large deviations are expected, whenever the curvature scale in the Kerr spacetime becomes large. Finally, the third principle is a requirement on the simplicity of the underlying theory, which should come with a single new-physics scale which determines the size of all effects.

Asymptotically safe gravity may be expected to respect all three principles, which is why the RG improvement procedure results in a specific example of a black-hole metric respecting these principles. The piece of information that is specific to asymptotic safety is the function $G(k)$, which becomes $G(r, \chi)$ by the RG improvement procedure. Going beyond asymptotic safety, the three principles require a modification of the strength of gravity and can thereby be incorporated by functions $G(r, \chi)$. Alternatively, in order to stay in a system of units where $G=1$, one can upgrade the mass parameter to a mass function, $M(r, \chi)$. These two descriptions are equivalent, because it is the product $G\cdot M$ that enters the classical Kerr metric.

To properly implement the locality principle, the use of horizon-penetrating coordinates is best. The reason is that in Boyer-Lindquist coordinates, which are otherwise a popular choice for modifications of the Kerr spacetime, the upgrade $M \rightarrow M(r)$ is usually viable, but the upgrade $M \rightarrow M(r, \chi)$ generically leads to divergences in curvature invariants at the location of the classical event horizon. The reason for these curvature divergences is the coordinate divergence at this location in the classical spacetime. To disappear at the level of curvature invariants, a delicate cancellation between the various divergent terms in the metric and its derivates must occur. This delicate cancellation is disturbed by the upgrade $M \rightarrow M(r, \chi)$.

The locality principle also generically leads to non-circularity of the spacetime, simply because the circularity conditions Eq.~\eqref{eq:circular} do not generically hold if $M = M(r, \chi)$.When one gives up the locality principle, one can, e.g., use the value of the Kretschmann scalar in the equatorial plane to determine the size of deviations from the Kerr spacetime at all angles. Then, the spacetime is generically circular and can therefore be transformed into Boyer-Lindquist coordinates by the standard transformation. Thus, such spacetimes are captured by the parameterization in \cite{Konoplya:2016jvv}.\\
Such non-local modifications are generically what arises when the Janis-Newman algorithm is used to construct axisymmetric counterparts of spherically symmetric black holes. Black-hole spacetimes constructed through the Janis-Newman algorithm \footnote{For which we are not aware of a reason why it should work beyond GR in the sense of upgrading a spherically symmetric \emph{solution} of the theory to an axisymmetric \emph{solution}; instead, beyond GR, it is just one particular way of reducing spherical symmetry to axisymmetry.} therefore generically appear to be in conflict with the locality principle.
\\

In turn, non-circularity means that popular parameterizations of spacetimes beyond GR \cite{1979GReGr..10...79B,Johannsen:2011dh,Johannsen:2013szh,Cardoso:2014rha,Konoplya:2016jvv} do not describe all such black holes, because they all make the assumption of circularity.\footnote{It is typically not made explicit.} Instead, the use of Boyer-Lindquist-type coordinates (which the above parameterizations typically use) requires the introduction of additional metric coefficients \cite{Delaporte:2022acp}.
\\

Finally, in \cite{Eichhorn:2021iwq,Eichhorn:2021etc} it was observed that the angular dependence of the mass function results in characteristic features of shadows of black holes, cf.~also Fig.~\ref{fig:image}. While it is an open question whether these features can also be produced through spacetimes that do not obey the locality principle, it is clear that local modifications that also obey regularity and simplicity, generically lead to such features. This is an example how the principled-parameterized approach establishes a (not necessarily one-to-one) connection between principles of fundamental physics and (in principle observable) image features of black-hole spacetimes. Asymptotically safe gravity is one example of a specific theory that is expected to obey a particular set of principles and therefore constitutes one example of a theory, which may be informed through observational constraints placed on black-hole spacetimes in the principled-parameterized approach.

\section{Summary, challenges and open questions}\label{sec:summary}
Asymptotic safety realizes a weakening of gravity at high scales, because it realizes scale symmetry in the UV. In turn, this implies that the Planck scale in the sense of the onset of a strong-coupling gravity becomes a low-energy ``mirage": the Planck scale is in fact the transition scale to the scaling regime, in which dimensionful quantities scale according to the canonical dimension. Thereby the Newton coupling scales quadratically towards zero, as the energy scale is increased.

This behavior is the basis for RG improved black holes. The RG improvement procedure posits that one can capture the leading quantum effects in a classical system by (i) substituting coupling constants by their scale dependent counterparts and (ii) identifying a suitable physical scale they depend on.\\
The procedure has been implemented for the Schwarzschild spacetime, the Kerr spacetime, and several spacetimes that model gravitational collapse. Ambiguities in the procedure concern mainly step (ii), because a suitable scale to describe the onset of quantum effects in black holes may be either a curvature scale or a geodesic distance in vacuum spacetimes, or the matter density in non-vacuum spacetimes. Despite these ambiguities, there is universality, namely in the resolution of curvature singularities for vacuum black holes and their weakening or resolution for gravitational-collapse spacetimes. Further, all physically distinct surfaces that characterize black-hole spacetimes (e.g., horizon, photon sphere, ergosphere) are more compact than in the classical case.

We argue that to most accurately model quantum effects in the Kerr spacetime, the angular dependence of curvature invariants has to be accounted for. This leads to spacetimes which fall outside popular parameterizations of spacetimes beyond GR \cite{Delaporte:2022acp}, because they do not feature a generalized Carter constant and break circularity.

Next, we address the -- critical! -- link to observations. We challenge the established wisdom that quantum-gravity effects are necessarily confined to the Planck length $\ell_{\rm Planck} \approx 10^{-35}\, \rm m$. Our challenge is based on the fact that the Planck length is the result of a simple dimensional estimate, which uses no dynamical information on quantum gravity whatsoever. It is further based on the assumption that quantum gravity only has one scale, and that this scale is ``natural", i.e., differs from $\ell_{\rm Planck}$ only by factors of order 1.
\\
In particular, in asymptotically safe quantum gravity, there are several free parameters, one of which is linked to higher-curvature terms. The free parameter can be used to set the coupling for these terms and might make them important already at significantly sub-planckian curvature scales.
\\
Given these considerations, the following strategy to probe quantum gravity is a promising one:
\begin{enumerate}
\item[a)] We remain agnostic and conscious that the expectation of quantum-gravity effects not above $\ell_{\rm Planck}\approx 10^{-35}\, \rm m$ is a purely theoretical one. We therefore keep the quantum-gravity scale as a free parameter and constrain it from observations as best possible. Generically, these constraints concern length scales much above the Planck scale. One may therefore also regard these constrains as constraints on classical modifications of GR.
\item[b)] We remain conservative in our theoretical assumptions and search for settings in which quantum-gravity effects at $\ell_{\rm Planck}\approx 10^{-35}\, \rm m$ results in observable effects. We find that the spin can serve as a lever arm to increase the imprint of quantum gravity. 
\end{enumerate}
 
For part a) of the strategy, the use of electromagnetic signatures is currently most advanced for asymptotic-safety inspired black holes. Both EHT observations as well as X-ray reflection spectroscopy of accretion disks can set limits on the quantum-gravity scale. These limits are, as is to be expected, very high above $\ell_{\rm Planck}$. The study of quasinormal modes still faces significant challenges, most importantly because it requires knowledge of the gravitational dynamics beyond GR, and the use of the perturbation equations from GR with an asymptotic-safety inspired black hole as a background is at best an approximation, but not guaranteed to be viable.
 
For part b) of the strategy, the next-generation Event Horizon Telescope may be in a position to discover whether a horizonless spacetime, achieved through a combination of Planck-scale effects with high-spin-effects, is a viable description of M$87^{\ast}$ and/or Sgr $A^{\ast}$. \\
 
The key outstanding question in this line of research is surely the derivation of a black-hole solution from the dynamical equations of asymptotically safe quantum gravity. These are currently not available, because the full effective action in asymptotically safe gravity has not yet been calculated. \\
A first step in this direction has recently been undertaken in \cite{Knorr:2022kqp}, where possible effective actions are constrained by demanding that certain regular black holes constitute a solution. Simultaneously, one can start by including the leading-order terms in the effective action, which are curvature-squared terms. While those are universally expected in quantum gravity, the values of the couplings are expected to satisfy a relation in asymptotic safety \cite{Eichhorn:2017egq,Pereira:2019dbn,Bonanno:2020bil,Reichert:2020mja,Pawlowski:2020qer,Eichhorn:2022jqj}. Therefore, studying black holes in curvature-squared gravity \cite{Lu:2015psa,Lu:2015cqa,Pravda:2016fue,Lu:2017kzi,Podolsky:2019gro,Held:2021pht,Daas:2022iid,Held:2022abx} may provide information on asymptotically safe gravity.
\\

RG-improved black holes also constitute one example within a broader line of research, namely the \emph{principled-parameterized} approach to black holes. In this approach, phenomenological models are constructed from classical spacetimes, by including modifications based on fundamental principles. For instance, RG-improved black holes satisfy regularity (i.e., no curvature singularities), locality (i.e., deviations from GR are parameterized by the size of the local curvature), simplicity (i.e., there is a single scale which sets deviations from GR). More broadly, modifications of GR may satisfy similar or different sets of principles, which can be incorporated in phenomenological models. In turn, one can constrain many such phenomenological models by observations.\\
Because the current status of black holes in many theories of quantum gravity is unsettled, with very few rigorous derivations of black-hole spacetimes from theories of quantum gravity (or, in the spinning case, even classical modifications of gravity), such a principled-parameterized approach is a promising strategy to make contact with observations. This can also guide observational searches, because one can extract where observational signatures may be present. Conversely, observational constraints may provide information on whether the fundamental principles we base our quantum-gravity theories on are indeed viable. Phenomenological models such as RG-improved black holes or more broadly black holes in the principled-parameterized approach therefore play a critical role, given the current state-of-the-art of the field.

\begin{acknowledgement}
AE is supported by a research grant (29405) from VILLUM FONDEN.
The work leading to this publication was supported by the PRIME programme of the German Academic Exchange Service (DAAD) with funds from the German Federal Ministry of Education and Research (BMBF). A.~Held acknowledges support by the Deutsche Forschungsgemeinschaft (DFG) under Grant No 406116891 within the Research Training Group RTG 2522/1.
\end{acknowledgement}


\bibliographystyle{utphys_sorted.bst}
\bibliography{bibliography}

\providecommand{\href}[2]{#2}\begingroup\raggedright\begin{thebibliography}{100}

\bibitem{Adeifeoba:2018ydh}
A.~Adeifeoba, A.~Eichhorn, and A.~Platania, ``{Towards conditions for
  black-hole singularity-resolution in asymptotically safe quantum gravity},''
  \href{http://dx.doi.org/10.1088/1361-6382/aae6ef}{{\em Class. Quant. Grav.}
  {\bfseries 35} no.~22, (2018) 225007},
  \href{http://arxiv.org/abs/1808.03472}{{\ttfamily arXiv:1808.03472 [gr-qc]}}.

\bibitem{1973blho.conf..215B}
J.~M. {Bardeen}, ``{Timelike and null geodesics in the Kerr metric.},'' in {\em
  Black Holes (Les Astres Occlus)}, pp.~215--239.
\newblock Jan., 1973.

\bibitem{Basu:2010nf}
S.~Basu and D.~Mattingly, ``{Asymptotic Safety, Asymptotic Darkness, and the
  hoop conjecture in the extreme UV},''
  \href{http://dx.doi.org/10.1103/PhysRevD.82.124017}{{\em Phys. Rev. D}
  {\bfseries 82} (2010) 124017},
  \href{http://arxiv.org/abs/1006.0718}{{\ttfamily arXiv:1006.0718 [hep-th]}}.

\bibitem{Benedetti:2009rx}
D.~Benedetti, P.~F. Machado, and F.~Saueressig, ``{Asymptotic safety in
  higher-derivative gravity},''
  \href{http://dx.doi.org/10.1142/S0217732309031521}{{\em Mod. Phys. Lett. A}
  {\bfseries 24} (2009) 2233--2241},
  \href{http://arxiv.org/abs/0901.2984}{{\ttfamily arXiv:0901.2984 [hep-th]}}.

\bibitem{1979GReGr..10...79B}
S.~{Benenti} and M.~{Francaviglia}, ``{Remarks on certain separability
  structures and their applications to general relativity},''
  \href{http://dx.doi.org/10.1007/BF00757025}{{\em General Relativity and
  Gravitation} {\bfseries 10} no.~1, (Jan., 1979) 79--92}.

\bibitem{Bonanno:2001hi}
A.~Bonanno and M.~Reuter, ``{Cosmology with selfadjusting vacuum energy density
  from a renormalization group fixed point},''
  \href{http://dx.doi.org/10.1016/S0370-2693(01)01522-2}{{\em Phys. Lett. B}
  {\bfseries 527} (2002) 9--17},
  \href{http://arxiv.org/abs/astro-ph/0106468}{{\ttfamily
  arXiv:astro-ph/0106468}}.

\bibitem{Bonanno:2006eu}
A.~Bonanno and M.~Reuter, ``{Spacetime structure of an evaporating black hole
  in quantum gravity},''
  \href{http://dx.doi.org/10.1103/PhysRevD.73.083005}{{\em Phys. Rev. D}
  {\bfseries 73} (2006) 083005},
  \href{http://arxiv.org/abs/hep-th/0602159}{{\ttfamily arXiv:hep-th/0602159}}.

\bibitem{Bonanno:2020bil}
A.~Bonanno, A.~Eichhorn, H.~Gies, J.~M. Pawlowski, R.~Percacci, M.~Reuter,
  F.~Saueressig, and G.~P. Vacca, ``{Critical reflections on asymptotically
  safe gravity},'' \href{http://dx.doi.org/10.3389/fphy.2020.00269}{{\em Front.
  in Phys.} {\bfseries 8} (2020) 269},
  \href{http://arxiv.org/abs/2004.06810}{{\ttfamily arXiv:2004.06810 [gr-qc]}}.

\bibitem{Bonanno:2020fgp}
A.~Bonanno, A.-P. Khosravi, and F.~Saueressig, ``{Regular black holes with
  stable cores},'' \href{http://dx.doi.org/10.1103/PhysRevD.103.124027}{{\em
  Phys. Rev. D} {\bfseries 103} no.~12, (2021) 124027},
  \href{http://arxiv.org/abs/2010.04226}{{\ttfamily arXiv:2010.04226 [gr-qc]}}.

\bibitem{Bonanno:2022jjp}
A.~Bonanno, A.-P. Khosravi, and F.~Saueressig, ``{Regular evaporating black
  holes with stable cores},'' \href{http://arxiv.org/abs/2209.10612}{{\ttfamily
  arXiv:2209.10612 [gr-qc]}}.

\bibitem{Bonanno:2016dyv}
A.~Bonanno, B.~Koch, and A.~Platania, ``{Cosmic Censorship in Quantum Einstein
  Gravity},'' \href{http://dx.doi.org/10.1088/1361-6382/aa6788}{{\em Class.
  Quant. Grav.} {\bfseries 34} no.~9, (2017) 095012},
  \href{http://arxiv.org/abs/1610.05299}{{\ttfamily arXiv:1610.05299 [gr-qc]}}.

\bibitem{Bonanno:2017zen}
A.~Bonanno, B.~Koch, and A.~Platania, ``{Gravitational collapse in Quantum
  Einstein Gravity},'' \href{http://dx.doi.org/10.1007/s10701-018-0195-7}{{\em
  Found. Phys.} {\bfseries 48} no.~10, (2018) 1393--1406},
  \href{http://arxiv.org/abs/1710.10845}{{\ttfamily arXiv:1710.10845 [gr-qc]}}.

\bibitem{Bonanno:1998ye}
A.~Bonanno and M.~Reuter, ``{Quantum gravity effects near the null black hole
  singularity},'' \href{http://dx.doi.org/10.1103/PhysRevD.60.084011}{{\em
  Phys. Rev. D} {\bfseries 60} (1999) 084011},
  \href{http://arxiv.org/abs/gr-qc/9811026}{{\ttfamily arXiv:gr-qc/9811026}}.

\bibitem{Bonanno:2000ep}
A.~Bonanno and M.~Reuter, ``{Renormalization group improved black hole
  space-times},'' \href{http://dx.doi.org/10.1103/PhysRevD.62.043008}{{\em
  Phys. Rev. D} {\bfseries 62} (2000) 043008},
  \href{http://arxiv.org/abs/hep-th/0002196}{{\ttfamily arXiv:hep-th/0002196}}.

\bibitem{Bonanno:2022rvo}
A.~Bonanno and F.~Saueressig, ``{Stability properties of Regular Black
  Holes},'' \href{http://arxiv.org/abs/2211.09192}{{\ttfamily arXiv:2211.09192
  [gr-qc]}}.

\bibitem{10.1093/mnras/107.5-6.410}
H.~Bondi, ``{Spherically Symmetrical Models in General Relativity},''
  \href{http://dx.doi.org/10.1093/mnras/107.5-6.410}{{\em Monthly Notices of
  the Royal Astronomical Society} {\bfseries 107} no.~5-6, (12, 1947)
  410--425}.

\bibitem{Borissova:2020knn}
J.~N. Borissova and A.~Eichhorn, ``{Towards black-hole singularity-resolution
  in the Lorentzian gravitational path integral},''
  \href{http://dx.doi.org/10.3390/universe7030048}{{\em Universe} {\bfseries 7}
  no.~3, (2021) 48}, \href{http://arxiv.org/abs/2012.08570}{{\ttfamily
  arXiv:2012.08570 [gr-qc]}}.

\bibitem{Borissova:2022jqj}
J.~N. Borissova, A.~Held, and N.~Afshordi, ``{Scale-Invariance at the Core of
  Quantum Black Holes},'' \href{http://arxiv.org/abs/2203.02559}{{\ttfamily
  arXiv:2203.02559 [gr-qc]}}.

\bibitem{Borissova:2022mgd}
J.~N. Borissova and A.~Platania, ``{Formation and evaporation of quantum black
  holes from the decoupling mechanism in quantum gravity},''
  \href{http://arxiv.org/abs/2210.01138}{{\ttfamily arXiv:2210.01138 [gr-qc]}}.

\bibitem{Broderick:2022tfu}
A.~E. Broderick {\em et~al.}, ``{The Photon Ring in M87*},''
  \href{http://dx.doi.org/10.3847/1538-4357/ac7c1d}{{\em Astrophys. J.}
  {\bfseries 935} (2022) 61}, \href{http://arxiv.org/abs/2208.09004}{{\ttfamily
  arXiv:2208.09004 [astro-ph.HE]}}.

\bibitem{Carballo-Rubio:2018pmi}
R.~Carballo-Rubio, F.~Di~Filippo, S.~Liberati, C.~Pacilio, and M.~Visser, ``{On
  the viability of regular black holes},''
  \href{http://dx.doi.org/10.1007/JHEP07(2018)023}{{\em JHEP} {\bfseries 07}
  (2018) 023}, \href{http://arxiv.org/abs/1805.02675}{{\ttfamily
  arXiv:1805.02675 [gr-qc]}}.

\bibitem{Carballo-Rubio:2021bpr}
R.~Carballo-Rubio, F.~Di~Filippo, S.~Liberati, C.~Pacilio, and M.~Visser,
  ``{Inner horizon instability and the unstable cores of regular black
  holes},'' \href{http://dx.doi.org/10.1007/JHEP05(2021)132}{{\em JHEP}
  {\bfseries 05} (2021) 132}, \href{http://arxiv.org/abs/2101.05006}{{\ttfamily
  arXiv:2101.05006 [gr-qc]}}.

\bibitem{Carballo-Rubio:2022kad}
R.~Carballo-Rubio, F.~Di~Filippo, S.~Liberati, C.~Pacilio, and M.~Visser,
  ``{Regular black holes without mass inflation instability},''
  \href{http://dx.doi.org/10.1007/JHEP09(2022)118}{{\em JHEP} {\bfseries 09}
  (2022) 118}, \href{http://arxiv.org/abs/2205.13556}{{\ttfamily
  arXiv:2205.13556 [gr-qc]}}.

\bibitem{Carballo-Rubio:2019nel}
R.~Carballo-Rubio, F.~Di~Filippo, S.~Liberati, and M.~Visser, ``{Opening the
  Pandora\textquoteright{}s box at the core of black holes},''
  \href{http://dx.doi.org/10.1088/1361-6382/ab8141}{{\em Class. Quant. Grav.}
  {\bfseries 37} no.~14, (2020) 14},
  \href{http://arxiv.org/abs/1908.03261}{{\ttfamily arXiv:1908.03261 [gr-qc]}}.

\bibitem{Cardoso:2014rha}
V.~Cardoso, P.~Pani, and J.~Rico, ``{On generic parametrizations of spinning
  black-hole geometries},''
  \href{http://dx.doi.org/10.1103/PhysRevD.89.064007}{{\em Phys. Rev. D}
  {\bfseries 89} (2014) 064007},
  \href{http://arxiv.org/abs/1401.0528}{{\ttfamily arXiv:1401.0528 [gr-qc]}}.

\bibitem{1991JMP....32.3135C}
J.~{Carminati} and R.~G. {McLenaghan}, ``{Algebraic invariants of the Riemann
  tensor in a four-dimensional Lorentzian space},''
  \href{http://dx.doi.org/10.1063/1.529470}{{\em Journal of Mathematical
  Physics} {\bfseries 32} no.~11, (Nov., 1991) 3135--3140}.

\bibitem{2002nmgm.meet..831C}
J.~{Carminati} and E.~{Zakhary},
  \href{http://dx.doi.org/10.1142/9789812777386\_0081}{``{Algebraic
  Completeness for the Invariants of the Riemann Tensor},''} in {\em The Ninth
  Marcel Grossmann Meeting}, V.~G. {Gurzadyan}, R.~T. {Jantzen}, and
  R.~{Ruffini}, eds., pp.~831--834.
\newblock Dec., 2002.

\bibitem{Casadio:2010fw}
R.~Casadio, S.~D.~H. Hsu, and B.~Mirza, ``{Asymptotic Safety, Singularities,
  and Gravitational Collapse},''
  \href{http://dx.doi.org/10.1016/j.physletb.2010.10.060}{{\em Phys. Lett. B}
  {\bfseries 695} (2011) 317--319},
  \href{http://arxiv.org/abs/1008.2768}{{\ttfamily arXiv:1008.2768 [gr-qc]}}.

\bibitem{Chen:2014jwq}
P.~Chen, Y.~C. Ong, and D.-h. Yeom, ``{Black Hole Remnants and the Information
  Loss Paradox},'' \href{http://dx.doi.org/10.1016/j.physrep.2015.10.007}{{\em
  Phys. Rept.} {\bfseries 603} (2015) 1--45},
  \href{http://arxiv.org/abs/1412.8366}{{\ttfamily arXiv:1412.8366 [gr-qc]}}.

\bibitem{Choptuik:1992jv}
M.~W. Choptuik, ``{Universality and scaling in gravitational collapse of a
  massless scalar field},''
  \href{http://dx.doi.org/10.1103/PhysRevLett.70.9}{{\em Phys. Rev. Lett.}
  {\bfseries 70} (1993) 9--12}.

\bibitem{Christodoulou:1999a}
D.~Christodoulou, ``The instability of naked singularities in the gravitational
  collapse of a scalar field,'' {\em Annals of Mathematics} {\bfseries 149}
  no.~1, (1999) 183--217. \url{http://www.jstor.org/stable/121023}.

\bibitem{Christodoulou:1999}
D.~Christodoulou, ``On the global initial value problem and the issue of
  singularities,'' \href{http://dx.doi.org/10.1088/0264-9381/16/12A/302}{{\em
  Classical and Quantum Gravity} {\bfseries 16} no.~12A, (Dec, 1999) A23}.
  \url{https://dx.doi.org/10.1088/0264-9381/16/12A/302}.

\bibitem{Coleman:1973jx}
S.~R. Coleman and E.~J. Weinberg, ``{Radiative Corrections as the Origin of
  Spontaneous Symmetry Breaking},''
  \href{http://dx.doi.org/10.1103/PhysRevD.7.1888}{{\em Phys. Rev. D}
  {\bfseries 7} (1973) 1888--1910}.

\bibitem{Daas:2022iid}
J.~Daas, K.~Kuijpers, F.~Saueressig, M.~F. Wondrak, and H.~Falcke, ``{Probing
  Quadratic Gravity with the Event Horizon Telescope},''
  \href{http://arxiv.org/abs/2204.08480}{{\ttfamily arXiv:2204.08480 [gr-qc]}}.

\bibitem{Daum:2010qt}
J.~E. Daum and M.~Reuter, ``{Renormalization Group Flow of the Holst Action},''
  \href{http://dx.doi.org/10.1016/j.physletb.2012.01.046}{{\em Phys. Lett. B}
  {\bfseries 710} (2012) 215--218},
  \href{http://arxiv.org/abs/1012.4280}{{\ttfamily arXiv:1012.4280 [hep-th]}}.

\bibitem{deAlwis:2019aud}
S.~de~Alwis, A.~Eichhorn, A.~Held, J.~M. Pawlowski, M.~Schiffer, and
  F.~Versteegen, ``{Asymptotic safety, string theory and the weak gravity
  conjecture},'' \href{http://dx.doi.org/10.1016/j.physletb.2019.134991}{{\em
  Phys. Lett. B} {\bfseries 798} (2019) 134991},
  \href{http://arxiv.org/abs/1907.07894}{{\ttfamily arXiv:1907.07894
  [hep-th]}}.

\bibitem{Delaporte:2022acp}
H.~Delaporte, A.~Eichhorn, and A.~Held, ``{Parameterizations of black-hole
  spacetimes beyond circularity},''
  \href{http://dx.doi.org/10.1088/1361-6382/ac7027}{{\em Class. Quant. Grav.}
  {\bfseries 39} no.~13, (2022) 134002},
  \href{http://arxiv.org/abs/2203.00105}{{\ttfamily arXiv:2203.00105 [gr-qc]}}.

\bibitem{DiFilippo:2022qkl}
F.~Di~Filippo, R.~Carballo-Rubio, S.~Liberati, C.~Pacilio, and M.~Visser, ``{On
  the Inner Horizon Instability of Non-Singular Black Holes},''
  \href{http://dx.doi.org/10.3390/universe8040204}{{\em Universe} {\bfseries 8}
  no.~4, (2022) 204}, \href{http://arxiv.org/abs/2203.14516}{{\ttfamily
  arXiv:2203.14516 [gr-qc]}}.

\bibitem{Dona:2013qba}
P.~Don\`a, A.~Eichhorn, and R.~Percacci, ``{Matter matters in asymptotically
  safe quantum gravity},''
  \href{http://dx.doi.org/10.1103/PhysRevD.89.084035}{{\em Phys. Rev. D}
  {\bfseries 89} no.~8, (2014) 084035},
  \href{http://arxiv.org/abs/1311.2898}{{\ttfamily arXiv:1311.2898 [hep-th]}}.

\bibitem{Donoghue:1994dn}
J.~F. Donoghue, ``{General relativity as an effective field theory: The leading
  quantum corrections},''
  \href{http://dx.doi.org/10.1103/PhysRevD.50.3874}{{\em Phys. Rev. D}
  {\bfseries 50} (1994) 3874--3888},
  \href{http://arxiv.org/abs/gr-qc/9405057}{{\ttfamily arXiv:gr-qc/9405057}}.

\bibitem{Donoghue:1993eb}
J.~F. Donoghue, ``{Leading quantum correction to the Newtonian potential},''
  \href{http://dx.doi.org/10.1103/PhysRevLett.72.2996}{{\em Phys. Rev. Lett.}
  {\bfseries 72} (1994) 2996--2999},
  \href{http://arxiv.org/abs/gr-qc/9310024}{{\ttfamily arXiv:gr-qc/9310024}}.

\bibitem{Donoghue:2019clr}
J.~F. Donoghue, ``{A Critique of the Asymptotic Safety Program},''
  \href{http://dx.doi.org/10.3389/fphy.2020.00056}{{\em Front. in Phys.}
  {\bfseries 8} (2020) 56}, \href{http://arxiv.org/abs/1911.02967}{{\ttfamily
  arXiv:1911.02967 [hep-th]}}.

\bibitem{Dymnikova:1992ux}
I.~Dymnikova, ``{Vacuum nonsingular black hole},''
  \href{http://dx.doi.org/10.1007/BF00760226}{{\em Gen. Rel. Grav.} {\bfseries
  24} (1992) 235--242}.

\bibitem{East:2012mb}
W.~E. East and F.~Pretorius, ``{Ultrarelativistic black hole formation},''
  \href{http://dx.doi.org/10.1103/PhysRevLett.110.101101}{{\em Phys. Rev.
  Lett.} {\bfseries 110} no.~10, (2013) 101101},
  \href{http://arxiv.org/abs/1210.0443}{{\ttfamily arXiv:1210.0443 [gr-qc]}}.

\bibitem{Eichhorn:2013xr}
A.~Eichhorn, ``{On unimodular quantum gravity},''
  \href{http://dx.doi.org/10.1088/0264-9381/30/11/115016}{{\em Class. Quant.
  Grav.} {\bfseries 30} (2013) 115016},
  \href{http://arxiv.org/abs/1301.0879}{{\ttfamily arXiv:1301.0879 [gr-qc]}}.

\bibitem{Eichhorn:2015bna}
A.~Eichhorn, ``{The Renormalization Group flow of unimodular f(R) gravity},''
  \href{http://dx.doi.org/10.1007/JHEP04(2015)096}{{\em JHEP} {\bfseries 04}
  (2015) 096}, \href{http://arxiv.org/abs/1501.05848}{{\ttfamily
  arXiv:1501.05848 [gr-qc]}}.

\bibitem{Eichhorn:2017egq}
A.~Eichhorn, ``{Status of the asymptotic safety paradigm for quantum gravity
  and matter},'' \href{http://dx.doi.org/10.1007/s10701-018-0196-6}{{\em Found.
  Phys.} {\bfseries 48} no.~10, (2018) 1407--1429},
  \href{http://arxiv.org/abs/1709.03696}{{\ttfamily arXiv:1709.03696 [gr-qc]}}.

\bibitem{Eichhorn:2018yfc}
A.~Eichhorn, ``{An asymptotically safe guide to quantum gravity and matter},''
  \href{http://dx.doi.org/10.3389/fspas.2018.00047}{{\em Front. Astron. Space
  Sci.} {\bfseries 5} (2019) 47},
  \href{http://arxiv.org/abs/1810.07615}{{\ttfamily arXiv:1810.07615
  [hep-th]}}.

\bibitem{Eichhorn:2020mte}
A.~Eichhorn, ``{Asymptotically safe gravity},'' in {\em {57th International
  School of Subnuclear Physics}: {In Search for the Unexpected}}.
\newblock 2, 2020.
\newblock \href{http://arxiv.org/abs/2003.00044}{{\ttfamily arXiv:2003.00044
  [gr-qc]}}.

\bibitem{Eichhorn:2022jqj}
A.~Eichhorn, ``{Status update: Asymptotically safe gravity-matter systems},''
  \href{http://dx.doi.org/10.1393/ncc/i2022-22029-4}{{\em Nuovo Cim. C}
  {\bfseries 45} no.~2, (2022) 29},
  \href{http://arxiv.org/abs/2201.11543}{{\ttfamily arXiv:2201.11543 [gr-qc]}}.

\bibitem{Eichhorn:2022fcl}
A.~Eichhorn, R.~Gold, and A.~Held, ``{Horizonless spacetimes as seen by present
  and next-generation Event Horizon Telescope arrays},''
  \href{http://arxiv.org/abs/2205.14883}{{\ttfamily arXiv:2205.14883
  [astro-ph.HE]}}.

\bibitem{Eichhorn:2021iwq}
A.~Eichhorn and A.~Held, ``{From a locality-principle for new physics to image
  features of regular spinning black holes with disks},''
  \href{http://dx.doi.org/10.1088/1475-7516/2021/05/073}{{\em JCAP} {\bfseries
  05} (2021) 073}, \href{http://arxiv.org/abs/2103.13163}{{\ttfamily
  arXiv:2103.13163 [gr-qc]}}.

\bibitem{Eichhorn:2021etc}
A.~Eichhorn and A.~Held, ``{Image features of spinning regular black holes
  based on a locality principle},''
  \href{http://dx.doi.org/10.1140/epjc/s10052-021-09716-2}{{\em Eur. Phys. J.
  C} {\bfseries 81} no.~10, (2021) 933},
  \href{http://arxiv.org/abs/2103.07473}{{\ttfamily arXiv:2103.07473 [gr-qc]}}.

\bibitem{Eichhorn:2022bbn}
A.~Eichhorn and A.~Held, ``{Quantum gravity lights up spinning black holes},''
  \href{http://arxiv.org/abs/2206.11152}{{\ttfamily arXiv:2206.11152 [gr-qc]}}.

\bibitem{Eichhorn:2022oma}
A.~Eichhorn, A.~Held, and P.-V. Johannsen, ``{Universal signatures of
  singularity-resolving physics in photon rings of black holes and horizonless
  objects},'' \href{http://arxiv.org/abs/2204.02429}{{\ttfamily
  arXiv:2204.02429 [gr-qc]}}.

\bibitem{Eichhorn:2022gku}
A.~Eichhorn and M.~Schiffer, ``{Asymptotic safety of gravity with matter},''
  \href{http://arxiv.org/abs/2212.07456}{{\ttfamily arXiv:2212.07456
  [hep-th]}}.

\bibitem{Falls:2012nd}
K.~Falls and D.~F. Litim, ``{Black hole thermodynamics under the microscope},''
  \href{http://dx.doi.org/10.1103/PhysRevD.89.084002}{{\em Phys. Rev. D}
  {\bfseries 89} (2014) 084002},
  \href{http://arxiv.org/abs/1212.1821}{{\ttfamily arXiv:1212.1821 [gr-qc]}}.

\bibitem{Falls:2010he}
K.~Falls, D.~F. Litim, and A.~Raghuraman, ``{Black Holes and Asymptotically
  Safe Gravity},'' \href{http://dx.doi.org/10.1142/S0217751X12500194}{{\em Int.
  J. Mod. Phys. A} {\bfseries 27} (2012) 1250019},
  \href{http://arxiv.org/abs/1002.0260}{{\ttfamily arXiv:1002.0260 [hep-th]}}.

\bibitem{Falls:2020qhj}
K.~Falls, N.~Ohta, and R.~Percacci, ``{Towards the determination of the
  dimension of the critical surface in asymptotically safe gravity},''
  \href{http://dx.doi.org/10.1016/j.physletb.2020.135773}{{\em Phys. Lett. B}
  {\bfseries 810} (2020) 135773},
  \href{http://arxiv.org/abs/2004.04126}{{\ttfamily arXiv:2004.04126
  [hep-th]}}.

\bibitem{Falls:2018ylp}
K.~G. Falls, D.~F. Litim, and J.~Schr\"oder, ``{Aspects of asymptotic safety
  for quantum gravity},''
  \href{http://dx.doi.org/10.1103/PhysRevD.99.126015}{{\em Phys. Rev. D}
  {\bfseries 99} no.~12, (2019) 126015},
  \href{http://arxiv.org/abs/1810.08550}{{\ttfamily arXiv:1810.08550 [gr-qc]}}.

\bibitem{Fehre:2021eob}
J.~Fehre, D.~F. Litim, J.~M. Pawlowski, and M.~Reichert, ``{Lorentzian quantum
  gravity and the graviton spectral function},''
  \href{http://arxiv.org/abs/2111.13232}{{\ttfamily arXiv:2111.13232
  [hep-th]}}.

\bibitem{Gies:2016con}
H.~Gies, B.~Knorr, S.~Lippoldt, and F.~Saueressig, ``{Gravitational Two-Loop
  Counterterm Is Asymptotically Safe},''
  \href{http://dx.doi.org/10.1103/PhysRevLett.116.211302}{{\em Phys. Rev.
  Lett.} {\bfseries 116} no.~21, (2016) 211302},
  \href{http://arxiv.org/abs/1601.01800}{{\ttfamily arXiv:1601.01800
  [hep-th]}}.

\bibitem{Gies:2022ikv}
H.~Gies and A.~S. Salek, ``{Asymptotically Safe Hilbert-Palatini Gravity in an
  On-Shell Reduction Scheme},''
  \href{http://arxiv.org/abs/2209.10435}{{\ttfamily arXiv:2209.10435
  [hep-th]}}.

\bibitem{Gubitosi:2018gsl}
G.~Gubitosi, R.~Ooijer, C.~Ripken, and F.~Saueressig, ``{Consistent early and
  late time cosmology from the RG flow of gravity},''
  \href{http://dx.doi.org/10.1088/1475-7516/2018/12/004}{{\em JCAP} {\bfseries
  12} (2018) 004}, \href{http://arxiv.org/abs/1806.10147}{{\ttfamily
  arXiv:1806.10147 [hep-th]}}.

\bibitem{Harst:2012ni}
U.~Harst and M.~Reuter, ``{The 'Tetrad only' theory space: Nonperturbative
  renormalization flow and Asymptotic Safety},''
  \href{http://dx.doi.org/10.1007/JHEP05(2012)005}{{\em JHEP} {\bfseries 05}
  (2012) 005}, \href{http://arxiv.org/abs/1203.2158}{{\ttfamily arXiv:1203.2158
  [hep-th]}}.

\bibitem{Hayward:2005gi}
S.~A. Hayward, ``{Formation and evaporation of regular black holes},''
  \href{http://dx.doi.org/10.1103/PhysRevLett.96.031103}{{\em Phys. Rev. Lett.}
  {\bfseries 96} (2006) 031103},
  \href{http://arxiv.org/abs/gr-qc/0506126}{{\ttfamily arXiv:gr-qc/0506126}}.

\bibitem{Held:2021vwd}
A.~Held, ``{Invariant Renormalization-Group improvement},''
  \href{http://arxiv.org/abs/2105.11458}{{\ttfamily arXiv:2105.11458 [gr-qc]}}.

\bibitem{Held:2019xde}
A.~Held, R.~Gold, and A.~Eichhorn, ``{Asymptotic safety casts its shadow},''
  \href{http://dx.doi.org/10.1088/1475-7516/2019/06/029}{{\em JCAP} {\bfseries
  06} (2019) 029}, \href{http://arxiv.org/abs/1904.07133}{{\ttfamily
  arXiv:1904.07133 [gr-qc]}}.

\bibitem{Held:2021pht}
A.~Held and H.~Lim, ``{Nonlinear dynamics of quadratic gravity in spherical
  symmetry},'' \href{http://dx.doi.org/10.1103/PhysRevD.104.084075}{{\em Phys.
  Rev. D} {\bfseries 104} no.~8, (2021) 084075},
  \href{http://arxiv.org/abs/2104.04010}{{\ttfamily arXiv:2104.04010 [gr-qc]}}.

\bibitem{Held:2022abx}
A.~Held and J.~Zhang, ``{Instability of spherically-symmetric black holes in
  Quadratic Gravity},'' \href{http://arxiv.org/abs/2209.01867}{{\ttfamily
  arXiv:2209.01867 [gr-qc]}}.

\bibitem{Hiscock:1980ze}
W.~A. Hiscock, ``{Models of Evaporating Black Holes},''
  \href{http://dx.doi.org/10.1103/PhysRevD.23.2813}{{\em Phys. Rev. D}
  {\bfseries 23} (1981) 2813}.

\bibitem{Jiang:2020mws}
J.~Jiang and Y.~Gao, ``{Investigating the gedanken experiment to destroy the
  event horizon of a regular black hole},''
  \href{http://dx.doi.org/10.1103/PhysRevD.101.084005}{{\em Phys. Rev. D}
  {\bfseries 101} no.~8, (2020) 084005},
  \href{http://arxiv.org/abs/2003.07501}{{\ttfamily arXiv:2003.07501
  [hep-th]}}.

\bibitem{Johannsen:2013szh}
T.~Johannsen, ``{Regular Black Hole Metric with Three Constants of Motion},''
  \href{http://dx.doi.org/10.1103/PhysRevD.88.044002}{{\em Phys. Rev. D}
  {\bfseries 88} no.~4, (2013) 044002},
  \href{http://arxiv.org/abs/1501.02809}{{\ttfamily arXiv:1501.02809 [gr-qc]}}.

\bibitem{Johannsen:2011dh}
T.~Johannsen and D.~Psaltis, ``{A Metric for Rapidly Spinning Black Holes
  Suitable for Strong-Field Tests of the No-Hair Theorem},''
  \href{http://dx.doi.org/10.1103/PhysRevD.83.124015}{{\em Phys. Rev. D}
  {\bfseries 83} (2011) 124015},
  \href{http://arxiv.org/abs/1105.3191}{{\ttfamily arXiv:1105.3191 [gr-qc]}}.

\bibitem{Joshi:2011rlc}
P.~S. Joshi and D.~Malafarina, ``{Recent developments in gravitational collapse
  and spacetime singularities},''
  \href{http://dx.doi.org/10.1142/S0218271811020792}{{\em Int. J. Mod. Phys. D}
  {\bfseries 20} (2011) 2641--2729},
  \href{http://arxiv.org/abs/1201.3660}{{\ttfamily arXiv:1201.3660 [gr-qc]}}.

\bibitem{Knorr:2021slg}
B.~Knorr, ``{The derivative expansion in asymptotically safe quantum gravity:
  general setup and quartic order},''
  \href{http://dx.doi.org/10.21468/SciPostPhysCore.4.3.020}{{\em SciPost Phys.
  Core} {\bfseries 4} (2021) 020},
  \href{http://arxiv.org/abs/2104.11336}{{\ttfamily arXiv:2104.11336
  [hep-th]}}.

\bibitem{Knorr:2022kqp}
B.~Knorr and A.~Platania, ``{Sifting quantum black holes through the principle
  of least action},''
  \href{http://dx.doi.org/10.1103/PhysRevD.106.L021901}{{\em Phys. Rev. D}
  {\bfseries 106} no.~2, (2022) L021901},
  \href{http://arxiv.org/abs/2202.01216}{{\ttfamily arXiv:2202.01216
  [hep-th]}}.

\bibitem{Koch:2013owa}
B.~Koch and F.~Saueressig, ``{Structural aspects of asymptotically safe black
  holes},'' \href{http://dx.doi.org/10.1088/0264-9381/31/1/015006}{{\em Class.
  Quant. Grav.} {\bfseries 31} (2014) 015006},
  \href{http://arxiv.org/abs/1306.1546}{{\ttfamily arXiv:1306.1546 [hep-th]}}.

\bibitem{Kofinas:2015sna}
G.~Kofinas and V.~Zarikas, ``{Avoidance of singularities in asymptotically safe
  Quantum Einstein Gravity},''
  \href{http://dx.doi.org/10.1088/1475-7516/2015/10/069}{{\em JCAP} {\bfseries
  10} (2015) 069}, \href{http://arxiv.org/abs/1506.02965}{{\ttfamily
  arXiv:1506.02965 [hep-th]}}.

\bibitem{Konoplya:2022hll}
R.~A. Konoplya, A.~F. Zinhailo, J.~Kunz, Z.~Stuchlik, and A.~Zhidenko,
  ``{Quasinormal ringing of regular black holes in asymptotically safe gravity:
  the importance of overtones},''
  \href{http://dx.doi.org/10.1088/1475-7516/2022/10/091}{{\em JCAP} {\bfseries
  10} (2022) 091}, \href{http://arxiv.org/abs/2206.14714}{{\ttfamily
  arXiv:2206.14714 [gr-qc]}}.

\bibitem{Konoplya:2016jvv}
R.~Konoplya, L.~Rezzolla, and A.~Zhidenko, ``{General parametrization of
  axisymmetric black holes in metric theories of gravity},''
  \href{http://dx.doi.org/10.1103/PhysRevD.93.064015}{{\em Phys. Rev. D}
  {\bfseries 93} no.~6, (2016) 064015},
  \href{http://arxiv.org/abs/1602.02378}{{\ttfamily arXiv:1602.02378 [gr-qc]}}.

\bibitem{Kumar:2019ohr}
R.~Kumar, B.~P. Singh, and S.~G. Ghosh, ``{Shadow and deflection angle of
  rotating black hole in asymptotically safe gravity},''
  \href{http://dx.doi.org/10.1016/j.aop.2020.168252}{{\em Annals Phys.}
  {\bfseries 420} (2020) 168252},
  \href{http://arxiv.org/abs/1904.07652}{{\ttfamily arXiv:1904.07652 [gr-qc]}}.

\bibitem{Lehners:2019ibe}
J.-L. Lehners and K.~S. Stelle, ``{A Safe Beginning for the Universe?},''
  \href{http://dx.doi.org/10.1103/PhysRevD.100.083540}{{\em Phys. Rev. D}
  {\bfseries 100} no.~8, (2019) 083540},
  \href{http://arxiv.org/abs/1909.01169}{{\ttfamily arXiv:1909.01169
  [hep-th]}}.

\bibitem{1933ASSB...53...51L}
G.~{Lema{\^\i}tre}, ``{L'Univers en expansion},'' {\em Annales de la
  Soci\&eacute;t\&eacute; Scientifique de Bruxelles} {\bfseries 53} (Jan.,
  1933) 51.

\bibitem{Li:2013kkb}
J.~Li and Y.~Zhong, ``{Quasinormal Modes for Electromagnetic Field Perturbation
  of the Asymptotic Safe Black Hole},''
  \href{http://dx.doi.org/10.1007/s10773-012-1476-0}{{\em Int. J. Theor. Phys.}
  {\bfseries 52} (2013) 1583--1587}.

\bibitem{Li:2013sea}
Z.~Li and C.~Bambi, ``{Destroying the event horizon of regular black holes},''
  \href{http://dx.doi.org/10.1103/PhysRevD.87.124022}{{\em Phys. Rev. D}
  {\bfseries 87} no.~12, (2013) 124022},
  \href{http://arxiv.org/abs/1304.6592}{{\ttfamily arXiv:1304.6592 [gr-qc]}}.

\bibitem{Litim:2013gga}
D.~F. Litim and K.~Nikolakopoulos, ``{Quantum gravity effects in Myers-Perry
  space-times},'' \href{http://dx.doi.org/10.1007/JHEP04(2014)021}{{\em JHEP}
  {\bfseries 04} (2014) 021}, \href{http://arxiv.org/abs/1308.5630}{{\ttfamily
  arXiv:1308.5630 [hep-th]}}.

\bibitem{Liu:2012ee}
D.-J. Liu, B.~Yang, Y.-J. Zhai, and X.-Z. Li, ``{Quasinormal modes for
  asymptotic safe black holes},''
  \href{http://dx.doi.org/10.1088/0264-9381/29/14/145009}{{\em Class. Quant.
  Grav.} {\bfseries 29} (2012) 145009},
  \href{http://arxiv.org/abs/1205.4792}{{\ttfamily arXiv:1205.4792 [gr-qc]}}.

\bibitem{Lu:2015cqa}
H.~Lu, A.~Perkins, C.~N. Pope, and K.~S. Stelle, ``{Black Holes in
  Higher-Derivative Gravity},''
  \href{http://dx.doi.org/10.1103/PhysRevLett.114.171601}{{\em Phys. Rev.
  Lett.} {\bfseries 114} no.~17, (2015) 171601},
  \href{http://arxiv.org/abs/1502.01028}{{\ttfamily arXiv:1502.01028
  [hep-th]}}.

\bibitem{Lu:2015psa}
H.~L\"u, A.~Perkins, C.~N. Pope, and K.~S. Stelle, ``{Spherically Symmetric
  Solutions in Higher-Derivative Gravity},''
  \href{http://dx.doi.org/10.1103/PhysRevD.92.124019}{{\em Phys. Rev. D}
  {\bfseries 92} no.~12, (2015) 124019},
  \href{http://arxiv.org/abs/1508.00010}{{\ttfamily arXiv:1508.00010
  [hep-th]}}.

\bibitem{Lu:2017kzi}
H.~L\"u, A.~Perkins, C.~N. Pope, and K.~S. Stelle, ``{Lichnerowicz Modes and
  Black Hole Families in Ricci Quadratic Gravity},''
  \href{http://dx.doi.org/10.1103/PhysRevD.96.046006}{{\em Phys. Rev. D}
  {\bfseries 96} no.~4, (2017) 046006},
  \href{http://arxiv.org/abs/1704.05493}{{\ttfamily arXiv:1704.05493
  [hep-th]}}.

\bibitem{Papapetrou:1966zz}
A.~Papapetrou, ``{Champs gravitationnels stationnaires a symetrie axiale},''
  {\em Ann. Inst. H. Poincare Phys. Theor.} {\bfseries 4} (1966) 83--105.

\bibitem{Pawlowski:2020qer}
J.~M. Pawlowski and M.~Reichert, ``{Quantum Gravity: A Fluctuating Point of
  View},'' \href{http://dx.doi.org/10.3389/fphy.2020.551848}{{\em Front. in
  Phys.} {\bfseries 8} (2021) 551848},
  \href{http://arxiv.org/abs/2007.10353}{{\ttfamily arXiv:2007.10353
  [hep-th]}}.

\bibitem{Pawlowski:2018swz}
J.~M. Pawlowski and D.~Stock, ``{Quantum-improved Schwarzschild-(A)dS and
  Kerr-(A)dS spacetimes},''
  \href{http://dx.doi.org/10.1103/PhysRevD.98.106008}{{\em Phys. Rev. D}
  {\bfseries 98} no.~10, (2018) 106008},
  \href{http://arxiv.org/abs/1807.10512}{{\ttfamily arXiv:1807.10512
  [hep-th]}}.

\bibitem{Penrose:1969pc}
R.~Penrose, ``{Gravitational collapse: The role of general relativity},''
  \href{http://dx.doi.org/10.1023/A:1016578408204}{{\em Riv. Nuovo Cim.}
  {\bfseries 1} (1969) 252--276}.

\bibitem{Penrose:1964wq}
R.~Penrose, ``{Gravitational collapse and space-time singularities},''
  \href{http://dx.doi.org/10.1103/PhysRevLett.14.57}{{\em Phys. Rev. Lett.}
  {\bfseries 14} (1965) 57--59}.

\bibitem{Penrose:1999vj}
R.~Penrose, ``{The question of cosmic censorship},''
  \href{http://dx.doi.org/10.1007/BF02702355}{{\em J. Astrophys. Astron.}
  {\bfseries 20} (1999) 233--248}.

\bibitem{Percacci:2017fkn}
R.~Percacci, \href{http://dx.doi.org/10.1142/10369}{{\em {An Introduction to
  Covariant Quantum Gravity and Asymptotic Safety}}}, vol.~3 of {\em 100 Years
  of General Relativity}.
\newblock World Scientific, 2017.

\bibitem{Pereira:2019dbn}
A.~D. Pereira, ``{Quantum spacetime and the renormalization group: Progress and
  visions},'' in {\em {Progress and Visions in Quantum Theory in View of
  Gravity}: {Bridging foundations of physics and mathematics}}.
\newblock 4, 2019.
\newblock \href{http://arxiv.org/abs/1904.07042}{{\ttfamily arXiv:1904.07042
  [gr-qc]}}.

\bibitem{Platania:2019kyx}
A.~Platania, ``{Dynamical renormalization of black-hole spacetimes},''
  \href{http://dx.doi.org/10.1140/epjc/s10052-019-6990-2}{{\em Eur. Phys. J. C}
  {\bfseries 79} no.~6, (2019) 470},
  \href{http://arxiv.org/abs/1903.10411}{{\ttfamily arXiv:1903.10411 [gr-qc]}}.

\bibitem{Podolsky:2019gro}
J.~Podolsk\'y, R.~\v{S}varc, V.~Pravda, and A.~Pravdova, ``{Black holes and
  other exact spherical solutions in Quadratic Gravity},''
  \href{http://dx.doi.org/10.1103/PhysRevD.101.024027}{{\em Phys. Rev. D}
  {\bfseries 101} no.~2, (2020) 024027},
  \href{http://arxiv.org/abs/1907.00046}{{\ttfamily arXiv:1907.00046 [gr-qc]}}.

\bibitem{Pravda:2016fue}
V.~Pravda, A.~Pravdova, J.~Podolsky, and R.~Svarc, ``{Exact solutions to
  quadratic gravity},''
  \href{http://dx.doi.org/10.1103/PhysRevD.95.084025}{{\em Phys. Rev. D}
  {\bfseries 95} no.~8, (2017) 084025},
  \href{http://arxiv.org/abs/1606.02646}{{\ttfamily arXiv:1606.02646 [gr-qc]}}.

\bibitem{Reichert:2020mja}
M.~Reichert, ``{Lecture notes: Functional Renormalisation Group and
  Asymptotically Safe Quantum Gravity},''
  \href{http://dx.doi.org/10.22323/1.384.0005}{{\em PoS} {\bfseries 384} (2020)
  005}.

\bibitem{Reuter:2010xb}
M.~Reuter and E.~Tuiran, ``{Quantum Gravity Effects in the Kerr Spacetime},''
  \href{http://dx.doi.org/10.1103/PhysRevD.83.044041}{{\em Phys. Rev. D}
  {\bfseries 83} (2011) 044041},
  \href{http://arxiv.org/abs/1009.3528}{{\ttfamily arXiv:1009.3528 [hep-th]}}.

\bibitem{Reuter:2004nx}
M.~Reuter and H.~Weyer, ``{Quantum gravity at astrophysical distances?},''
  \href{http://dx.doi.org/10.1088/1475-7516/2004/12/001}{{\em JCAP} {\bfseries
  12} (2004) 001}, \href{http://arxiv.org/abs/hep-th/0410119}{{\ttfamily
  arXiv:hep-th/0410119}}.

\bibitem{Reuter:2003ca}
M.~Reuter and H.~Weyer, ``{Renormalization group improved gravitational
  actions: A Brans-Dicke approach},''
  \href{http://dx.doi.org/10.1103/PhysRevD.69.104022}{{\em Phys. Rev. D}
  {\bfseries 69} (2004) 104022},
  \href{http://arxiv.org/abs/hep-th/0311196}{{\ttfamily arXiv:hep-th/0311196}}.

\bibitem{Reuter:2019byg}
M.~Reuter and F.~Saueressig, {\em {Quantum Gravity and the Functional
  Renormalization Group}: {The Road towards Asymptotic Safety}}.
\newblock Cambridge University Press, 1, 2019.

\bibitem{Rincon:2020iwy}
A.~Rinc\'on and G.~Panotopoulos, ``{Quasinormal modes of an improved
  Schwarzschild black hole},''
  \href{http://dx.doi.org/10.1016/j.dark.2020.100639}{{\em Phys. Dark Univ.}
  {\bfseries 30} (2020) 100639},
  \href{http://arxiv.org/abs/2006.11889}{{\ttfamily arXiv:2006.11889 [gr-qc]}}.

\bibitem{2003GReGr..35.1909T}
E.~{Teo}, ``{Spherical Photon Orbits Around a Kerr Black Hole},''
  \href{http://dx.doi.org/10.1023/A:1026286607562}{{\em General Relativity and
  Gravitation} {\bfseries 35} no.~11, (Nov., 2003) 1909--1926}.

\bibitem{tolman1934effect}
R.~C. Tolman, ``Effect of inhomogeneity on cosmological models,'' {\em
  Proceedings of the National Academy of Sciences} {\bfseries 20} no.~3, (1934)
  169--176.

\bibitem{Torres:2017gix}
R.~Torres, ``{Non-singular quantum improved rotating black holes and their
  maximal extension},'' \href{http://dx.doi.org/10.1007/s10714-017-2236-5}{{\em
  Gen. Rel. Grav.} {\bfseries 49} no.~6, (2017) 74},
  \href{http://arxiv.org/abs/1702.03567}{{\ttfamily arXiv:1702.03567 [gr-qc]}}.

\bibitem{Torres:2014pea}
R.~Torres and F.~Fayos, ``{Singularity free gravitational collapse in an
  effective dynamical quantum spacetime},''
  \href{http://dx.doi.org/10.1016/j.physletb.2014.04.038}{{\em Phys. Lett. B}
  {\bfseries 733} (2014) 169--175},
  \href{http://arxiv.org/abs/1405.7922}{{\ttfamily arXiv:1405.7922 [gr-qc]}}.

\bibitem{Torres:2014gta}
R.~Torres, ``{Singularity-free gravitational collapse and asymptotic safety},''
  \href{http://dx.doi.org/10.1016/j.physletb.2014.04.010}{{\em Phys. Lett. B}
  {\bfseries 733} (2014) 21--24},
  \href{http://arxiv.org/abs/1404.7655}{{\ttfamily arXiv:1404.7655 [gr-qc]}}.

\bibitem{Torres:2017ygl}
R.~Torres, ``{Nonsingular black holes, the cosmological constant, and
  asymptotic safety},''
  \href{http://dx.doi.org/10.1103/PhysRevD.95.124004}{{\em Phys. Rev. D}
  {\bfseries 95} no.~12, (2017) 124004},
  \href{http://arxiv.org/abs/1703.09997}{{\ttfamily arXiv:1703.09997 [gr-qc]}}.

\bibitem{Torres:2015aga}
R.~Torres and F.~Fayos, ``{On the quantum corrected gravitational collapse},''
  \href{http://dx.doi.org/10.1016/j.physletb.2015.05.078}{{\em Phys. Lett. B}
  {\bfseries 747} (2015) 245--250},
  \href{http://arxiv.org/abs/1503.07407}{{\ttfamily arXiv:1503.07407 [gr-qc]}}.

\bibitem{Uehling:1935}
E.~A. Uehling, ``{Polarization effects in the positron theory},''
  \href{http://dx.doi.org/10.1103/PhysRev.48.55}{{\em Phys. Rev.} {\bfseries
  48} (1935) 55--63}.

\bibitem{Vaidya:1951zz}
P.~Vaidya, ``{The Gravitational Field of a Radiating Star},'' {\em Proc. Natl.
  Inst. Sci. India A} {\bfseries 33} (1951) 264.

\bibitem{Vaidya:1951zza}
P.~C. Vaidya, ``{Nonstatic Solutions of Einstein's Field Equations for Spheres
  of Fluids Radiating Energy},''
  \href{http://dx.doi.org/10.1103/PhysRev.83.10}{{\em Phys. Rev.} {\bfseries
  83} (1951) 10--17}.

\bibitem{Vaidya:1966zza}
P.~C. Vaidya, ``{An Analytical Solution for Gravitational Collapse with
  Radiation},'' \href{http://dx.doi.org/10.1086/148692}{{\em Astrophys. J.}
  {\bfseries 144} (1966) 943}.

\bibitem{Wald:1997wa}
R.~M. Wald,
  \href{http://dx.doi.org/10.1007/978-94-017-0934-7_5}{``{Gravitational
  collapse and cosmic censorship},''} pp.~69--85.
\newblock 10, 1997.
\newblock \href{http://arxiv.org/abs/gr-qc/9710068}{{\ttfamily
  arXiv:gr-qc/9710068}}.

\bibitem{Xie:2021bur}
Y.~Xie, J.~Zhang, H.~O. Silva, C.~de~Rham, H.~Witek, and N.~Yunes, ``{Square
  Peg in a Circular Hole: Choosing the Right Ansatz for Isolated Black Holes in
  Generic Gravitational Theories},''
  \href{http://dx.doi.org/10.1103/PhysRevLett.126.241104}{{\em Phys. Rev.
  Lett.} {\bfseries 126} no.~24, (2021) 241104},
  \href{http://arxiv.org/abs/2103.03925}{{\ttfamily arXiv:2103.03925 [gr-qc]}}.

\bibitem{Yang:2022yvq}
S.-J. Yang, Y.-P. Zhang, S.-W. Wei, and Y.-X. Liu, ``{Destroying the event
  horizon of a nonsingular rotating quantum-corrected black hole},''
  \href{http://dx.doi.org/10.1007/JHEP04(2022)066}{{\em JHEP} {\bfseries 04}
  (2022) 066}, \href{http://arxiv.org/abs/2201.03381}{{\ttfamily
  arXiv:2201.03381 [gr-qc]}}.

\bibitem{1997GReGr..29..539Z}
E.~{Zakhary} and C.~B.~G. {McIntosh}, ``{A Complete Set of Riemann
  Invariants},'' \href{http://dx.doi.org/10.1023/A:1018851201784}{{\em General
  Relativity and Gravitation} {\bfseries 29} no.~5, (May, 1997) 539--581}.

\bibitem{Zhang:2018xzj}
Y.~Zhang, M.~Zhou, and C.~Bambi, ``{Iron line spectroscopy of black holes in
  asymptotically safe gravity},''
  \href{http://dx.doi.org/10.1140/epjc/s10052-018-5875-0}{{\em Eur. Phys. J. C}
  {\bfseries 78} no.~5, (2018) 376},
  \href{http://arxiv.org/abs/1804.07955}{{\ttfamily arXiv:1804.07955 [gr-qc]}}.

\bibitem{Zhou:2020eth}
B.~Zhou, A.~B. Abdikamalov, D.~Ayzenberg, C.~Bambi, S.~Nampalliwar, and
  A.~Tripathi, ``{Shining X-rays on asymptotically safe quantum gravity},''
  \href{http://dx.doi.org/10.1088/1475-7516/2021/01/047}{{\em JCAP} {\bfseries
  01} (2021) 047}, \href{http://arxiv.org/abs/2005.12958}{{\ttfamily
  arXiv:2005.12958 [astro-ph.HE]}}.

\bibitem{Zhou:2022yio}
T.~Zhou and L.~Modesto, ``{Geodesic incompleteness of some popular regular
  black holes},'' \href{http://arxiv.org/abs/2208.02557}{{\ttfamily
  arXiv:2208.02557 [gr-qc]}}.

\end{thebibliography}\endgroup


\end{document}